\newcommand{\pdv}[2]{\frac{\partial #1}{\partial #2}}
\newcommand{\BigD}[1]{\pdv{\vec{#1}}{t}+(\vec{u}\cdot\vec{\nabla})\vec{#1}}

\newcommand{\rot}{\vec{\nabla}\times}

\newcommand{\Tfrac}[2]{\left(\frac{#1}{#2}\right)}
\newcommand{\divi}{\vec{\nabla}\cdot}
\newcommand{\vgrad}[1]{\vec{\nabla}{#1}}

\newcommand{\er}{\vec{e_{r}}}
\newcommand{\ethe}{\vec{e_{\theta}}}
\newcommand*\diff{\mathop{}\!\mathrm{d}}

\DeclareMathAlphabet\mathbfcal{OMS}{cmsy}{b}{n}

\newcommand{\upo}{\vec{u_{p}}}
\newcommand{\Bpo}{\vec{B_{p}}}

\newcommand{\spacio}{\hspace{4pt}}

\newcommand{\mean}[1]{\left\langle #1 \right\rangle}

\newcommand{\vmean}[1]{\left\langle \vec{#1} \right\rangle}

\newcommand{\de}{\delta}
\newcommand{\rhomean}[1]{\left\langle #1 \right\rangle_\rho}
\newcommand{\phimean}[1]{\left\langle #1 \right\rangle_\varphi}
\newcommand{\tdiff}[1]{\frac{1}{\Delta t}\left[ #1 \right]_{t_1}^{t_2}}

\newcommand{\VA}{V_\mathrm{A}}
\newcommand{\VbA}{\bm{V}_\mathrm{A}}
\newcommand{\VAz}{V_{\mathrm{A}z}}
\newcommand{\VAy}{V_{\mathrm{A}\varphi}}
\documentclass{aa}  

\usepackage{graphicx}
\usepackage{upgreek}

\usepackage{txfonts}
\usepackage{color}
\usepackage{bm}
\begin{document}

   \title{Magnetic outflows from turbulent accretion disks}
 \subtitle{I. Vertical structure \& secular evolution}

   \author{J. Jacquemin-Ide
          \and
          G. Lesur
          \and
          J. Ferreira
          }

   \institute{ Univ. Grenoble Alpes, CNRS, IPAG, 38000 Grenoble, France 
             }

   \date{Received September 2020;}

 
  \abstract
   {Astrophysical disks are likely embedded in an ambient vertical magnetic field generated by its environment. This ambient field is known to drive magneto-rotational turbulence in the disk bulk, but it is also responsible for launching magnetised outflows at the origin of astrophysical jets. Yet, the interplay between turbulence and outflows is not understood. In particular, the vertical structure and long-term (secular) evolution of such a system lack quantitative predictions. It is, nevertheless, this secular evolution which is proposed to explain time variability in many accreting systems such as FuOr, X-ray binaries, and novae like systems.}
   {We seek to constraint the structure and long-term evolution of turbulent astrophysical disks subject to magnetised outflows in the non-relativistic regime. More specifically we aim to characterise the mechanism driving accretion, the dynamics of the disk atmosphere, the role played by the outflow, and the long-term evolution of mass and magnetic flux distributions.}
   {We computed and analysed global 3D ideal magnetohydrynamic (MHD) simulations of an accretion disk threaded by a large-scale  magnetic field. We measured the turbulent state of the system by Reynolds averaging the ideal MHD equations and evaluate the role of the turbulent terms in the equilibrium of the system. We then computed the transport of mass, angular momentum, and magnetic fields in the disk to characterise its secular evolution. Finally, we performed a parameter exploration survey in order to characterise how the transport properties depend on the disk properties.}
   {We find that weakly magnetised disks drive jets that carry a small fraction of the disk angular momentum away. The mass-weighted accretion speed remains subsonic, although there is always an upper turbulent atmospheric region where transsonic accretion takes place. We {show} that this turbulence is driven by a strongly magnetised version of the magneto-rotational instability. The internal disk structure therefore appears drastically different from the conventional hydrostatic picture. {We expect that the turbulent atmosphere region will lead to non-thermal features in the emission spectra from compact objects}.
   In addition, we show that the disk is subject to a secular viscous-type instability, which leads to the formation of long-lived ring-like structures in the disk surface density distribution. This instability is likely connected to the magnetic field transport. 
   Finally, we show that for all of the parameters explored, the ambient magnetic field is always dragged {inward} in the disk at a velocity which increases with the disk magnetisation. Beyond a threshold on the latter, the disk undergoes a profound radial readjustment. It leads to the formation of an inner accretion-ejection region with a supersonic mass-weighted accretion speed and where the magnetic field distribution becomes steady and reaches a magnitude near equipartition with the thermal pressure.  This inner structure shares many properties with the jet emitting disk (JED) model. Overall, these results pave the way for quantitative self-consistent secular disk models.}
  {}


   \maketitle
%

\section{Introduction}

The emission of jet-like outflows from accretion disks is ubiquitous in the Universe, as both jets and disks are observed around a variety of astrophysical objects and on a wide range of spatial scales.  Jet-like outflows are observed being emitted from active galactic nuclei (AGN) and quasar (\citealt{merloni_fundamental_2003} and references therein; \citealt{blandford2019} ), X-ray binaries \citep{mirabel_sources_1999,corbel2000,gallo_universal_2003,gallo_black_2005}, and possibly also nova-like cataclysmic variables \citep{coppejans_novalike_2015} and protoplanetary disks \citep{Burrows1996,hirth_spatial_1997,ray_hst_1996,dougados_t_2000,bally_observations_2007}. All of these jet-like outflows exhibit several common properties: They are highly collimated, supersonic and highly correlated with the measured accretion rate in their emitting disk \citep{cabrit_forbidden-line_1990,hartigan_disk_1995,serjeant_radio-optical_1998,markoff_exploring_2003,ferreira_which_2006}. This shows that accretion and ejection are intrinsically linked. Recently, an additional component consisting of a more massive and slower outflow, called winds,  have been observed around the same objects: AGN \citep{Cicone2014},  X-ray binaries \citep{tetarenko_strong_2018,jimenez-ibarra2019}, and proto-planetary disks \citep{tabo2017,devalon2020}. Whether winds and jets share a common source is still an open debate.

The question of accretion in astrophysical disks is tightly linked to that of angular momentum transport: Mass accretion is allowed only if angular momentum is removed from the accreted material.
In this framework, it is customary to separate two mechanisms through which accretion can occur: Turbulent torques,  which transport angular momentum radially in the disk bulk \citep{shak73}, and laminar torques, usually associated with outflows  which carry angular momentum vertically away from the disk \citep{blan82}. In the simple hypothesis of a Keplerian disk embedded in a large-scale vertical magnetic field, both mechanisms naturally emerge. First, the seminal work of \cite{balb91} showed that the magneto-rotationnal instability (MRI) is triggered, leading to vigorous magnetohydrodynamic (MHD) turbulence and radial transport of angular momentum. Second, large scale magnetic fields spontaneously lead to the launching of an outflow \citep{ward93,ferr95,konigl_disk_2000,Ferreira_theory_2002,Pudritz07}, which carries angular momentum away very efficiently. Historically, these two processes have been studied separately, but it is now becoming clear that these mechanisms are interdependent and can even be considered as two faces of the same basic physical process \citep{lesu13}.

The launching of outflows has been studied using self-similar models \citep{ferr95,Li1995,ferr97} or simplified 2.5D simulations \citep{murp10,Stepa2016}. These effective models generically rely on azimuthally-averaged equations of motions, in which turbulence is modelled by an effective viscosity and resistivity. While they have been successful in elucidating the constraints needed for steady ejection \citep{ferr95} or the relation between disk structure and jet properties \citep{ferr97,Stepa2016,jacquemin-ide_magnetically-driven_2019}; their use of {ad hoc} turbulent profile limits their predictive power as they are dependent on the prescription used to model turbulence (see \citealt{jacquemin-ide_magnetically-driven_2019} for a discussion).

MRI-driven turbulence has been studied using both shearing box simulations \citep{hawl95} and more recently global simulations \citep{sorathia10}. Most of these simulations focused on the configuration where there is no mean field threading the disk (i.e. no external source of magnetic field), thereby excluding the possibility of a magnetically-driven outflow. In these 'zero net flux simulations', it is found that the MRI drives 3D MHD turbulence, which results in an efficient radial angular momentum transport and therefore accretion. It is only recently that, thanks to robust numerical methods and more abundant numerical resources, stratified shearing box models with a mean field have been developed. These models show both turbulent radial torques {and} magnetically-driven outflows \citep{Suzuki2009,from13,bai13}. These models however suffer from several drawbacks due to the local nature of the shearing box model: the mass loss rate depends on the vertical domain size and the flow top/down symmetry is unphysical \citep{from13}. This implies that global curvature effects, absent from the shearing box model, are mandatory to obtain physically valid outflow solutions.

To this end, global numerical simulations combining MRI-driven turbulence and magnetically-driven outflows have been computed \citep{suzu14,Zhu_Stone,mish2020}. These simulations are known to exhibit an exotic global configuration. For instance, the stationary state found by \cite{Zhu_Stone} possess 3 distinct features: a turbulent disk, a powerfully accreting atmosphere, mostly driven by a laminar Maxwell torque, and a tenuous wind that does not transport a considerable amount of angular momentum and mass. 

These numerical results are in flagrant contradiction with usual two dimensional 'effective' models of wind-emitting disk. In these models, accretion typically occurs in the disk bulk and angular momentum escapes through a magnetised outflow launched from the disk surface \citep{murp10,Stepa2016}.
Even though some effective models predict accretion preferentially located within the disk surface \citep[$\simeq 3h$: ][]{guilet2013,jacquemin-ide_magnetically-driven_2019}, where $h$ is the disk pressure scale-height, none of these models produce accretion in the disk atmosphere at $z\sim 10 h$, as is observed in the numerical simulations of \cite{Zhu_Stone} and \cite{mish2020}. Furthermore, \cite{Zhu_Stone} (see their Fig.~10) and \cite{mish2020} (see their Fig.~7) show that this accreting atmosphere features turbulent torques at high magnetisation, high above the disk.  The dynamical role and the origin of this 'second' turbulence it not known as of today.

While most of the literature has focused on mass and angular momentum transport, the question of magnetic field transport is also a key element if one is to predict the secular evolution of a disk subject to a magnetised outflow. It is this secular evolution which then leads to interesting observational evidence, like for example eruptions in X-ray binaries \citep{Ferreira_2006Xray,marc18b,marcel2019} or dwarf novae \citep{Scepi2019,Scepi2020}. This question was tackled in the seminal work of \cite{lubow1994}, who showed how the magnetic field could be advected towards the inner regions of the disk by the accretion flow. However, they also show that in the case where accretion is driven by turbulent torques, the field advection is efficiently counter-balanced by field diffusion due to MRI turbulence. \cite{rothstein2008}  later showed that the vertical structure plays a quintessential role on the secular transport of the magnetic field.
Indeed, the upper layers of the disk, that are magnetically dominated and have a lower conductivity, could enhance the transport of magnetic field inwards. They showed that in the case of wind driven accretion it is possible to accumulate magnetic flux in the inner regions. \cite{guilet2014} later showed that even in a purely viscous case, the vertical structure can help the transport of magnetic field towards the inner regions. However, all of these results rely on simplified models of turbulent transport, which are not entirely self-consistent. It is now mandatory to confirm the possibility of field transport from direct numerical simulations as it would then guide future theoretical and observational developments on this problem. 

In this work we aim at understanding the dynamics of a disk subject to MRI-driven turbulence and magnetised winds, and link our understanding to the question of the secular evolution. The paper is organised as follows. Section 2 provides the fundamental equations, and the numerical methods for the simulations and the analysis. Section 3 describes in details our fiducial simulation: its vertical structure, the transport properties and secular evolution of the magnetic field, and the characterisation of the outflow. Section 4 focuses on the secular evolution of the density distribution and the formation of ring-like structures. In section 5 we vary the field strength and disk aspect ratio to study their impact on the processes discussed in section 3 and 4. We finally put in perspective our results and discuss their astrophysical implications in section 6.

\section{Physical and numerical framework}
\subsection{Governing equations}
The analysis done in this paper is performed using spherical coordinates ($r$,$\theta$,$\varphi$). Even though this choice may lead to more cumbersome expressions, the spherical symmetry of the system justifies it. For convenience, we also define the cylindrical radius $R=r\sin\theta$ and the cylindrical height $z=r\cos\theta$.
We assume the disk is sufficiently ionised to be described in the ideal MHD approximation. In this limit the equations describing a magnetised fluid around a compact object of mass $M$ can be written as
\begin{align}
    \label{Eq:Mass_Con}
    &\pdv{\rho}{t} + \divi\left[\rho \vec{u}\right] = 0,\\
    \label{Eq:Mom_Con}
    &\rho\left(\BigD{u}\right) = -\vgrad P + \rho\vgrad \Phi_G + \frac{1}{c}\vec{J}\times \vec{B},\\
    \label{Eq:Energy}
    &\pdv{P}{t} + \vec{u}\cdot\vgrad P + \Gamma P \divi \vec{u} = -\Lambda\\
    \label{Eq:Induc}
    &\pdv{\vec{B}}{t} = \rot\left[\vec{u}\times\vec{B}\right],
\end{align}
where $\rho$, $\vec{u}$, $P$, and $\vec{B}$ are respectively the plasma density, velocity, thermal pressure, and  magnetic field, $\Phi_{G}= - GM/r$ is the gravitational potential, $\Gamma$ is the heat capacity ratio which is fixed to $1.001$ and $\vec{J}=c\rot\vec{B}/4\pi$ is the current. To close Eq.~(\ref{Eq:Mass_Con}-\ref{Eq:Induc}), we assume an ideal equation of state.  We relax our system to a locally isothermal temperature profile $T=T(R)\propto 1/R$ using a prescribed cooling
\begin{equation*}
    \Lambda = \frac{P}{T}\frac{T-T_\mathrm{eff}}{\tau_\mathrm{cool}}
\end{equation*}
where $T$ is the temperature of the plasma, $T_\mathrm{eff}$ is the relaxation temperature and $\tau_\mathrm{cool}$ is the cooling time scale.
To avoid the development of the vertical shear instability \citep[VSI,][]{Nelson2013} we use a thermal relaxation method on a fixed timescale. The VSI would otherwise appear in a strictly locally isothermal model. Hence we relax the temperature of our system on a time scale $\tau_\mathrm{cool}$, equal to 0.1 times the local keplerian time scale. This cooling can be rewritten as
\begin{equation*}
    \Lambda = \frac{P-\rho c_s^2(R)}{\tau_\mathrm{cool}}
\end{equation*}
which leads to an {effective} equation of state of the form
\begin{equation}
    P = c_s^2(R) \rho,
    \label{eq:local_isotherm}
\end{equation}
where $c_s$ is the local isothermal sound speed
\begin{equation}
    c_s = h\Omega_K
\end{equation}
defined from the Keplerian angular velocity $\Omega_K=\sqrt{GM/R^3}$ and the disk geometrical scale height $h$. Finally, we define the poloidal velocity and magnetic field as 
\begin{align*}
    \upo &= u_r\er+u_\theta\ethe\\
    \Bpo &= B_r\er + B_\theta\ethe.
\end{align*}
\subsection{Numerical method}
\label{sec:Method}
\begin{table*}[t]
\centering{
\begin{tabular}{|l|l|l|l|l|l|l|l|}
\hline
Name &  $\beta_{\mathrm{ini}}$ & $\epsilon$ & $r$ domain &  $\varphi$ domain & Resolution   & $t_1$ & $t_2=T_\mathrm{end}$          \\ \hline
SB4  & $10^4$                    & $0.1$             & $[1,200]$                & $[-\pi/4,\pi/4]$               & {[}850,176L+64S,62{]}  & $ 923T_{\mathrm{in}}$ &  $ 1114T_{\mathrm{in}}$  \\ \hline
SB3  & $10^3$                    & $0.1$             & $[1,200]$                & $[-\pi/4,\pi/4]$               & {[}850,176L+64S,62{]} & $ 923T_{\mathrm{in}}$ &  $ 1114T_{\mathrm{in}}$     \\ \hline
SB2  & $10^2$                    & $0.1$             & $[1,200]$                & $[-\pi/4,\pi/4]$               & {[}850,176L+64S,62{]}  & $ 1719T_{\mathrm{in}}$ &  $ 1910T_{\mathrm{in}}$    \\ \hline
SEp  & $10^4$                    & $0.05$            & $[1,100]$                & $[-\pi/4,\pi/4]$               & {[}1475, 96L+70S+2L, 62{]}& $ 923T_{\mathrm{in}}$ &  $ 1114T_{\mathrm{in}}$ \\ \hline
S2pi & $10^4$                    & $0.1$             & $[1,200]$                & $[0,2\pi]$                     & {[}850,176L+64S,256{]} & $ 923T_{\mathrm{in}}$ &  $ 1034T_{\mathrm{in}}$    \\ \hline
\end{tabular}
\caption{Parameters, grid spacing and radial extension of the simulations discussed in the paper. The latitudinal grid is not homogeneous, with grid spacing varing linearly (L) close to the disk and geometrically (S=stretched) close to the poles. We define $T_\mathrm{in} = T_K(R_\mathrm{in})$ where $T_K(R) = 2\pi/\Omega_K(R)$.
              }
\label{tab:simu}
              }
         
\end{table*}
To solve Eq.~(\ref{Eq:Mass_Con}-\ref{eq:local_isotherm}) on a spherical grid, we use the conservative Godunov-type code PLUTO \citep{Migno2007}. Constrained transport \citep{evans1988} ensures the solenoidal condition for the magnetic field is satisfied at machine precision. We use a second order Runge-Kutta method for the time step combined with a HLLD type solver with linear reconstruction to handle Riemann problems.
   \begin{figure}[h!]
   \centering
   \includegraphics[width=\hsize]{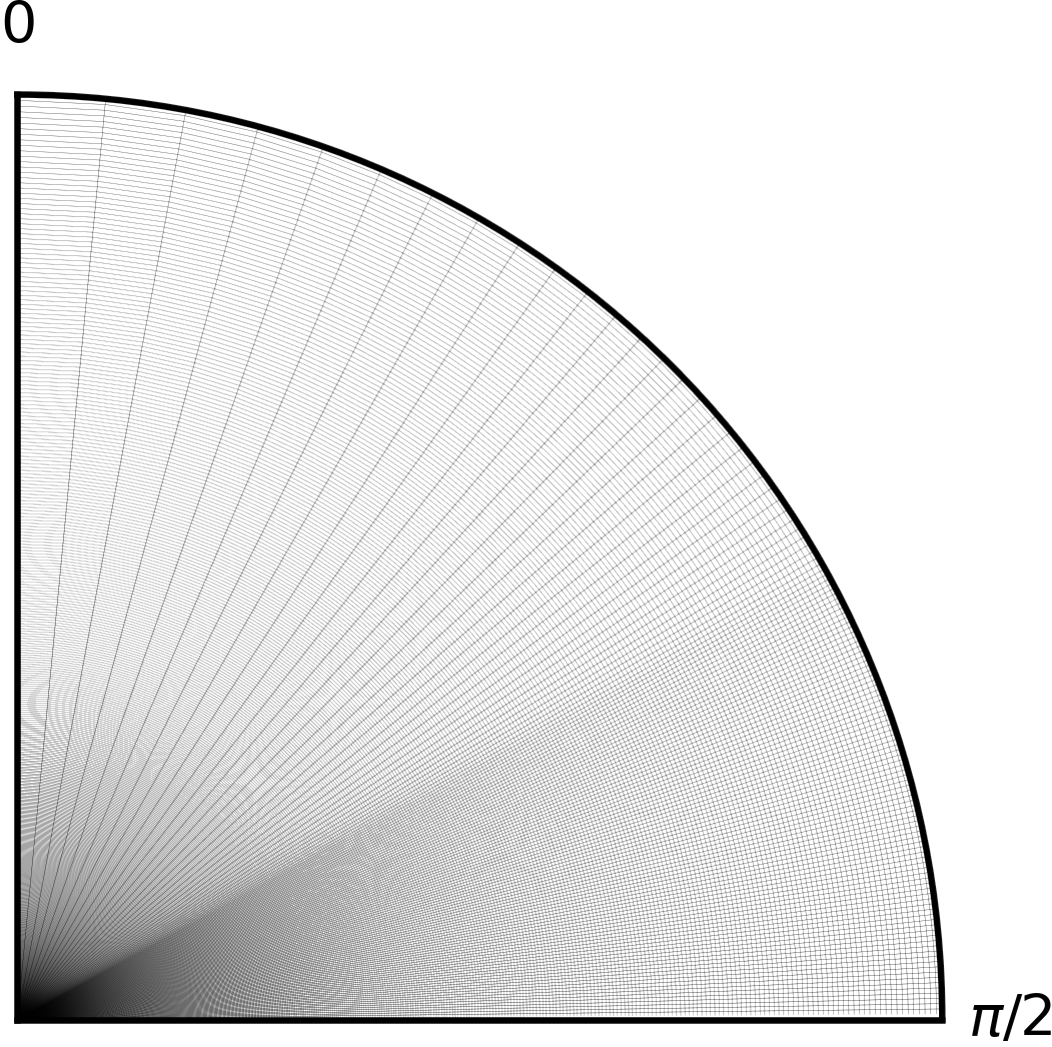}
      \caption{Computational Grid 
              }
         \label{Fig:Num_Grid}
   \end{figure}
The extent of our radial and toroidal domain varies from one simulation to the next, while our latitudinal extent is always $[0,\pi]$. We summarise the  domain properties, as well as the main control parameters for each simulations in Tab.~(\ref{tab:simu}). In the following, we use the innermost radius $r_{\mathrm{in}}$ as the length unit, so that $r_{\mathrm{in}}=R_{\mathrm{in}}=1$ in all simulations. 

We use periodic boundary conditions in the toroidal direction and zero-gradient in the radial direction (mass flow into the numerical domain is forbidden).  As for the latitudinal direction we implement the same polar ($z$-axis) boundary condition as \cite{Zhu_Stone}.

We follow \cite{Nelson2013} for the hydrostatic initial condition. The initial density and temperature profiles are defined at the disk mid-plane as:
\begin{align*}
    \rho_i(R,z=0) &= \rho_i(R=R_{\mathrm{in}},z=0) \Tfrac{R_{\mathrm{min}}}{R_{\mathrm{in}}}^p\\
    T_i(R) &= T_i(R=R_{\mathrm{in}}) \Tfrac{R_{\mathrm{min}}}{R_{\mathrm{in}}}^q
\end{align*}
where, for convenience, we use the cylindrical coordinates and we define $R_{\mathrm{min}} = \max\left(R,R_{\mathrm{in}}\right)$ to avoid singularities at the pole. As explained above, the temperature is constant on cylinders and remains so during the entire the simulation thanks to the {ad hoc} cooling function. 

We can write the solution of Eq.~(\ref{Eq:Mass_Con}-\ref{eq:local_isotherm}) assuming a hydrostatic equilibrium (neglecting all velocities but the azimuthal one) and no magnetic fields using the initial density and temperature at the disk mid-plane defined above. We get
\begin{equation*}
    \rho_i(R,z) = \rho_i(R,z=0) \exp{\left[\frac{\Omega_K^2(R_{\mathrm{min}})R_{\mathrm{min}}^3}{c_s^2(R_{\mathrm{min}})}\left(\frac{1}{\sqrt{R_{\mathrm{min}}^2+z^2}}-\frac{1}{R_{\mathrm{min}}}\right)\right]}
\end{equation*}
and 
\begin{equation*}
    u_{\varphi,i}(R,z) = V_K \left[(q+1) - q\frac{R_{\mathrm{min}}}{\sqrt{R_{\mathrm{min}}^2+z^2}}+\frac{c_s^2(R_{\mathrm{min}})}{V_K^2(R_{\mathrm{min}})}(p+q)\right]^{1/2}
\end{equation*}
where we have defined  the Keplerian velocity $V_K=\Omega_K R$. 
We then use the Keplerian angular velocity and the sound speed to define the disk scale height
\begin{equation}
    h=c_s/\Omega_K = \epsilon R,
\end{equation}
where $\epsilon$ is the disk geometrical thickness. 

Following \cite{Zhu_Stone}, we assume a large scale vertical magnetic field that is constant with respect to $z$ but follows a power law in $R$. In order to get a self-similar initial condition, we choose a radially constant initial plasma beta,
\begin{equation}
    \beta_{\mathrm{ini}} = \frac{8\pi P_i}{B_{z,i}^2},
\end{equation}
in the disk midplane. The magnetic field radial profile is then related to that of temperature and density:
\begin{equation}
    B_{z,i}(R) = B_{i} \Tfrac{R_{\mathrm{min}}}{R_{\mathrm{in}}}^{\frac{p+q}{2}} 
\end{equation}
where
\begin{equation}
    B_i = \sqrt{8\pi P_i(R=R_{\mathrm{in}},z=0)/\beta_{\mathrm{ini}}}.
\end{equation} 
To ensure that $\divi \vec{B}=0$ at $t=0$, we initialise the magnetic field using the magnetic potential $\vec{A}$. We can solve the equation $\vec{B}=\rot\vec{A}$ to find 
\begin{align*}
    A_\varphi=\left\{
                \begin{array}{ll}
                  \frac{RB_{i}}{2}\spacio \mathrm{for}\,\, R\leq R_{\mathrm{in}},\\
                  B_{i} \Tfrac{R_{\mathrm{min}}}{R_{\mathrm{in}}}^{\frac{p+q}{2}}\frac{2R}{(p+q)+4}+ B_{i}\frac{R_{\mathrm{in}}^2}{R}\left(\frac{1}{2}-\frac{2}{(p+q)+4}\right) \mathrm{for}\,\, R>R_{\mathrm{in}}.
                \end{array}
              \right.
\end{align*}

Finally, we choose $p=-3/2$ and $q=-1$ for all our simulations, in the same way as \cite{mish2020}.

To handle runaway magnetisations close to the polar boundary we implement a density floor in our simulation
\begin{equation*}
    \rho_{\mathrm{fl}} = \mathrm{max}\left( \rho_{\mathrm{fl},0}\Tfrac{R_{\mathrm{min}}}{R_{\mathrm{in}}}^{p}\frac{1}{z^2+a_{\mathrm{fl}}\epsilon^2 R_{\mathrm{min}}^2} ,10^{-9}\right)
\end{equation*}
where $\rho_{\mathrm{fl},0}=10^{-7}$ and $a_{\mathrm{fl}}\sim1$. Finally, to minimise integration time, we enforce a maximal Alfv\'enic velocity $V_{A,\mathrm{max}} = V_K(R_{\mathrm{in}})$, where $V_A=B/\sqrt{4\pi\rho}$ is computed with the instantaneous magnetic components. This maximal Alfv\'enic velocity is imposed by changing the local density and not the magnetic field, hence it acts as an additional density floor. In our simulations, this numerical artifice is only noticeable close to the axis as well as near to the inner boundary, $r\sim r_{\mathrm{in}}$, which are ignored in our analysis.

We use a nonuniform grid (Fig.~\ref{Fig:Num_Grid}): in the radial direction the grid points are logarithmically spaced,  while in the latitudinal direction the grid is linearly spaced up to $|z|=5.5h$ and then geometrically stretched up to the axis. The grid is identical for all the simulation except for the simulation at $\epsilon=0.05$ where the grid is spread differently. For this simulation the grid is linearly spaced up to $|z|=3h$ and then stretched up until $|\pi/2-\theta|>\pi/2-0.35$ where the spacing becomes linear again for two grid points close to the polar boundary. This particular grid maximises the resolution within the disk and its surface as well as providing a buffer zone close to the polar boundary where the CFL condition is the smallest. Finally, in the toroidal direction the grid is linearly spaced. The resolutions for the different simulations per region and direction are displayed in Tab.~(\ref{tab:simu}). This choice of resolution allow us to resolve the MRI within the disk thanks to roughly 16 grid points per vertical scale height. 

We integrate our simulations up to  $T_\mathrm{end}$, see Tab.(\ref{tab:simu}). We define our reference time scale $T_\mathrm{in} = T_K(R_\mathrm{in})$, corresponding to orbits of the innermost radii, where $T_K(R) = 2\pi/\Omega_K(R)$.

\subsection{Diagnostics and turbulent decomposition}
\subsubsection{Averaging procedure}
\label{sec:Ave_proc}
To quantify the turbulent state of the accreting system, we need to define a set of averages. This allows us to decompose our quantities in turbulent, $\de X$, and laminar, $\mean{X}$, terms. We use three kinds of average, on $\varphi$, on ($\varphi$, $t$), and on ($\varphi$, $t$) density weighted:
\begin{align*}
    \phimean{X} &= \frac{1}{\Delta \varphi} \int\limits_{\varphi_1}^{\varphi_2} X \diff \varphi,\\
    \mean{X} &=\frac{1}{\Delta \varphi} \int\limits_{\varphi_1}^{\varphi_2} \frac{1}{t_2-t_1} \int\limits_{t_1}^{t_2} X \diff \varphi \diff t, \\
    \rhomean{X} &= \frac{1}{\Delta \varphi \mean{\rho}} \int\limits_{\varphi_1}^{\varphi_2} \frac{1}{t_2-t_1} \int\limits_{t_1}^{t_2} \rho X \diff \varphi \diff t = \frac{1}{\mean{\rho}}\mean{X\rho}.
    \end{align*}
In these definitions, $\varphi_1$ and $\varphi_2$ are taken as the boundaries of the whole computational domain (Tab.~(\ref{tab:simu})). It is important to note that $\mean{X}\neq\rhomean{X}$, since we define our turbulent fluctuations from the $\langle\cdot\rangle$ average
\begin{equation}
    X = \mean{X} + \de X.
\end{equation}
 With this definition, the quantities $\mean{X}(R,z)$ are equivalent to the ones used in 2d effective models such as self-similar solutions (see \cite{jacquemin-ide_magnetically-driven_2019}). Hence, they allow for a direct comparison to said models and provide guidance for turbulence closures. Let us emphasise  however that $\rhomean{\de X}\neq 0$, which leads to additional complications when dealing with high order correlation functions. Finally, we note that 
 \begin{equation}
    \label{Eq:dev_azi}
      \mean{\pdv{X}{\varphi}}=0
 \end{equation}
 since the boundary condition in the $\varphi$ direction is periodic. The averages are taken from time $t_1$ to $t_2$, with a temporal resolution of $\Delta t = 1/\Omega_{K}(R_{\mathrm{in}})\simeq 0.16 T_\mathrm{in}$ and then azimuthally averaged over the whole domain (Tab.~\ref{tab:simu}).
\subsubsection{Turbulent decomposition}
In order to study the impact of turbulence on the vertical structure of the accretion-ejection system, we use the averages defined above to rewrite the ideal MHD equations as a set of Reynolds-averaged MHD equations. In this approach, turbulence then appears in the averaged equation as an additional dynamical term influencing the mean flow. This is actually an extension of the famous \cite{shak73} $\alpha$-disk approach which can be derived as a Reynolds-averaged model for MHD turbulence \citep{bp99}. While the $\alpha$ disk approach only focuses on angular momentum transport, our objective is to generalise this to the whole set of ideal MHD equations.
To illustrate our procedure, we show its application to the equation of angular momentum transport below. A complete derivation for the whole set of ideal MHD equations can be found in Appendix \ref{A:RA_MHD}.

We project and rearrange (using Eq.\ref{Eq:Mass_Con}) the equation for toroidal momentum transport to find the equation of angular momentum transport
\begin{equation}
    \pdv{Ru_\varphi \rho}{t} + \divi{R\left[\rho u_\varphi \vec{u}-\frac{B_\varphi \vec{B}}{4\pi}\right]} = -\pdv{}{\varphi}\left(\frac{B^2}{8\pi}+P\right).
\end{equation}
We decompose the different quantities into fluctuating (turbulent)  plus mean (averaged) terms. { We also call mean terms, laminar terms, in the same way as \cite{beth17} \citep[see also][]{mish2020}}.
We then combine it with the continuity equation (Eq.~\ref{Eq:Mass_Con}) and average the result with respect to time and $\varphi$. We eventually get

\begin{align}
    \label{eq:moy_AngMom}
    &\mean{\rho}\rhomean{\vec{u_p}}\cdot\vgrad R\mean{u_\varphi} +\nonumber \\ &\divi R \left[\mean{\rho}\rhomean{\de u_\varphi \vec{\de u_p}}
    -\frac{1}{4\pi}\mean{\de B_\varphi \vec{\de B_p}} - \frac{1}{4\pi}\mean{ B_\varphi} \mean{\vec{B_p}}\right]=0.
\end{align}
For the sake of brevity, we have neglected the term with the least dynamical importance and assumed stationarity, the equation can be found in full form in Appendix \ref{A:RA_MHD}. We then define a fluctuating stress tensor as well as a laminar stress tensor:
\begin{equation}
    \vec{\mathcal{T}} = \vec{\mathcal{T}_\mathrm{tu}} + \vec{\mathcal{T}_\mathrm{la}},
\end{equation}
where we define
\begin{equation}
    \vec{\mathcal{T}_\mathrm{la}} = - \frac{1}{4\pi}\mean{ B_\varphi} \mean{\vec{B_p}},
\end{equation}
and
\begin{equation}
    \vec{\mathcal{T}_\mathrm{tu} }= \mean{\rho}\rhomean{\de u_\varphi \vec{\de u_p}} -\frac{1}{4\pi}\mean{\de B_\varphi \vec{\de B_p}}.
\end{equation}
We can then define the usual turbulent alpha viscosity as
\begin{align}
    \alpha_{\mathrm{ef}} &= \frac{\mathcal{T}_{r\varphi}}{\mean{P}},\\
    \alpha_{\mathrm{ef}} &= \alpha_\mathrm{la} + \alpha_\mathrm{tu}
\end{align}
where $\alpha_\mathrm{la}$ is related to the laminar torque and $\alpha_\mathrm{tu}$ to the turbulent one.
The torques have different origins as one is a consequence of the turbulence while the latter is a consequence of the mean field geometry.

It is important to note that MRI is expected to induce a radial turbulent torque while an MHD wind is expected to induce a latitudinal and vertical laminar torque. 
However, we see in Eq.~(\ref{eq:moy_AngMom}) that the fluctuating turbulent and mean laminar terms have  both a radial and a vertical component.
Furthermore, in the present work we find\footnote{similarly to \cite{Zhu_Stone}} that the radial laminar torque can be important in the angular momentum budget. It is not clear whether such a torque is a consequence of the wind or the turbulence. This calls into question the {a priori} distinction of wind vs turbulent torques and may require a change of nomenclature within the community.

 \begin{figure*}
    \centering{
   \includegraphics[width=0.495\hsize,height=0.45\vsize]{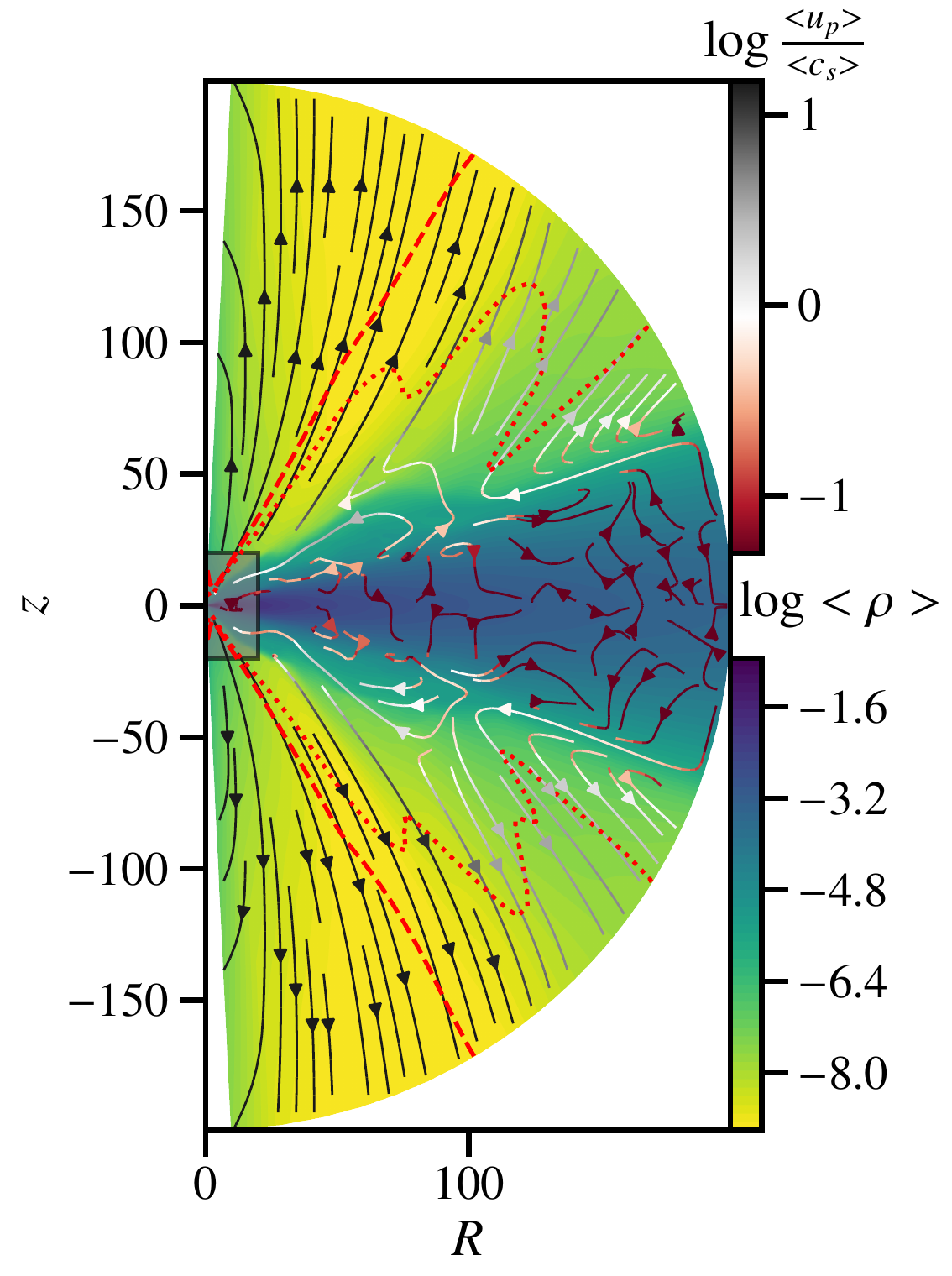}
   \includegraphics[width=0.485\hsize,height=0.45\vsize]{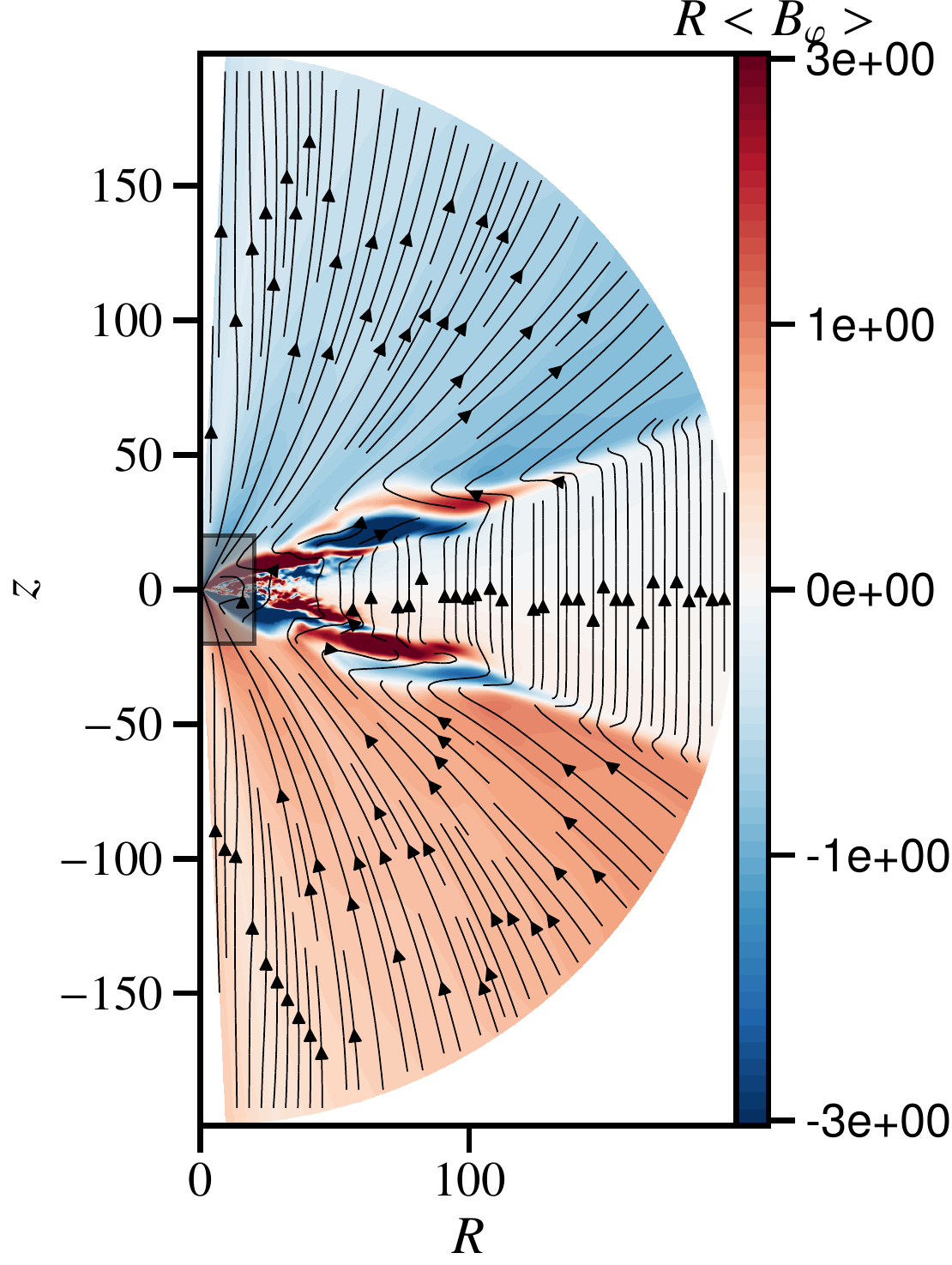}\\
   \includegraphics[width=0.495\hsize,height=0.45\vsize]{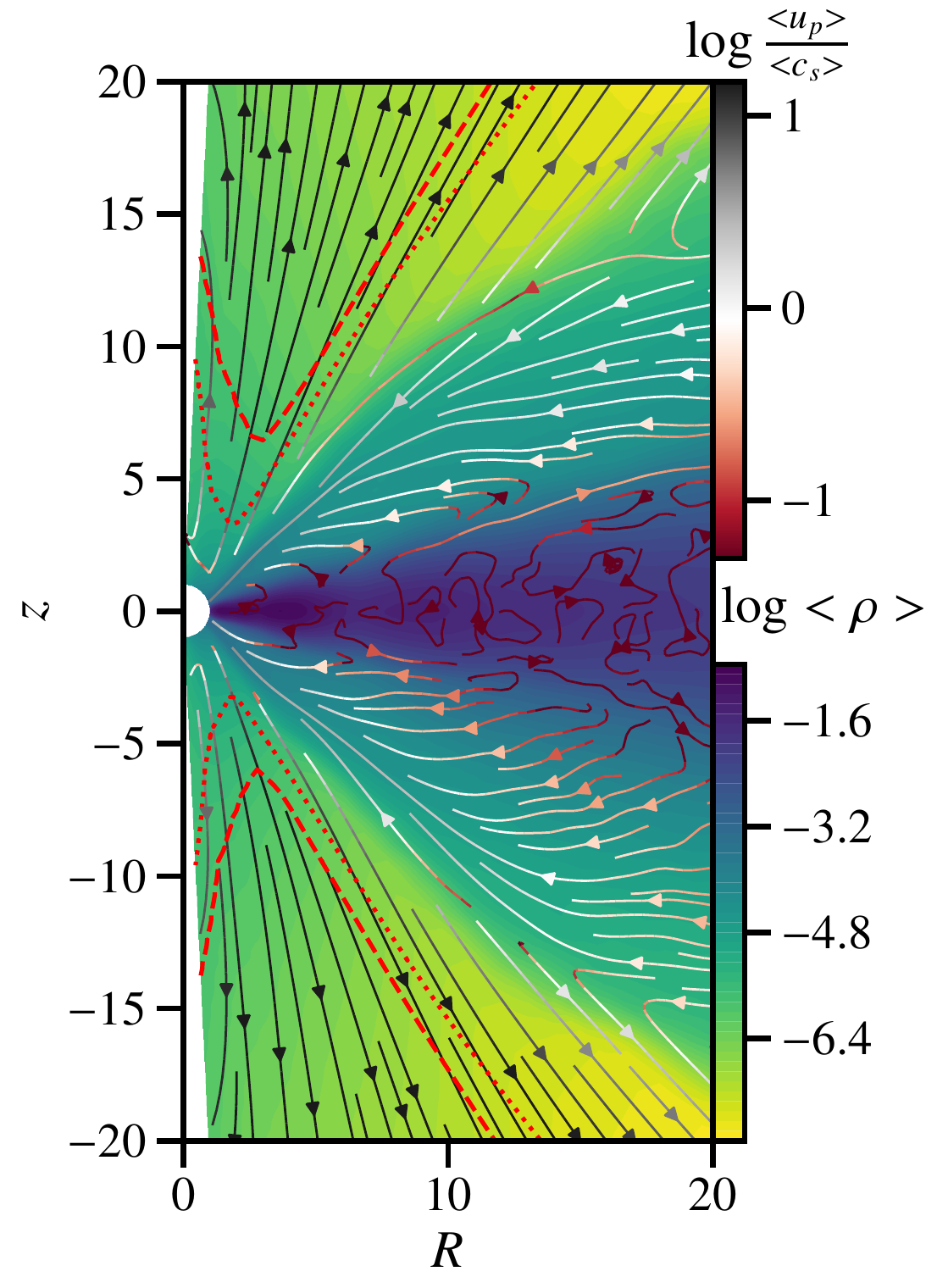}
   \includegraphics[width=0.485\hsize,height=0.45\vsize]{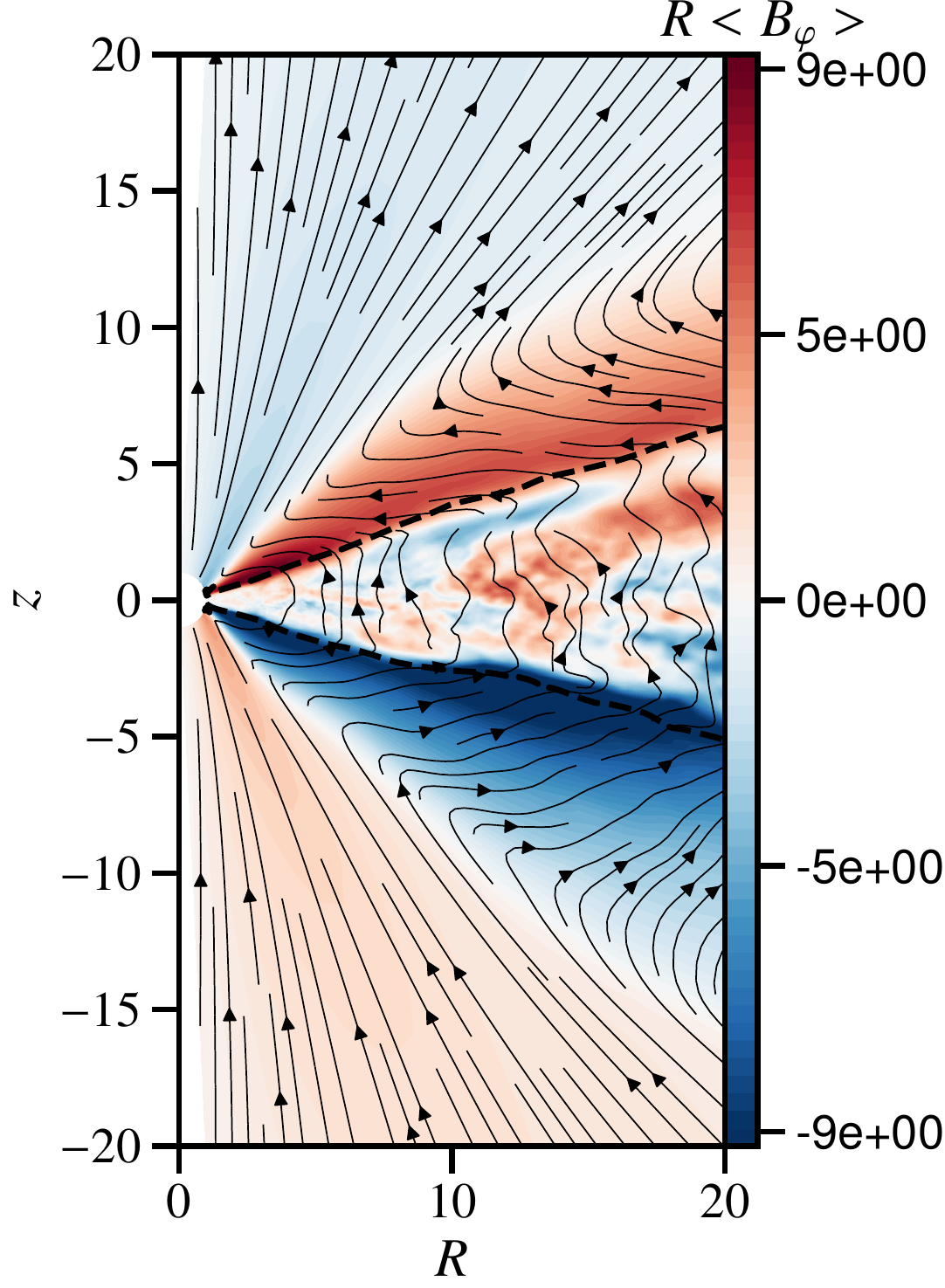}}
      \caption{(top,left) Gas density and mean poloidal stream lines. The red dotted line corresponds to the Alfv\'enic surface,  
      and the red dashed line corresponds to the fast magneto-sonic surface. The colour of the poloidal stream lines correspond to the logarithm of their magnitude normalised to the sound speed.
      (top,right) $RB_\varphi$ normalised to $B_iR_{in}$; and mean poloidal field lines. The grey square corresponds to the zoomed in region the bottom figure.
      (bottom) same as top but zoomed in the greyed region. The black dashed line indicates the surface where $\mean{\beta}=8\pi\mean{P}/\mean{B}^2=1$.
              }
         \label{Fig:field_stream}
 \end{figure*}

\section{Simulation SB4, the fiducial case}
\label{sec:fiducial}
\subsection{Global picture}
\label{sec:Global_picture}
We first present a global picture of our fiducial simulations. We show in figure \ref{Fig:field_stream} the  mean poloidal stream lines and field lines, the Alfv\'en surface and the fast magneto-sonic surface of the fiducial $\beta_{\mathrm{ini}}=10^4$ run. We define the Alfv\'en surface as the location where $\mean{u_p}=\mean{V_{ap}}$, where $V_{ap} = \mean{B_p}/\sqrt{4\pi \mean{\rho}}$ and the fast magneto-sonic surface where $\mean{u_p}=\mean{V_{fm}}$, with $V_{fm}^2=\frac{1}{2}(c_s^2+V_a^2+|V_a^2-c_s^2|)$, having defined the total Alfv\'enic velocity $V_a=\mean{B}/\sqrt{4\pi\mean{\rho}}$. 
 
 In the top panel of Fig.~(\ref{Fig:field_stream}) we show the whole domain of our numerical simulation. We find that the upper and lower hemispheres of the system are approximately symmetric. The fast magneto-soni (FM) surface is conical up to the border of our domain. In contrast, the Alfv\'enic surface remains conical up to a certain radii where its shape is modified, demonstrating that the outer radii of the domain has not yet reached a stationary state. This is corroborated by the lack of toroidal field within the outer regions of the disk. 
 
 In Fig.~(\ref{Fig:field_stream}) (top right) we can clearly see some collimation happening for the inner field lines, close to the end of their trajectory. This may be the sign of some intrinsic MHD property, as it has been also observed in self-similar jet solutions \citep{ferr97,jacquemin-ide_magnetically-driven_2019}.
However, it is unclear whether this is a physical or a numerical effect, akin to a bias of our boundary conditions or a consequence of the density floor. Nevertheless, a study of collimation in our numerical models is outside the scope of this paper.
 
 Since the FM surface corresponds to the point where the outflow becomes causally disconnected from the disk, we can use this surface to estimate the last radius that has achieved stationarity. We do this by computing the anchoring radius of the last field line that traverses the FM surface, this corresponds to $R \sim 35$. Hence, we focus the rest of our analysis to a domain within that radii, the stationary inner regions (grey rectangle in top of Fig.~(\ref{Fig:field_stream})). 

We can then identify three zones shown in the bottom panels of Fig.~(\ref{Fig:field_stream}):

Firstly, A turbulent disk where the density is maximal and the poloidal flow is disorganised, indicating a turbulent flow. The toroidal component of the magnetic field exhibits similar disorganised structure, while the average poloidal field lines are remarkably straight.

Secondly, An accreting atmosphere where the poloidal flow is laminar and  transonic. The field lines in this region seem to be curved by the poloidal flow. In addition, a strong and laminar toroidal field emerges.

Finally, a super fast magneto-sonic outflow is ejected from the atmosphere, in the wind region. The poloidal field lines are approximately aligned with the outflow direction.

\begin{figure}
   \includegraphics[width=\hsize]{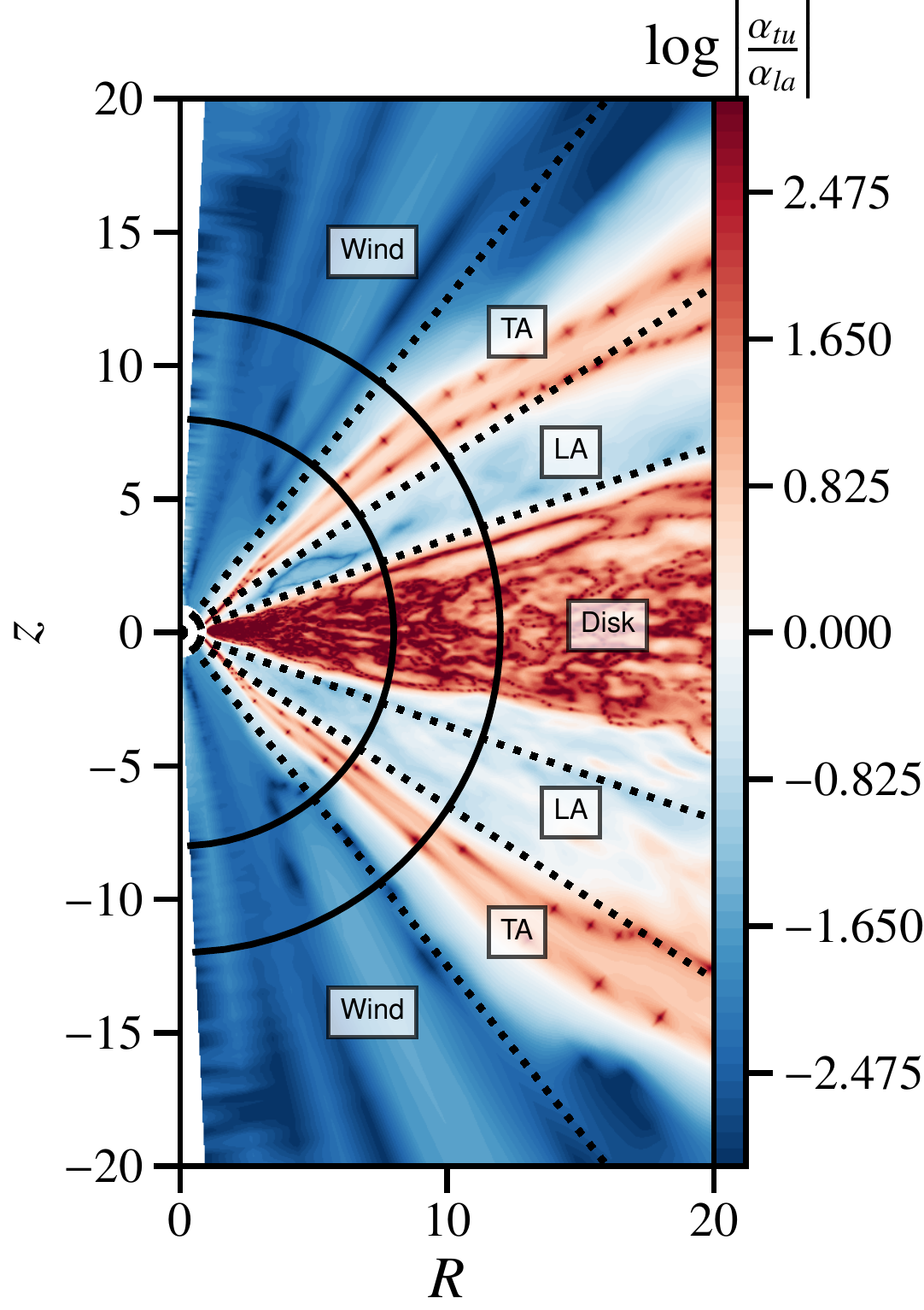}\\
   \includegraphics[width=0.9\hsize]{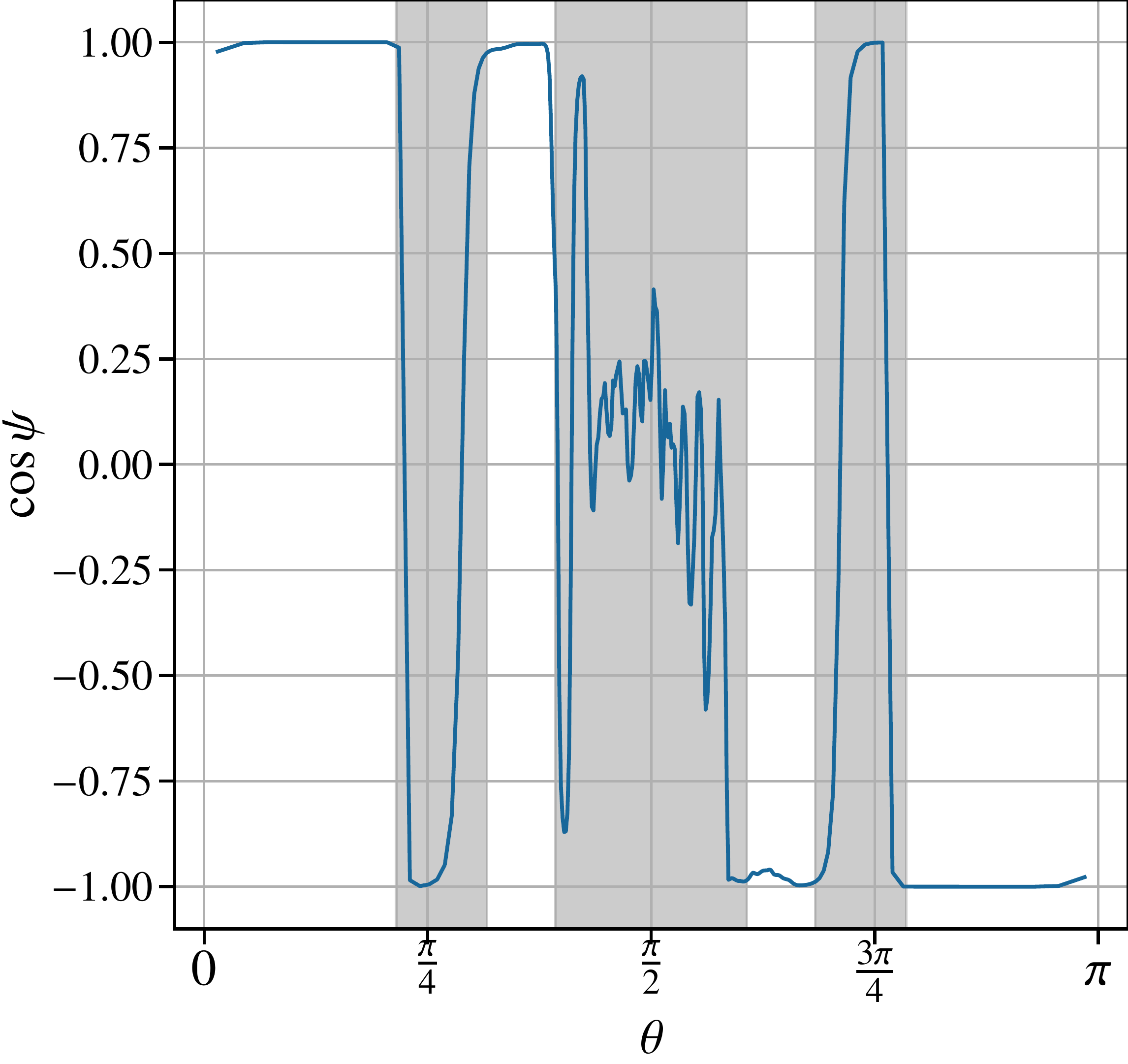}
      \caption{(top) Ratio $\alpha_\mathrm{la}/\alpha_\mathrm{tu}$ denoting turbulent (red) and laminar (blue) regions. The dotted lines delimit these regions and represent $|z|=3.5h$, $|z|=6.5h$ and $|z|=12.5h$. The two different circles denote the radii used for the calculation of the fluxes ($r_1=8$, $r_2=14$), see section \ref{sec:transport} and section \ref{sec:Comp_flux}. TA and LA correspond for turbulent and laminar atmosphere, respectively.
      (bottom) Cosine of the angle $\psi$ between the mean poloidal velocity and the mean poloidal magnetic field as a function of the latitudinal coordinate. The grey zones correspond to the turbulent regions in the top panel, delimited by the dotted lines. 
              }
         \label{Fig:ratio}
 \end{figure}

The region we identify as the accreting atmosphere exhibits turbulence which is not obvious from Fig.~\ref{Fig:field_stream}. To quantify the dynamical importance of this turbulence, we measure the ratio $\alpha_\mathrm{tu}/\alpha_\mathrm{la}$ (Fig.~\ref{Fig:ratio}, top). As expected, the disk is clearly dominated by the turbulent stress component. As we move through the accreting atmosphere from the disk surface, the stress becomes dominated by its laminar component before turning back to a turbulent stress region, as in the disk. Even higher still, in the wind region, the torque is laminar. The fact that the atmosphere is divided into two sub-regions, dominated respectively by a laminar and a turbulent stress, invites us to define a laminar and a turbulent atmosphere (LA and TA), which we need to delimit more precisely.

\begin{figure}
 \centering{
   \includegraphics[width=\hsize]{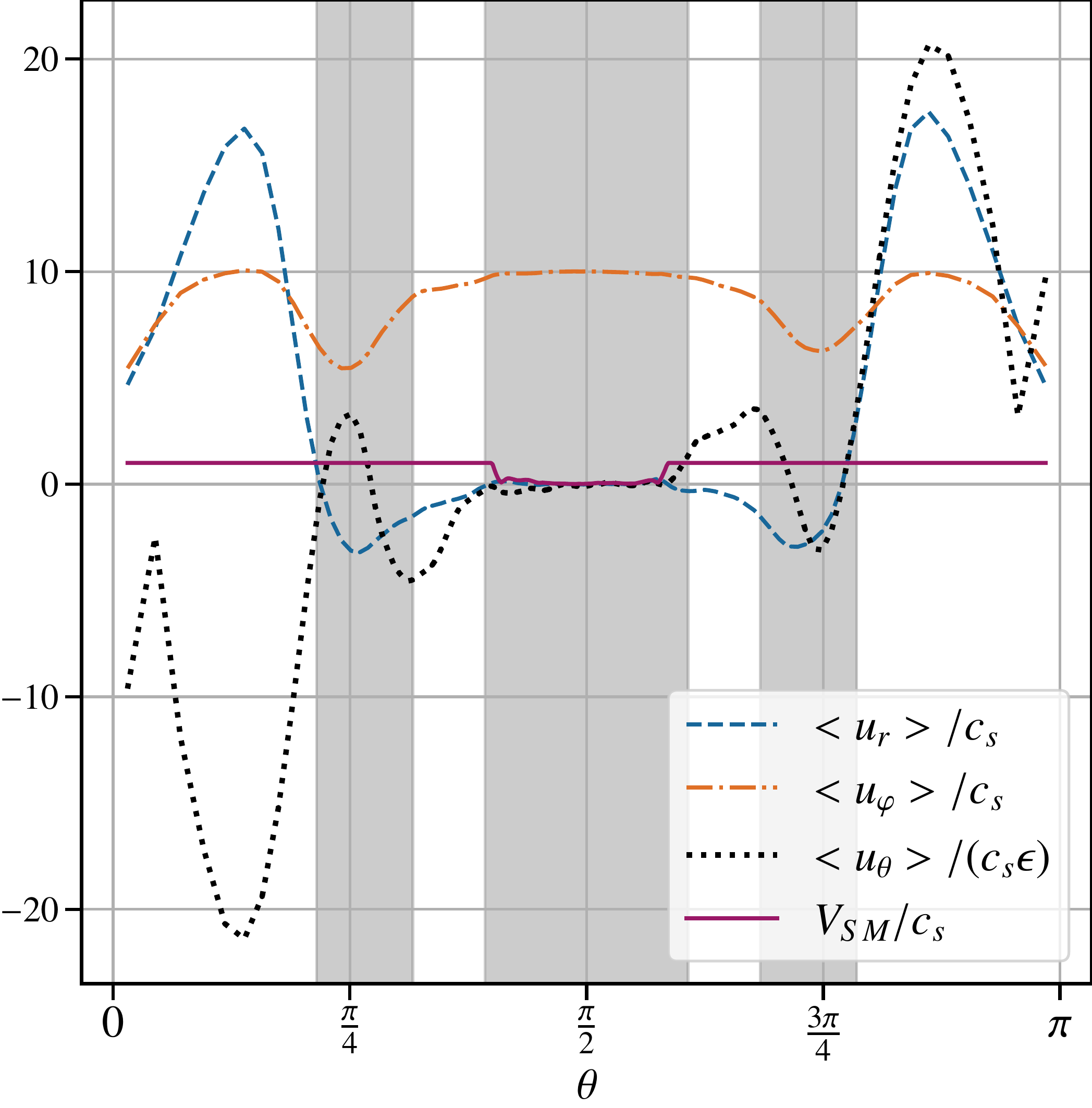}\\
   \includegraphics[width=\hsize]{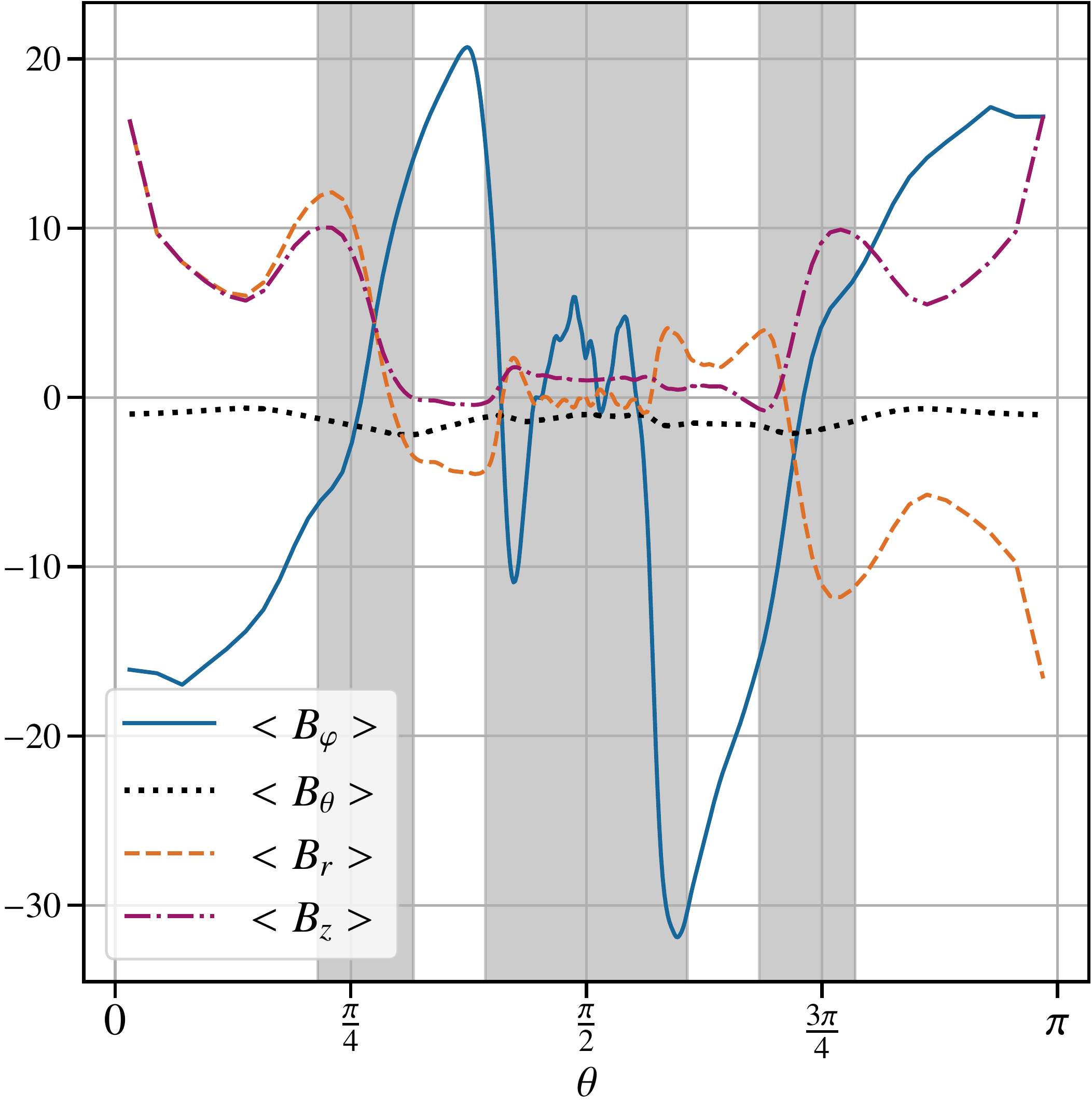}\\
      \caption{(top): Velocity profiles as functions of the latitudinal coordinate, $\theta$, normalised to the local sound speed. 
      (bottom): mean magnetic field profiles normalised to the vertical magnetic field in the disk mid-plane, $\mean{B_z}(r,\theta=\pi/2)$ as functions of the latitudinal coordinate. 
      The grey zones correspond to the turbulent regions in figure \ref{Fig:ratio}. All profiles are radially averaged between $r_1=8$ and $r_2=14$ after being normalised.  
              }
               \label{Fig:regin_def}
    }
\end{figure}

Under the assumptions of ideal MHD and steady state, the average poloidal velocity and magnetic field should be perfectly aligned \citep[e.g.~][]{mes61}. Hence, the value of the angle $\psi$ between the averaged poloidal stream and field lines,
\begin{equation*}
    \cos \psi = \frac{\mean{\upo}\cdot\mean{\Bpo}}{\mean{u_p}\mean{B_p}},
\end{equation*}
identifies the regions where the turbulent terms are important. In Fig.~\ref{Fig:ratio} (bottom) we plot this angle as a function of the latitudinal coordinate. We also provide in Fig.~\ref{Fig:regin_def} the behaviour of the mean velocity and mean magnetic fields as functions of the latitudinal coordinate.

In the disk, the mean poloidal velocities and field lines are approximately perpendicular $(\cos(\psi)=0)$: the disk is turbulent. We identify the disk surface as the location where $\cos \psi=\pm1$. In the following, we define the disk as  $|z|<3.5h$ (the central shaded area in Fig.~\ref{Fig:regin_def} and Fig.~\ref{Fig:ratio} (bottom)), consistently with Fig.~\ref{Fig:ratio} (top). The disk surface coincides with the location where the average radial velocity becomes negative (Fig.~\ref{Fig:regin_def}), marking the beginning of the accreting atmosphere.

Above the disk, the average poloidal stream and field are aligned ($\cos(\psi)=\pm1$). This defines the laminar atmosphere previously identified with the stress, located at altitudes $3.5h<|z|\lesssim6.5h$.
As we move to even higher altitudes, $\cos(\psi)$ changes sign twice (Fig.~\ref{Fig:ratio}). This suggests that ideal-MHD is broken in this region, which we identify as the turbulent atmosphere, located at altitudes $6.5h<|z|\lesssim12.5h$. This region ends at the surface delimiting the base of the wind, where $\cos(\psi)=\pm1$, which coincides with the slow magneto-sonic (SM) surface defined as $\mean{u_p}=V_{SM}$, where $V_{SM}^2=\frac{1}{2}(c_s^2+V_a^2-|V_a^2-c_s^2|)$ (Fig.~(\ref{Fig:regin_def},top)). 

It is important to note that $V_{SM}$ is indeed constant in units of $c_s$ since the magnetic field is dominant inside the atmosphere (laminar and turbulent), $\mean{\beta}=8\pi\mean{P}/\mean{B}^2<1$. However, the mean poloidal magnetic field stays infra-thermal within the laminar atmosphere, $\mean{\beta_p}>1$, where 
\begin{equation}
    \mean{\beta_p}=\frac{8\pi \mean{P}}{\mean{B_p}^2},
\end{equation}
but becomes supra-thermal  within the turbulent atmosphere, $\mean{\beta_p}<1$.

\subsection{Transport properties}
\subsubsection{Mass transport}
\label{sec:transport}
Using the regions defined in section \ref{sec:Global_picture}, we compute the fluxes through the boundaries of each region, delimited radially by $r_1=8$ and $r_2=14$.
We then analyse the transport of mass and angular momentum within our system.
The analysis is carried out in detail in appendices \ref{sec:Mass_flux} and \ref{sec:Ang_flux}. In this section, we summarise the main results as they have already been described in details by \cite{Zhu_Stone}.

Within the disk the mass is latitudinally transported towards the atmosphere and radially transported outwards, as described by \cite{Zhu_Stone}. Moreover, the total mass within the disk section delimited by $r_1$ and $r_2$ decreases with time. This is the consequence of the secular self-organisation of the radial disk structure. Indeed, a closer look at the radial density profiles reveals the formation of ring-like structures. Our chosen disk section turns out to be located on a gap. The mechanism behind the formation of these structures is discussed in section \ref{sec:ring}. 

In the accreting atmosphere the mass is transported radially inwards, the gas is accreted. In the laminar region  the material falls in along the mean poloidal field.  Most of the actual accretion happens in the turbulent atmosphere, where material slips though the magnetic field thanks to a turbulent 'effective' resistivity. In the latitudinal direction the atmosphere receives mass from the disk and looses a negligible amount to the outflow. In contrast to the disk, the transport of mass within the atmosphere is in steady state, as no mass excess or deficit is observed.

\subsubsection{Angular momentum transport and accretion}
Within the disk, angular momentum is transported radially outwards through turbulent stresses. Moreover, the disk receives angular momentum from the atmosphere through laminar latitudinal stresses, which explains the observed decretion in the disk \citep[see also][]{Zhu_Stone}. The accreting atmosphere losses angular momentum and transfer it mostly to the disk through a laminar stress. As previously shown by \cite{Zhu_Stone}, the wind contribution to the angular momentum budget is essentially negligible. 

To understand the global behaviour of the accretion flow in our simulation, we compute the mass weighted accretion velocity,
\begin{equation}
\label{eq:glob_acc}
    v_{\mathrm{acc}} = \frac{\int\limits_{\theta_{\mathrm{SM},1}}^{\theta_{\mathrm{SM},2}}r \sin\theta \, \mean{u_r \rho} \diff{\theta}.}{\int\limits_{\theta_{\mathrm{SM},1}}^{\theta_{\mathrm{SM},2}}r \sin\theta \, \mean{\rho} \diff{\theta}.},
\end{equation}
where $\theta_{\mathrm{SM},1}$ and $\theta_{\mathrm{SM},2}$ are respectively the angles where the flow becomes super SM in the upper and the lower hemisphere. These surfaces coincide with the end of the turbulent atmosphere. The mass weighted accretion velocity is related to the mass accretion rate $\dot{M}_\mathrm{acc}$, indeed $v_\mathrm{acc}$ is the velocity at which mass is transported. We then time average this quantity between $t_a=318 T_\mathrm{in}$ and $t_b=955T_\mathrm{in}$ with temporal resolution of $\Delta t = 0.8T_\mathrm{in}$.  We find
\begin{equation}
    v_{\mathrm{acc}}(R) \simeq - 1.1 \times 10^{-3} V_K(R),
\end{equation}
valid for  $R \in [2,14]$, which is clearly subsonic. The accretion velocity, $v_{\mathrm{acc}}$, follows a radial dependency very close to Keplerian.  The negative sign shows that accretion in the turbulent atmosphere dominates over the disk decretion and self-organisation. Moreover, the accretion time scale we deduce from this: $t_{\mathrm{acc}} \sim 9.1\times10^{2}/\Omega_K(R) = 145 T_K$ is substantially longer than than the local dynamical timescale of MRI or ejection which are of the order of $T_K$. Hence, there is a clear timescale separation between local dynamics and accretion.

\subsubsection{Magnetic flux transport}
The question of magnetic flux transport in disks is a corner stone of accretion theory \citep{lubow1994,rothstein2008,guilet2014}. To characterise this transport, we introduce the magnetic flux threading the disk as
\small
\begin{equation}
    \label{eq:Psi_def}
    \Psi(R,t) = \int\limits_{0}^{\pi/2} r^2 \sin\theta \phimean{B_r}(r=R_\text{in},\theta,t) \diff\theta + \int\limits_{R_{\mathrm{in}}}^{R}R\phimean{B_z}(r,\theta=\pi/2,t) \,\diff{R}.
\end{equation}
\normalsize
which corresponds to an integration on a surface that goes from the axis ($z=1$ and $R=0$) down to the inner boundary of the disk ($z=0$ and $R=R_{in}$) and then to radius $R$ in the disk midplane. 

\begin{figure}[h!]
   \centering
   \includegraphics[width=\hsize]{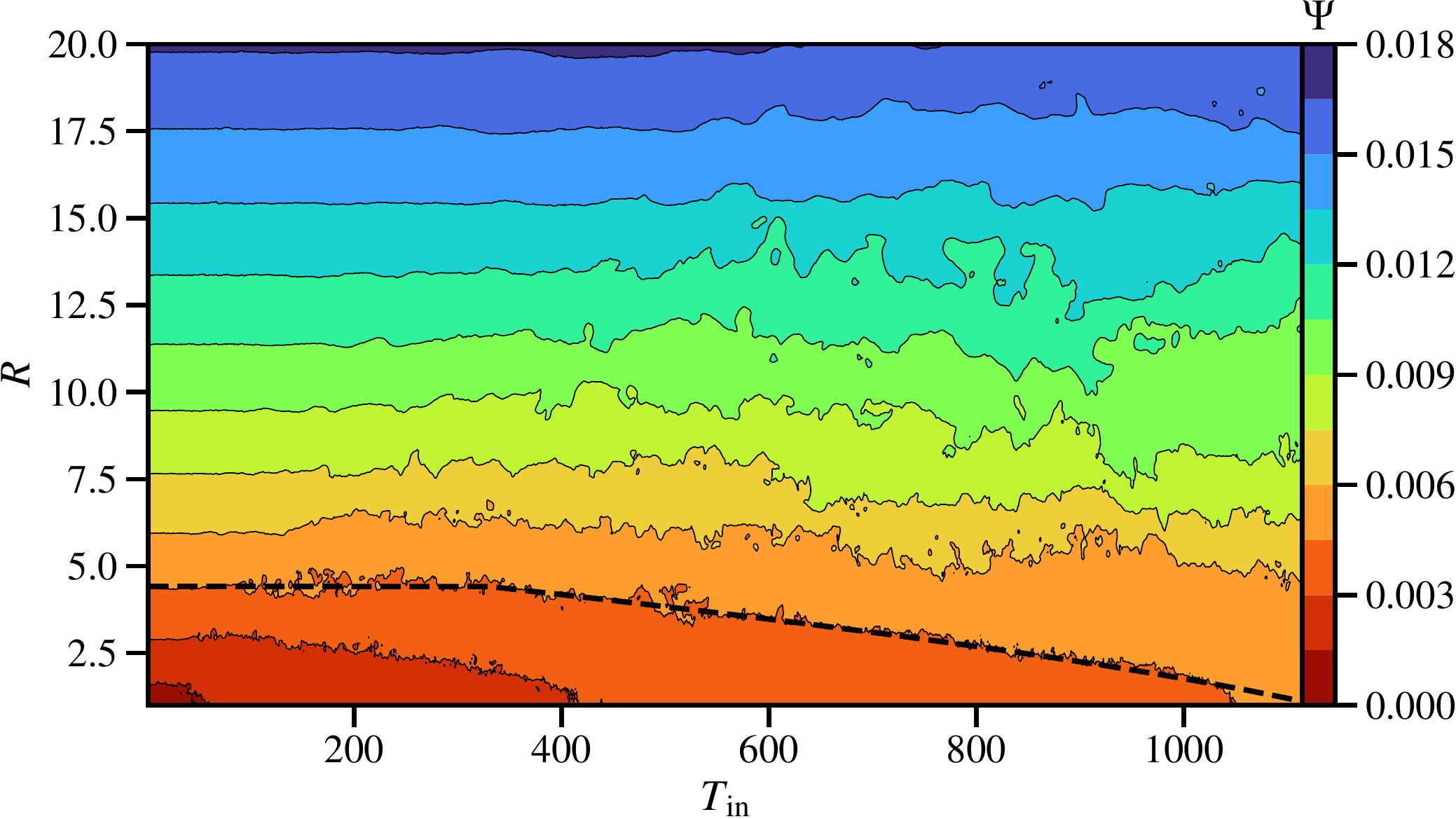}
      \caption{Magnetic flux threading the disk midplane $\Psi$, computed between the axis and $R$, as a function of time in innermost orbital units. Every contour line represents a field line. The dashed line shows an example of a fit using Eq.(\ref{eq:fit}) with $R_i=4.4$, $t_i=328T_\mathrm{in}$ and $\frac{v_\Psi}{V_K}=-1\times 10^{-3}$.
              }
         \label{Fig:SB4_Psi}
 \end{figure}

In our fiducial simulation, we first confirm that $\Psi(R=R_{\mathrm{max}})$ is constant during the entire run, indicating that the total flux within the numerical domain is conserved. We show $\Psi$ in Fig.~(\ref{Fig:SB4_Psi}) as a function of the radial coordinate. In this figure, each contour line corresponds to the midplane footpoint of a poloidal magnetic field line. We see that the field lines are advected inwards. Eventually, the magnetic flux accumulates in the central hole of the domain, but is not lost.

Given that magnetic flux accumulates towards the disk centre, it is desirable to measure an effective velocity of magnetic field transport, $v_\Psi$, from our simulation. We compute this effective advection speed assuming that the magnetic flux can be modelled following a simple advection equation, as proposed by \cite{guilet2012}
\begin{equation}
    \label{eq:Psi_transport}
    {\pdv{\Psi}{t}} + v_{\Psi}\pdv{\Psi}{R} = 0.
\end{equation}
This equation is {only} a phenomenological equation designed to evaluate the typical advection speed of the flow. Using the above definition, we compute $v_\Psi$ and then average it between $t_a=318 T_\mathrm{in}$ and $t_b=955T_\mathrm{in}$, which provides
\begin{equation}
    v_{\Psi} \sim - 1 \times 10^{-3} V_K,
\end{equation}
valid for $R \in[2,11]$. For $R>11$ we cannot conclude on magnetic flux advection, since we did not integrate for long enough to get a measurable deviation from the initial condition. As expected from Fig.~(\ref{Fig:SB4_Psi}) the advection velocity is negative within the inner regions of the disk. Moreover, the advection velocity follows a radial dependency close to Keplerian. We cross-checked the numerical value of the magnetic field advection velocity by fitting the contours of Fig.~(\ref{Fig:SB4_Psi}) with 
\begin{equation}
\label{eq:fit}
    R_\Psi(t) = R_i\left[1- \frac{3\Omega_K(R_i)}{2}\frac{v_\Psi}{V_K}(t-t_i)\right]^{\frac{2}{3}}
\end{equation}
where $R_\Psi$ is the position of the contour, $R_i$ is its initial position and $t_i$ is the time the system needs to stabilise  the transient due to the initial condition and start advecting magnetic flux, a delay visible in Fig.~(\ref{Fig:SB4_Psi}). Equation (\ref{eq:fit}) is the integrated form of an equation describing a field line advected at a velocity proportional to the local Keplerian velocity, $\dot{R}_\Psi=v_\Psi(R)$. These fits show a reasonable agreement with the data (Fig.
~\ref{Fig:SB4_Psi}, dashed line), which suggests that magnetic field lines are advected in a self-similar manner once the local equilibrium is reached ($t_i \sim 30 T_K(R)$).

 Since $ v_{\Psi} \sim v_{\mathrm{acc}}$, the time scale for the magnetic field transport, namely $t_\Psi = 10^{3}/\Omega_K(R) = 160 T_K(R)$, is comparable to the accretion time scale and substantially longer than the dynamical time scale.

\subsection{Dynamical equilibrium}
\label{sec:dyn_equi}
\begin{figure*}[t]
  \centering
  \includegraphics[width=0.48\hsize]{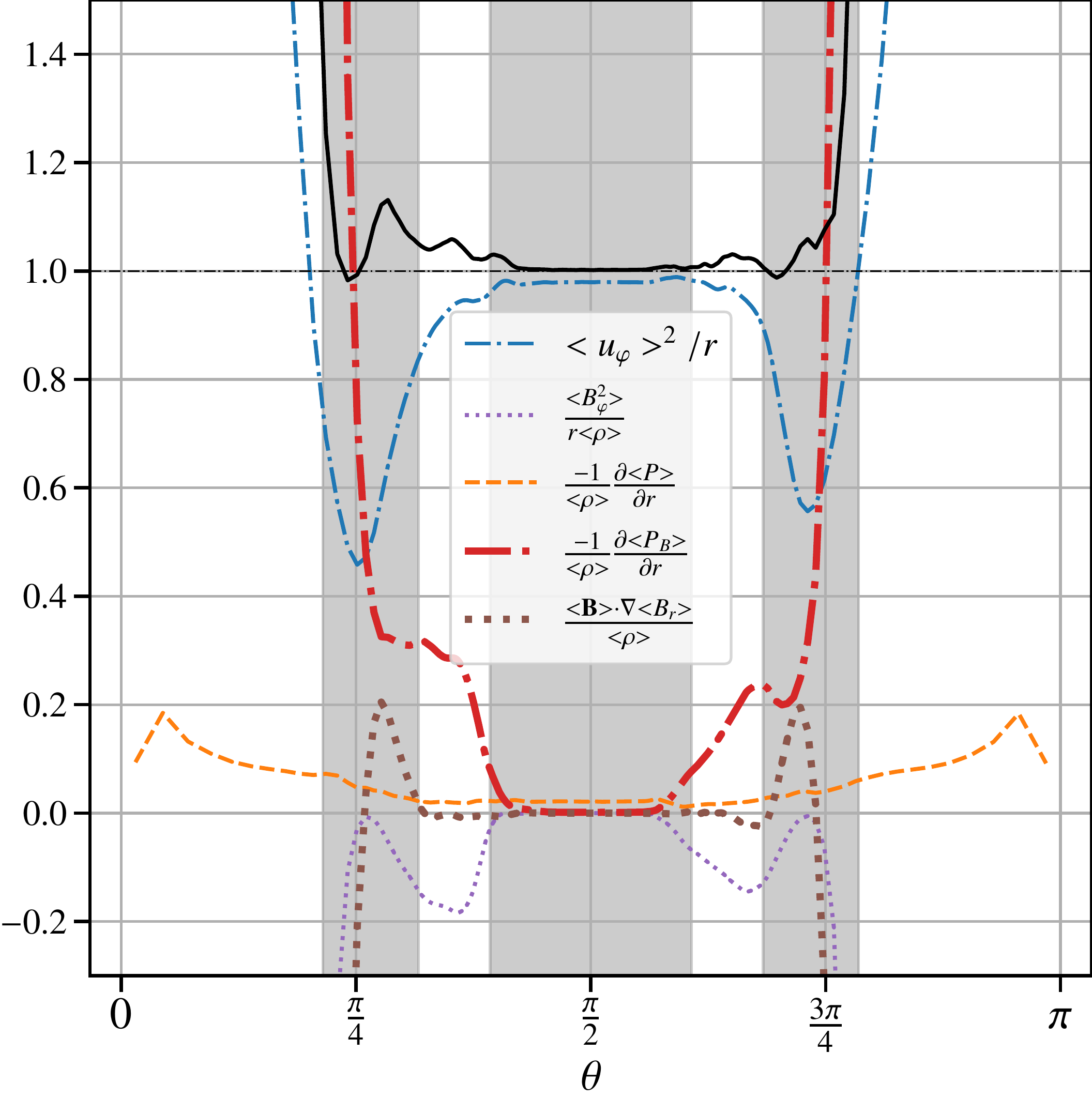}
  \includegraphics[width=0.48\hsize]{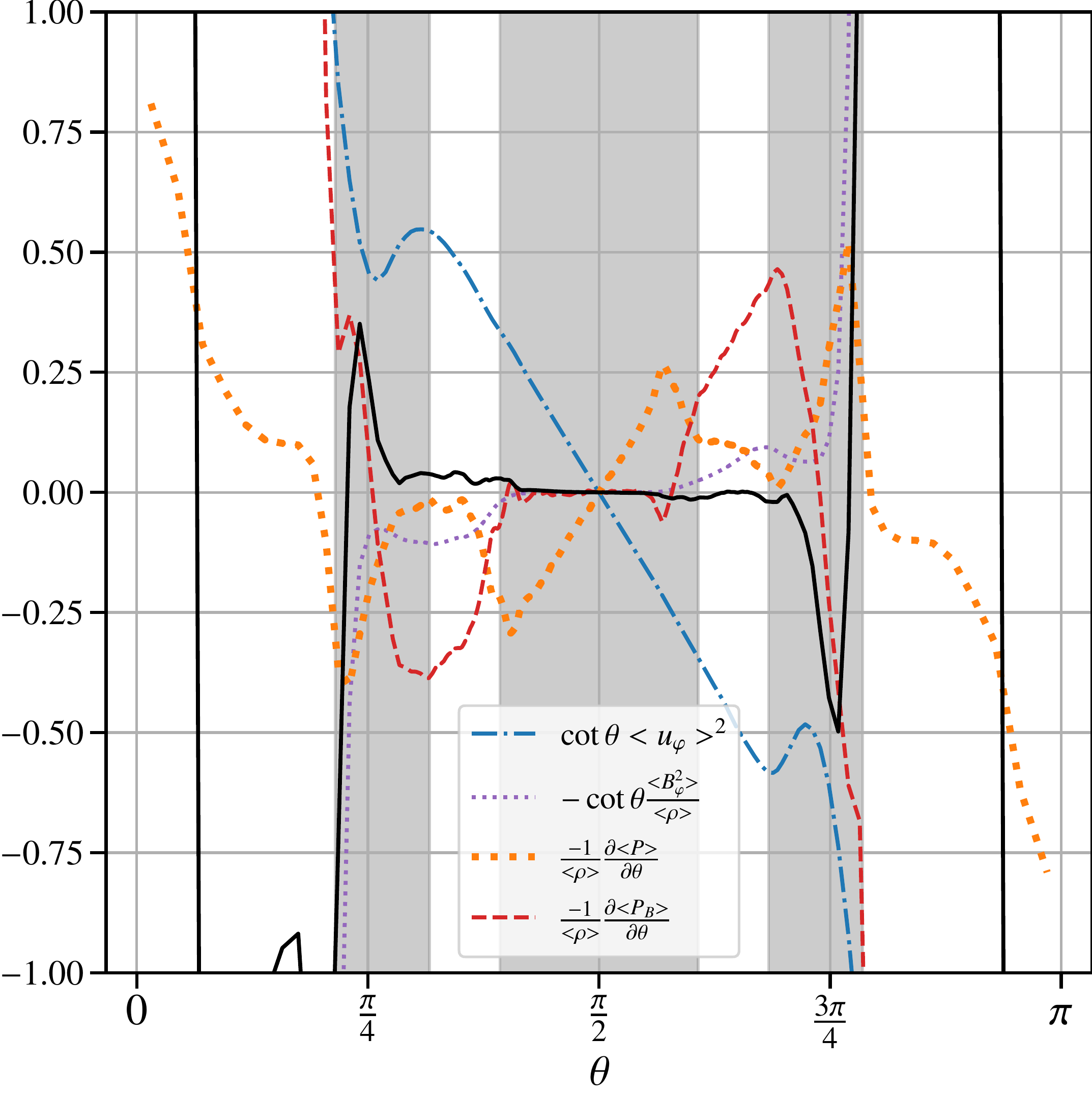}
      \caption{Left: Radial equilibrium (eq.~\ref{eq:rad_equi}). Right: Latitudinal equilibrium (eq.~\ref{eq:lati_equi}) as a function of the latitudinal coordinate $\theta$. The terms are averaged between $r_1=8$ and $r_2=14$ after being normalised to $g$ (left) or $gr$ (right). If the system is in equilibrium the sum of the terms, the black solid line , should be equal to $1$ (left) or $0$ (right).
      The shaded areas denotes the turbulent regions.
              }
         \label{Fig:equilibrium}
 \end{figure*}

Assuming the total poloidal flow acceleration is negligible we get the following radial and latitudinal equilibria from Eq.~(\ref{eq:MradMom})
\begin{align}
\nonumber    \frac{\mean{u_\varphi}^2}{r}-\frac{\mean{B_\varphi}^2+ \mean{\de B_\varphi^2}}{4\pi r\mean{\rho}} -\frac{1}{\mean{\rho}}\pdv{\mean{P}}{r}-\frac{1}{\mean{\rho}}\pdv{\mean{P_B}}{r}\\
        \label{eq:rad_equi}
    \quad +\frac{1}{4\pi\mean{\rho}}\vmean{B_p}\cdot\vgrad{\mean{B_r}} = g,\\
   \label{eq:lati_equi}
    \cot\theta\left[\mean{u_\varphi}^2-\frac{\mean{B_\varphi}^2 +\mean{\de B_\varphi^2}}{4\pi\mean{\rho}}\right]-\frac{1}{\mean{\rho}}\pdv{\mean{P}}{\theta}-\frac{1}{\mean{\rho}}\pdv{\mean{P_B}}{\theta}=0,
\end{align}
where $g = GM/r^2$ and $\mean{P_B} = \left(\mean{B}^2+\mean{\de B^2}\right)/8\pi$ is the magnetic pressure, which contains a laminar, $\mean{B}^2$, and a turbulent, $\mean{\de B^2}$, contribution. We neglect the tension due to the latitudinal magnetic field in Eq.~(\ref{eq:lati_equi}) due to its small impact in the latitudinal equilibrium. Figure \ref{Fig:equilibrium} shows the latitudinal profile of the radial and latitudinal equilibria, respectively normalised to $g$ and $gr$.
 
We find that the disk region is in radial and latitudinal equilibrium thanks to the balance between the centrifugal force, ${\mean{u_\varphi}^2}/{r}$ or $\cot\theta\mean{u_\varphi}^2$, the gravity and the thermal pressure gradient. Hence, the disk is in hydrostatic balance as is expected from a weakly magnetised disk.

{As we enter} the accreting atmosphere the thermal pressure gradient becomes close to negligible, {when compared with the magnetic forces}. The radial and latitudinal equilibriums are then enforced by a balance between gravity, the centrifugal force, the magnetic pressure gradient and the hoop stress, $\mean{B_\varphi^2}/(4\pi \mean{\rho} r)$ or $\cot\theta \mean{B_\varphi^2}/(4\pi \mean{\rho})$. 
The radial equilibrium is maintained up to the base of the outflow ($\theta\simeq \pi/2\pm \pi/4$). From this point on, MHD acceleration is no longer negligible and Eqs.~(\ref{eq:rad_equi},\ref{eq:lati_equi}) break down.

\begin{figure}[h!]
   \centering
   \includegraphics[width=\hsize]{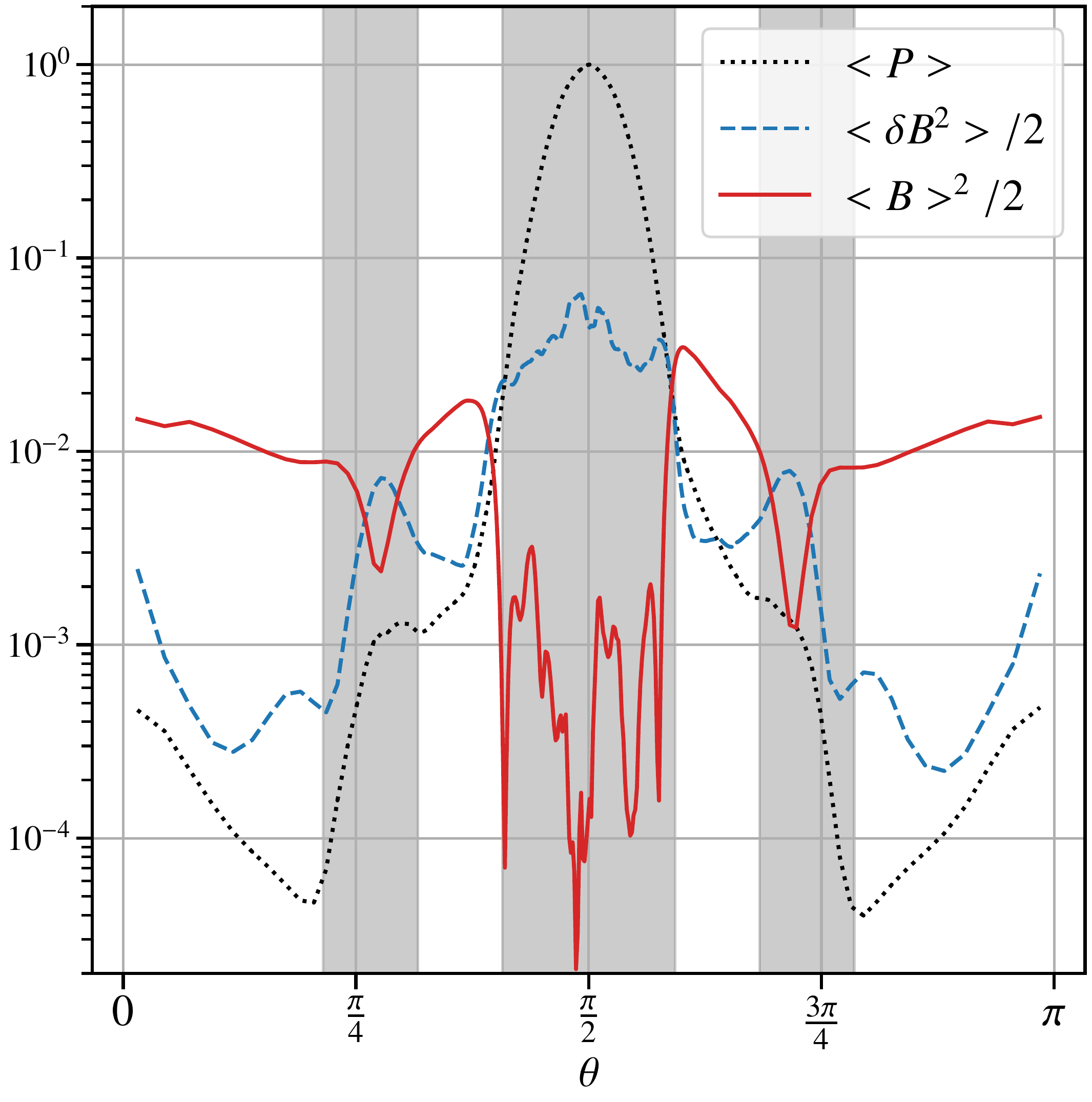}
      \caption{Comparison of the different pressure terms normalised to the thermal pressure within the disks mid-plane as functions of the latitudinal coordinate. The shaded areas denote the turbulent regions.
              }
         \label{Fig:pressure}
 \end{figure}

Within the laminar and turbulent atmosphere, the system is no longer in hydrostatic equilibrium, but in magnetostatic equilibrium. This  magnetostatic equilibrium emerges from a combination of the laminar and turbulent magnetic pressure (Fig.~\ref{Fig:pressure}).
This situation is similar to what was proposed by \cite{begelman2015}, a disk in magnetostatic equilibrium where the magnetic pressure emerges as a consequence of turbulence.
Indeed, within this region, we see the emergence of a powerful toroidal field, $20$ times the vertical field at the disk mid-plane ( Fig.~\ref{Fig:regin_def} bottom), that changes sign within the turbulent atmosphere. 

The strong toroidal field within the laminar atmosphere is consistent with the fact that the laminar atmosphere is in ideal MHD. In this region, accretion drags the poloidal field lines inwards. Since the fluid is plunging into a region with higher angular momentum, the toroidal speed increases, which leads to enhanced shear for the toroidal magnetic field, increasing its magnitude \citep{Zhu_Stone}. 

The emergent toroidal field is consumed by the dynamical equilibrium. Indeed, it is the gradient of the magnetic pressure that supports the turbulent atmosphere \citep{mish2020} and achieves latitudinal equilibrium within the accreting atmosphere. Hence, as we move further up into the turbulent atmosphere the toroidal magnetic field decreases (Fig.~\ref{Fig:regin_def}). Within the turbulent atmosphere the toroidal field changes sign, which might be surprising at first sight. However, such a change of sign is mandatory for launching an MHD outflow, as first realised by \cite{ferr95}. This allows to switch from the underlying  braking torque to an accelerating MHD torque that drives the outflow. It corresponds also to a radial electric current which is circulating outwardly at the TA layer. The origin of this current is the electromotive force driven by the rotating disk across the large scale $B_z$ field (see also  \citealt{ferr97} and \citealt{jacquemin-ide_magnetically-driven_2019} for more details on the electric current vertical profile).

Overall, it is clear from Fig.~(\ref{Fig:pressure}) that the dynamical importance of the turbulent pressure cannot be understated. Indeed, \cite{ferr95} showed that the mass loaded onto the field lines depends crucially on the disk pressure which pushes material up to the 'surface', and loads the field lines. It is clear from Fig.~(\ref{Fig:pressure}) that the laminar magnetic pressure is compressing the disk surface, as shown by \cite{ferr95}. However, the effect of this compression seems nonexistent in Fig.~(\ref{Fig:equilibrium}). This is a consequence of the presence of a important turbulent magnetic pressure in the disk. Indeed, the compression of the disk surface due to the laminar magnetic pressure, $\mean{B}^2$, is compensated by the turbulent magnetic pressure, $\mean{\de B^2}$ within the disk (Fig.~\ref{Fig:pressure}). This ensures that material can flow relatively freely from the disk to the atmosphere and enhances the mass loading into the laminar atmosphere region. The same process is at play for the transition between the turbulent atmosphere and the outflow. However, in this case it is the combined effect of thermal pressure and the turbulent magnetic pressure (Fig.~\ref{Fig:pressure}) that allows the loading of the magnetic field lines at the base of the outflow. The reason for the reappearance of the turbulent pressure within the turbulent atmosphere will be described in section \ref{sec:dis_turb}. 

\subsection{Super fast magneto-sonic wind}
\label{sec:wind}
In the regions where the flow is approximately laminar and steady-state, it is possible to define a set of MHD invariants which consists of conserved physical quantities along the field lines. These invariants characterise  MHD outflows and allow direct comparisons to be made with other outflow solutions.

Under these conditions, Eq.~(\ref{Eq:Mass_Con}) reduces to:
\begin{equation}
    \label{eq:Inv_mass}
    \mean{\vec{u_p}} =  \frac{\Tilde{\eta}}{4\pi\mean{\rho}}\mean{\vec{B_p}},
\end{equation}
where  $\Tilde{\eta}$ measures how much mass is being loaded onto the field lines. It can be transformed into a dimensionless mass loading parameter 
$\kappa\equiv \Tilde{\eta}R_{SM}\Omega_K(R_{SM})/B_{z,SM}$, where $R_{SM}$ is the anchoring radius of the magnetic field line and the SM stands for quantities evaluated at the slow magneto-sonic surface. 

Following the same regime of approximation, we can also write the $\varphi$ component of the induction equation as
\begin{equation}
    \label{eq:Inv_om}
    \Omega_{\star} = \mean{\Omega} -\Tilde{\eta}\frac{\mean{B_\varphi}}{4\pi\mean{\rho} R} ,
\end{equation}
where $\Omega_{\star}$ can be interpreted as the angular velocity of the magnetic field lines. This can be recast into a dimensionless number $\omega=\Omega_{\star}/\Omega_K(R_{SM})$. 

Similarly, the angular momentum invariant is deduced from Eq.~(\ref{eq:MAngMom})
\begin{equation}
\label{eq:Inv_ang}
    L = \Omega_\star R_A^2 =  \mean{\Omega} R^2 - \frac{R}{\Tilde{\eta}}\mean{B_\varphi},
\end{equation}
where $R_A$ is the Alfv\'en radius of the anchored field line, which is the radius where the flow following the field line becomes super Alfv\'enic. We can then define the dimensionless invariant $\lambda = L/(R\Omega_K)|_{SM}$, known as the magnetic lever arm. It is a measure of the angular and poloidal acceleration of the wind by the disk mediated by the magnetic field. 

Finally, by projecting Eq.~(\ref{Eq:Mom_Con}) along the poloidal magnetic surface we can define the Bernoulli invariant  
\begin{equation}
    \label{eq:Inv_ene}
    E =\frac{\mean{u}^2}{2}+\Phi_G - \Omega_\star \frac{R\mean{B_\varphi}}{\Tilde{\eta}}+ w,
\end{equation}
 where $w$ is the thermal energy term which can be written as
\begin{equation}
    w = \int \frac{\vgrad{\mean{P}}}{\mean{\rho}}\cdot\diff{\vec{l}},
\end{equation}
where $l$ is the coordinate along the field line. It is important to note that, when heating is allowed, $w$ is a function of the full stream line as a result of its integral form, therefore it cannot be computed from a single point within the field line as the other terms. 

However, in our case, the heating contribution $w$ is negligble so that the Bernoulli invariant is fully determined at the SM (launching) point.
We  normalise this invariant with respect to the gravitational energy at the mid-plane defining $ e = \frac{E}{E_{K,SM}}$, where $E_{K,SM}=\Omega_K^2(R_{SM})R_{SM}^2/2$. 

\begin{figure}[h!]
    \centering{
   \includegraphics[width=\hsize]{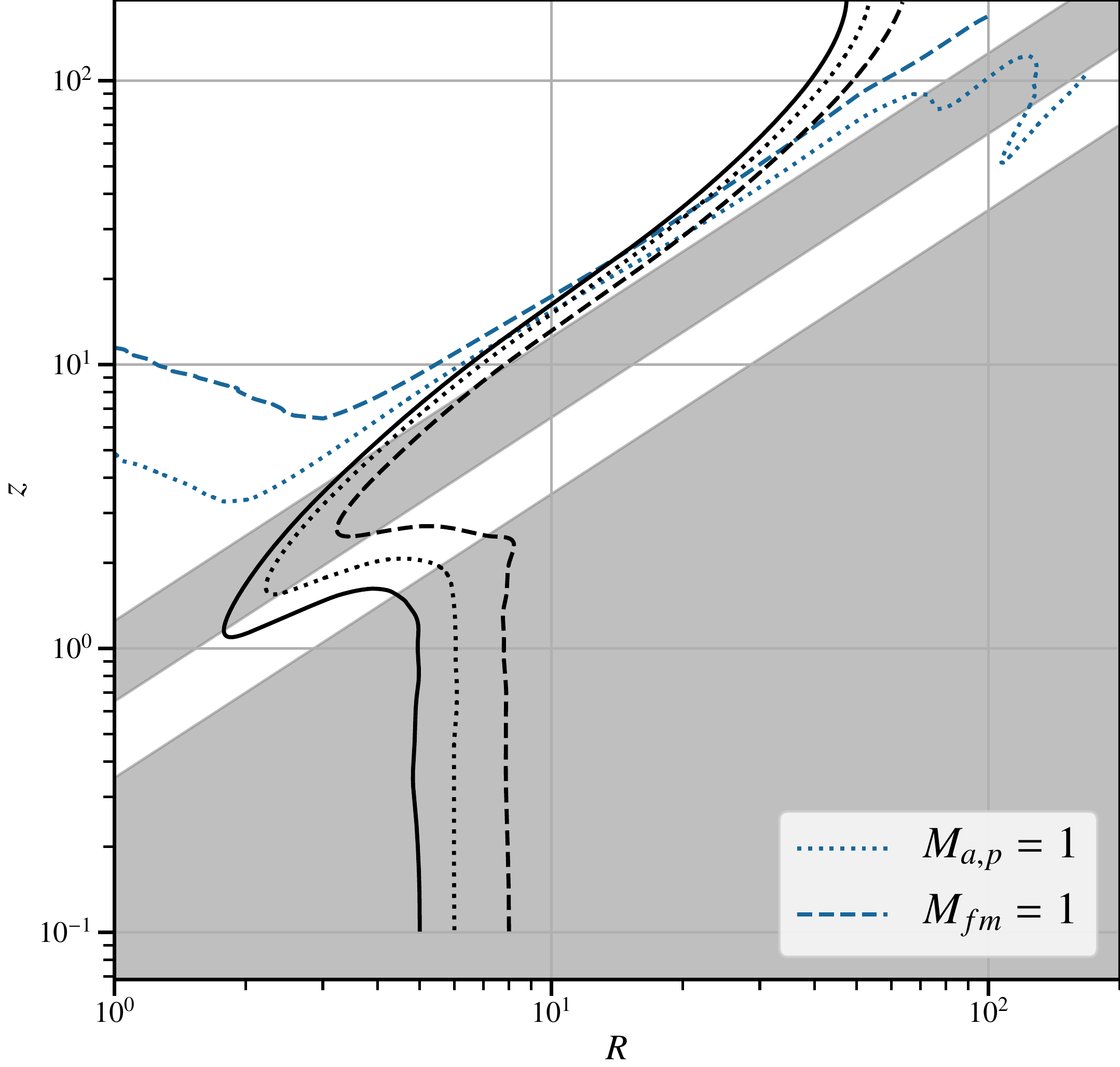}}\\
   \includegraphics[width=0.95\hsize]{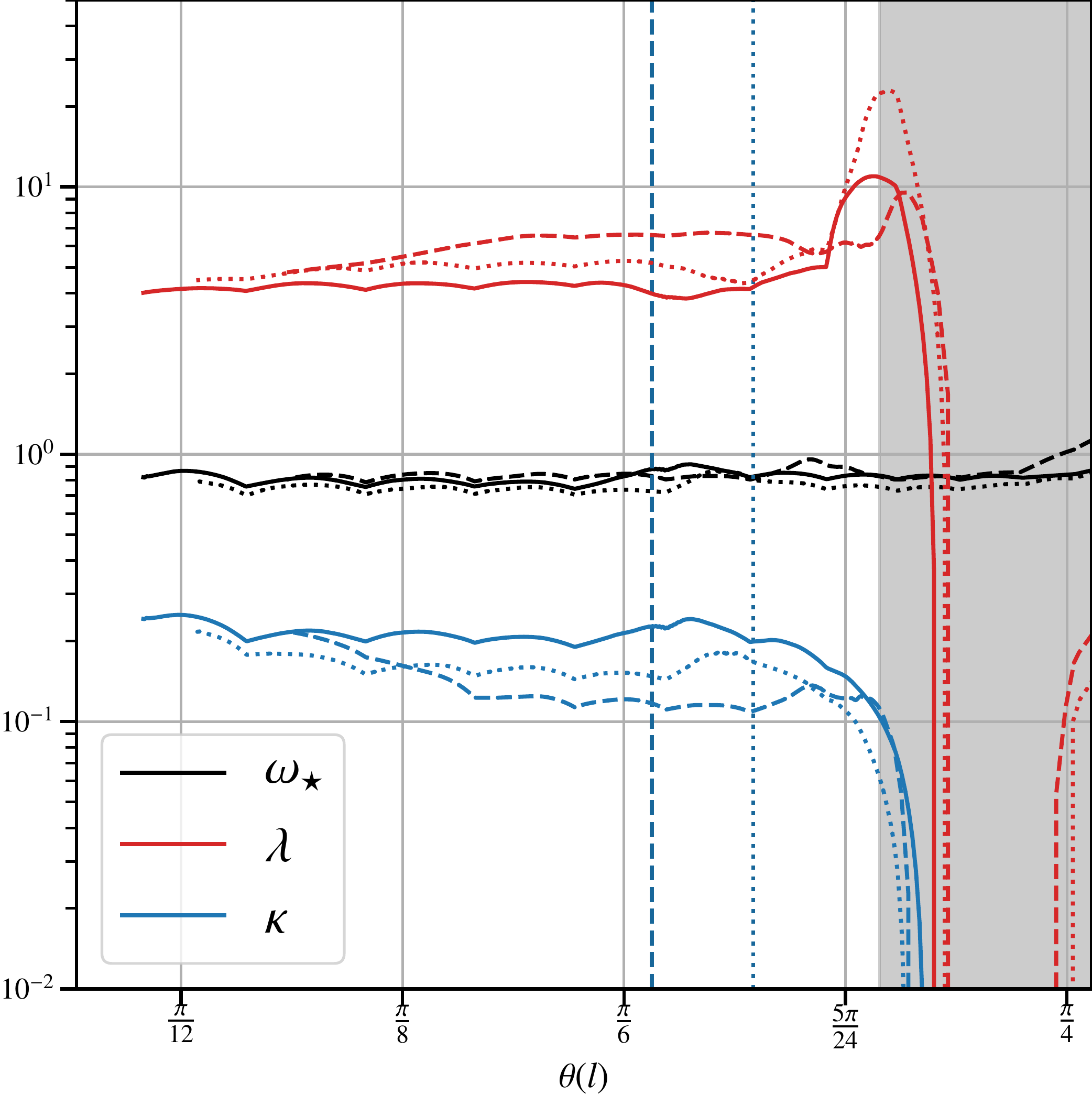}\\
      \caption{Top: mean poloidal magnetic field lines in the ($R$,$z$) plane. The blue lines define the different critical surfaces (see section \ref{sec:Global_picture}) and they are represented as vertical lines in the lower panel. The grey zones determine the turbulent zones defined in section \ref{sec:Global_picture}. The lower zone corresponds to the disk and the upper zone corresponds to the turbulent atmosphere.
      Bottom: MHD invariants calculated along the field lines of the upper panel as functions of the latitudinal coordinate. The line style of the MHD invariants has a one to one correspondence with the field line where the invariant was calculated. The grey zone corresponds to the end of the turbulent atmosphere. 
     }
         \label{Fig:Invariants}
 \end{figure}
We follow three different field lines originating at $R_0=[5,6,8]$ respectively (Fig.~\ref{Fig:Invariants}, top).  We then compute the ideal MHD invariants using Eq.~(\ref{eq:Inv_mass}-\ref{eq:Inv_ene}) across the 3 different field lines anchored at the radii $R_{SM}\simeq[3.5,4,7.5]$.

We show in figure \ref{Fig:Invariants} (bottom) the MHD invariants computed along the field lines. We find that the invariants are approximately constant once the outflow leaves the turbulent atmosphere. The field lines closest to the central object are the ones that exhibit the least variability in their invariants. This is to be expected, as not all anchored field lines are created equal: the farther they are from the central object the longer their MHD invariants will take to converge. 

Several features of the wind can be deduced from its MHD invariants (see also Tab.\ref{tab:inv}): 

Firstly, the rotation invariant $\omega\lesssim 1$, indicates that field lines are rotating close to the Keplerian speed of their anchoring radius, or slightly slower. This is consistent with the values expected in self-similar models \citep{jacquemin-ide_magnetically-driven_2019}.

Secondly, the angular momentum invariant $\lambda \simeq 4$---$6$, so the wind is effectively free ($\lambda>3/2$). This provides $R_A^2/R_{SM}^2=\lambda/\omega\sim8$ which is similar to the value found in \cite{Zhu_Stone}. The magnetic lever arm can be used to estimate the terminal velocity of the outflow using $u_{p\infty}\simeq R_{SM}\Omega_K(R_{SM}) \sqrt{2\lambda-3}$ \citep{blan82}. Using this expression we find $u_{p\infty}\sim 3R_{SM}\Omega_K(R_{SM})$.

Finally, the mass loading invariant $\kappa\sim 0.1$ implies that the energetic content of the wind at the base of the outflow is magnetically dominated instead of being kinematically dominated, consistent with a jet-like outflow. Since the laminar atmosphere is in ideal MHD, we can also compute the value of the mass loading invariant within it, finding $\kappa\sim 8$. The value of $\kappa$ is an order of magnitude larger within the laminar atmosphere that within the outflow. This is consistent with a laminar atmosphere that is so heavily loaded with mass that it falls towards the central object. It is that fall that leads to the generation of a large scale toroidal magnetic field within this region.

\begin{figure}
    \centering
   \includegraphics[width=\hsize]{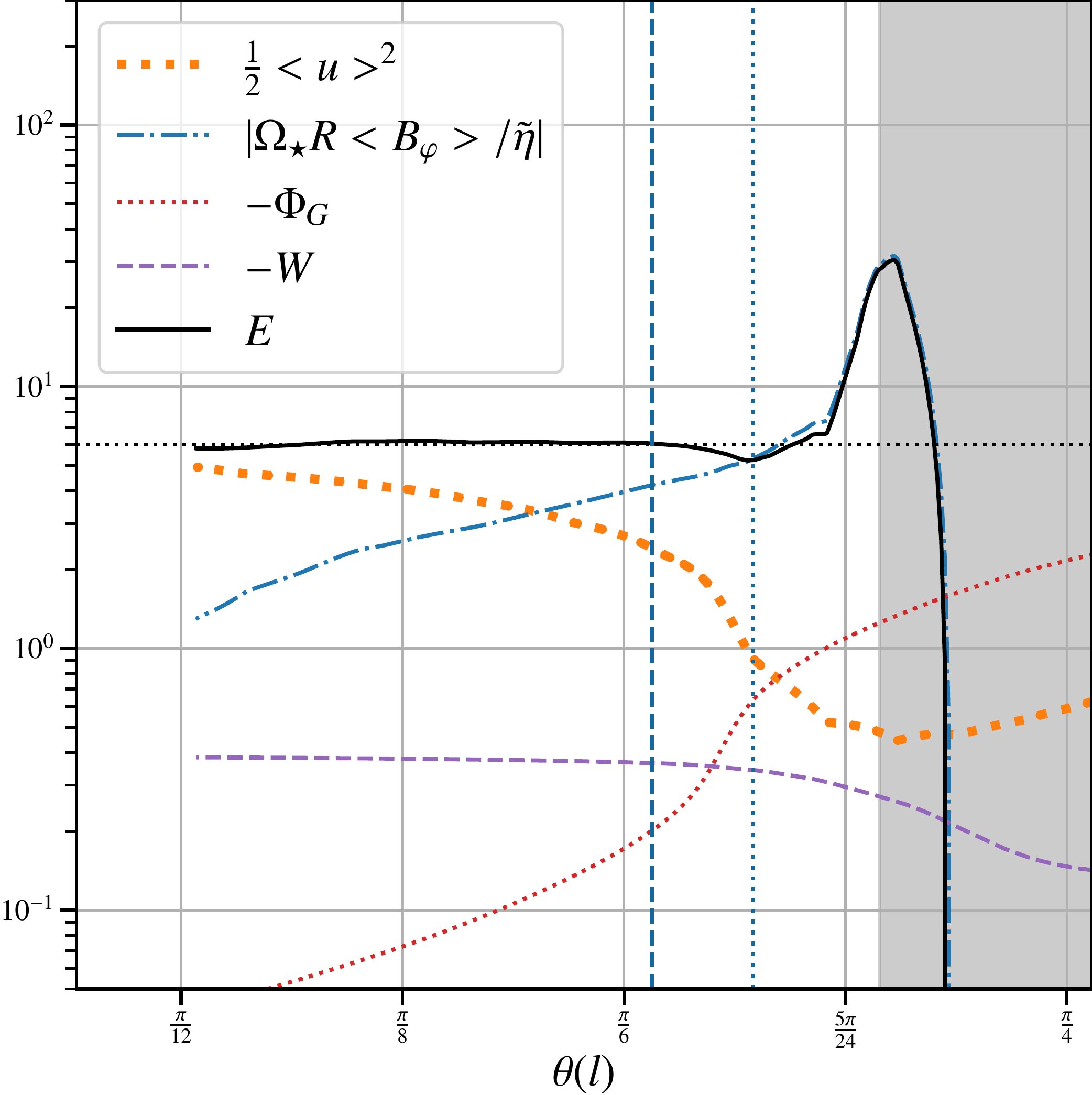}
    \caption{Bernoulli invariant and its components for the field line originating at $R_0=6$ as functions of the latitudinal coordinate $\theta$. All quantities are normalised to $E_{K,SM}=\Omega_K^2(R_{SM})R_{SM}^2/2$. The shaded grey area corresponds to the turbulent atmosphere, the end of the turbulent atmosphere coincides with the SM surfaces. The vertical lines correspond to the different critical surfaces. The vertical dashed line is the fast magneto-sonic surface while the vertical dotted line is the Alfv\'enic surface, consistent with the notation figure \ref{Fig:Invariants}.}
    \label{fig:Bernoulli}
\end{figure}

The Bernoulli invariant allows us to characterise the wind energetics and its drivers. Fig.(\ref{fig:Bernoulli}) shows the different contributions. The conservation of the Bernoulli invariant is striking, especially when compared to the other invariants. The positive sign of the Bernoulli invariant indicates that the flow is free from the potential well and continues its propagation up to infinity, consistently with $\lambda>3/2$. When the flow reaches the end of our simulation box, its energy content is dominated by the kinetic component and the magnetic energy has been mostly consumed, so the outflow is close to its asymptotic state. This confirms what was deduced from the magnetic lever arm $\lambda$, mainly that $u_{p\infty}\sim 3R_{SM}\Omega_K(R_{SM})$, since $e\sim6$.

It is also clear from figure \ref{fig:Bernoulli} that the outflow is cold, i.e the thermal pressure term $w$ is negligible in the outflow energetics. Indeed, at the wind launching point, the outflow is dominated by the magnetic energy, and even the gravitational energy is negligible at this location. Finally, we also compute the invariants on the south pole of our system but originating on the same radii, $R_0=[5,6,8]$, and find very similar values, confirming the symmetric nature of the system.

\cite{jacquemin-ide_magnetically-driven_2019} propose a criterion to distinguish between jets and winds, which essentially quantifies whether the outflow can self collimate thanks to magnetic forces ('jet') or not ('wind'). They use the definition of the jet magnetisation
\begin{equation}
    \sigma = \left . \frac{- \Omega_* R \mean{B_\varphi} \mean{B_p}}{\left(\frac{\mean{u}^2}{2}+H\right) \mean{\rho}\mean{ u_p} 4\pi} \right |_{SM},
\end{equation}
defined as the ratio of the MHD poynting flux to the kinetic plus thermal energy flux at the base of the outflow. If $\sigma>1$ the outflow can be categorized as a jet, in the contrary it is a wind.
Using figure \ref{fig:Bernoulli}, we derive  $\sigma\sim8$ which is compatible with a jet like outflow and is consistent with the observed collimation. 

\subsection{Origin of turbulence in the atmosphere}
\label{sec:dis_turb}
As described in section \ref{sec:Global_picture} the accreting atmosphere is divided into two distinct regions: a turbulent and a laminar atmosphere. This configuration, where turbulence is localised in a layer above the disk is pretty surprising and has never been addressed so far. It suggests that within the laminar atmosphere, the source of turbulence must first be quenched and then reinvigorated further up, within the turbulent atmosphere. Because the turbulent atmosphere is strongly magnetised ($\mean{\beta_p} \lesssim 1$), several MHD instabilities could in principle be invoked to explain this. However, since the main source of free energy for turbulence is sill the shear, the MRI remains a very serious contender.

\begin{figure}
    \centering
   \includegraphics[width=\hsize]{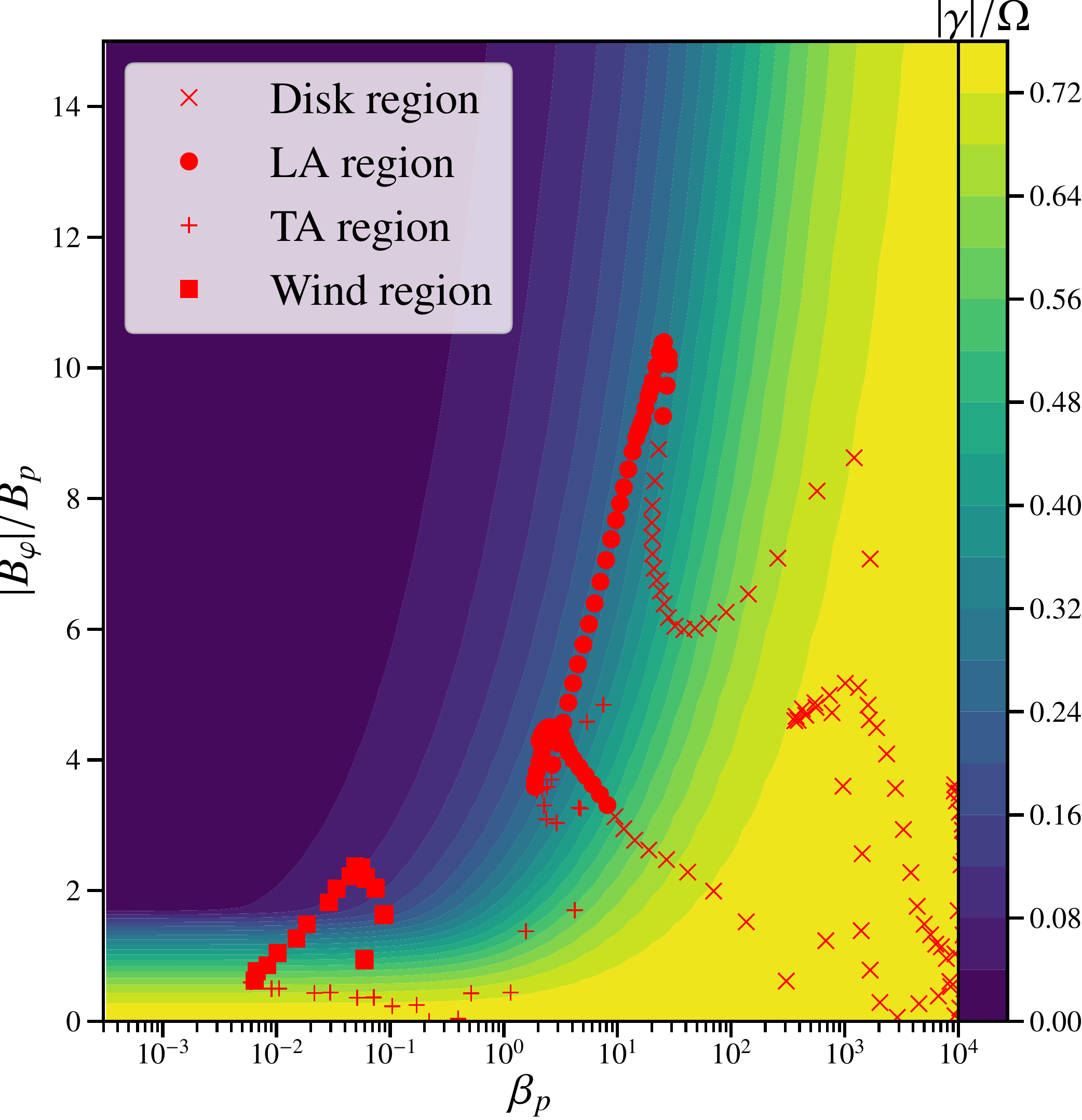}
    \caption{Maximum growth rate of the compressible MRI as a function of the poloidal plasma beta and the ratio between the toroidal and poloidal magnetic field, calculated using Eq.~(\ref{Eq:theMeat}). The red points correspond to the values of the ratio of the magnetic field and the poloidal plasma beta for different regions within our simulation averaged between $r_1=8$ and $r_2=12$ (see Fig.\ref{Fig:regin_def}). We find that laminar regions are characterised by reduced MRI growth rates ($\lesssim 0.25\Omega_K$).
    }
    \label{fig:growth}. 
\end{figure}

It is widely believed that the MRI is a weak field instability and that it is quenched once $\mean{\beta_p} \lesssim 1$. This condition is however only geometrical and based on thin disks, requiring MRI modes to be confined within the vertical thickness of the disk (set by the hydrostatic equilibrium). If one considers a system with a size not determined by thermal pressure, this geometrical constraint vanishes and the MRI can in principle exist for low $\mean{\beta_p}$. This is the situation studied by \cite{kim2000}, who confirmed that in the limit $\mean{\beta_p}\rightarrow 0$ the MRI still exists. However, in contrast to the weak field regime, the MRI can be suppressed by a too strong toroidal field. The stability condition in this low beta limit reads
\begin{equation}
\label{Eq:MRI_quench}
    \frac{\mean{B_\varphi}^2}{\mean{B_z}^2}>\frac{3}{4} \quad \mathrm{with}\quad \mean{\beta_p}\rightarrow 0 \quad \Rightarrow \quad \mathrm{stability}.
\end{equation} 
This condition is verified in the laminar atmosphere, (Fig.~\ref{Fig:regin_def}, bottom), indicating that it should be stable to the magneto-rotational instability.
For the sake of completeness we derive Eq.~(\ref{Eq:MRI_quench}) as well as the dispersion relation for the compressible MRI, Eq.~(\ref{Eq:theMeat}), in Appendix \ref{A:MRI_lowbeta}.  We then use this dispersion relation to compute the maximum growth rate of the MRI as a function of the poloidal plasma beta and the ratio between the toroidal and poloidal magnetic field (Fig.~\ref{fig:growth}). To compare this to our numerical results, we use the profiles of the magnetic field components and pressure in our simulations to characterise the value of the MRI growth rate within each region. From Fig.~(\ref{fig:growth}) we find that the regions where MRI grows at a sufficient rate ($\gtrsim 0.25\Omega_K$) correspond to the zones described as turbulent in section \ref{sec:Global_picture}, while the laminar regions show lower growth rates. We thus conclude that MRI could be the main driver of the turbulence within the turbulent atmosphere. 

As explained in section \ref{sec:dyn_equi} the emergence of a powerful toroidal field is consistent with the laminar atmosphere being in ideal MHD. Furthermore, the decrease of the toroidal magnetic field within the turbulent atmosphere is also consistent with the magnetostatic equilibrium of the atmosphere. Hence, we can deduce that the re-ignition of turbulence is required for the establishment of a stationary magnetostatic equilibrium.

The re-activation of turbulence leads to anomalous resistivities within the turbulent atmosphere that allows accretion through the poloidal field lines and a steady state transition between accretion and ejection \citep{ferr95}. It is therefore an essential ingredient for launching the wind.

\section{Formation of ring like structures.}
\label{sec:ring}
\begin{figure}[h!]
  \centering
  \includegraphics[width=\hsize]{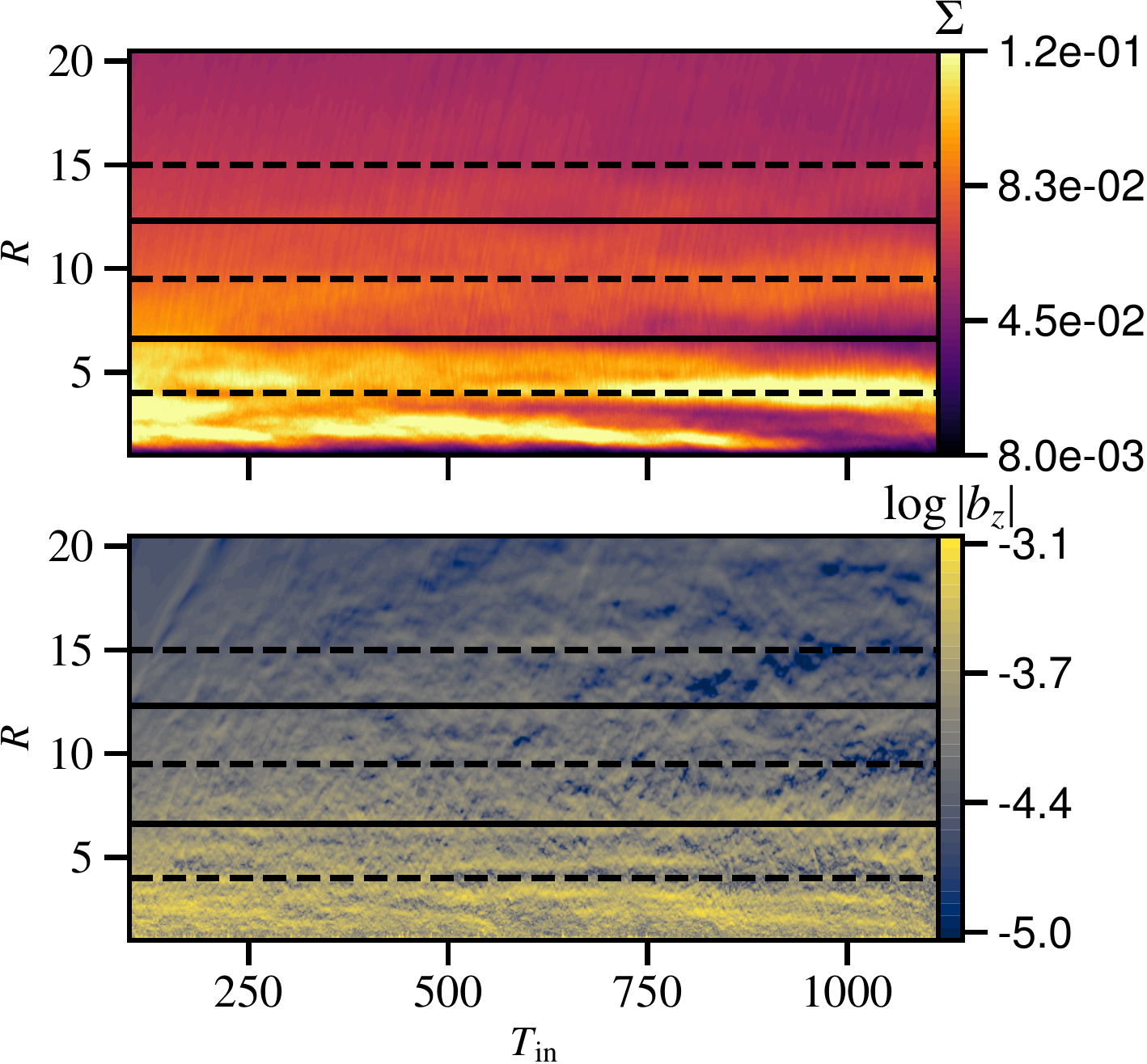}
      \caption{Space time diagram of the column density $\Sigma$ (up) and the vertically averaged vertical magnetic field $b_z$ (down) as functions of the time in units of orbits at the inner most radius and of the cylindrical radial coordinate. The solid black lines represent the gaps located at $R\simeq[6.6,12.3]$ while the dashed black lines represent the rings located at $R\simeq[4,9.5,15]$.
              }
         \label{Fig:Rspatime}
 \end{figure}

We discussed in section \ref{sec:transport} that the disk region has not achieved a rigorous steady state. Indeed, the plasma within the disk is seen to concentrate into ring like structures, a special type of self-organisation. 
To understand the formation and evolution of the ring like structures, it is useful to vertically and azimuthally average the equation of mass conservation. Equation (\ref{Eq:Mass_Con}) leads to
\begin{equation}
\label{eq:con_mas_vert}
   \pdv{\Sigma}{t} + \frac{1}{r^2}\pdv{r^2\overline{\phimean{\rho u_r}}}{r} + \left[ \sin\theta\phimean{\rho u_\theta}\right]_{\theta_1}^{\theta_2} = 0,
\end{equation}
where $\Sigma = \overline{\phimean{\rho}}$ is the column density of the disk and the vertical and azimuthal average $\overline{X}$ is defined in Appendix \ref{sec:Comp_flux}.  We compute the vertical averages within the latitudinal wedge $\theta_{2,1} =\pi/2 \pm \arctan(3.5\epsilon)$, equivalent to the disk region (section \ref{sec:Global_picture}).
We show $\Sigma$ in a space time diagram in the top panel of Fig.~(\ref{Fig:Rspatime}).

We concentrate our analysis on the gaps, minimum of $\Sigma$, located at $R\simeq[6.6,12.3]$ as well as the rings, maximum of $\Sigma$ located at  $R\simeq[9.5,15]$. Indeed, the ring located at $R\simeq2$, and to a certain extent the one located at $R\simeq4$, are within close range of the inner boundary which could bias their analysis. Nonetheless, we find that the innermost ring-like structure ($R\sim4$) starts to emerge at around 200  Keplerian orbits of the innermost radii (25 local orbits) while the farther rings (at $R\sim10$) appear at  800 Keplerian orbits of the innermost radii (25 local orbits). Hence, we can conclude that the local time scale of self-organisation $t_{so}(R)\simeq 25 T_K(R) = 160/\Omega_K(R)$.

Once formed, the ring structures survive for the entire duration of the simulation. They also exhibit a weak radial inward migration, more evidently seen for the $R\sim2$ ring. We measure the migration speed of the  $R\sim2$ ring to be $v_{\mathrm{ring}}\simeq5\times 10^{-4} V_K(R)$ by measuring the radial velocity of its density maximum. This velocity is twice smaller than $v_\mathrm{acc}$. It is unclear whether this drift is a bias of the boundary condition or a consequence of the intrinsic physics.  It is important to note that while $R\sim4$ may seem to be migrating inwards we were not able to measure any radial drift. 

The magnetic field is probably linked to the production of ring-like structures.
To understand the role of the magnetic field in the self-organisation mechanism, we compute the vertically average vertical field  $b_z = \overline{\phimean{B_Z}}/(7h)$ and show its space time diagram in  Fig. \ref{Fig:Rspatime}, bottom. We find that the vertical magnetic field is accumulating within the gaps, and is anti-correlated with the rings. However, the magnetic structure is not as stable with time as the density structure.

To understand if the self-organisation of the disk into ring-like structures is a consequence of radial or latitudinal transport of mass, we need to turn back to Eq.~(\ref{eq:con_mas_vert}).
We average it with respect to time and divide it by the time average of the column density, $\mean{\Sigma} = \overline{\mean{\rho}}$, and the Keplerian angular velocity to find the dimensionless equation
\begin{equation}
\label{eq:Dsig}
    \frac{\Delta \Sigma}{\mean{\Sigma} \Delta t\Omega_K}+ \frac{1}{\mean{\Sigma}r^2\Omega_K}\pdv{r^2\overline{\mean{\rho u_r}}}{r} +\frac{\dot{\sigma}_w}{\mean{\Sigma}\Omega_K}=0
\end{equation}
 where we define $\dot{\sigma}_w =\left[ \sin\theta\mean{\rho u_\theta}\right]_{\theta_1}^{\theta_2}$ as the wind mass loss rate. To simplify the comparison of each term, we have normalised equation (\ref{eq:Dsig}) to the Keplerian frequency.
 
\begin{figure}[h!]
  \centering
  \includegraphics[width=\hsize]{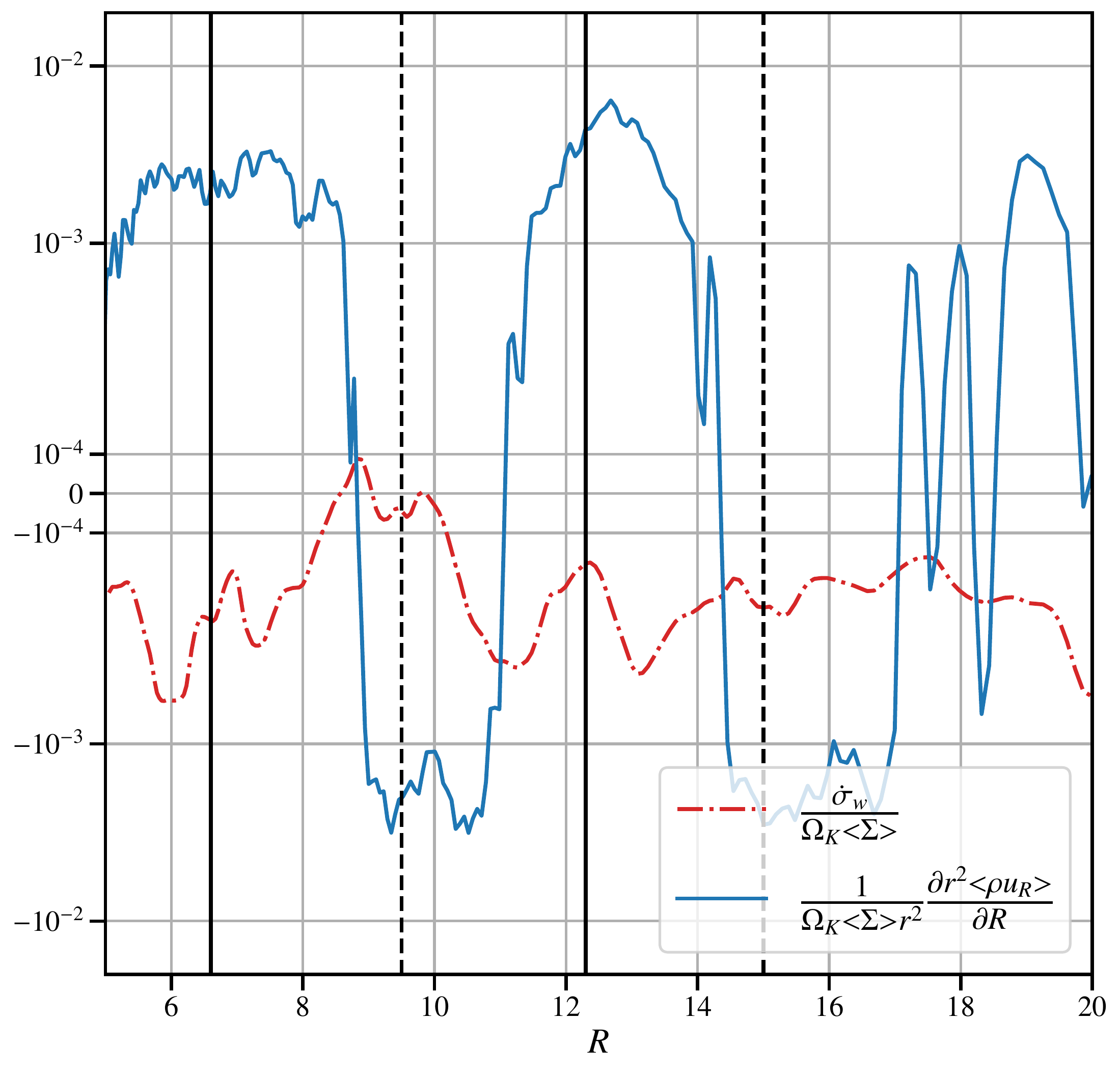}
      \caption{Radial (blue) and vertical (red) contributions to the mass evolution rate in local $\Omega_K$ units as a function of the cylindrical radii, $R$. A positive (negative) term leads to the formation of local gap (ring)
      The solid black lines represent the gaps located at $R\simeq[6.6,12.3]$ while the dashed black lines represent the rings located at $R\simeq[9.5,15]$. We use symmetrical log-scale that becomes linear within the interval $[5\times10^{-5},-5\times10^{-5}]$.
              }
         \label{Fig:div_Macc}
 \end{figure}

Figure (\ref{Fig:div_Macc}) shows the wind mass loss rate as well as the derivative of the radial mass flux. We see that the  wind mass loss rate, $\dot{\sigma}_w$, is essentially negligible, while the radial term, $\partial_r r^2\overline{\mean{\rho u_r}}$ shows a good correlation with the ring structures: the radial locations where the radial term is positive (divergent mass flow) correspond to minima of the column density, while the radial locations where the radial term is negative (convergent mass flow) correspond precisely to maximal column densities. We have checked that the mean radial mass flow forming these structures is driven by the radial stress, following the vertical average of Eq.~(\ref{Eq:AngMom}). This implies that the origin of the self-organisation is due to $\alpha_\mathrm{tu}$. 

 From Fig.~(\ref{Fig:div_Macc}) we can probe the time scale, $t_{so}$, on which this self-organisation takes places by looking at the amplitude of the radial divergence of the radial velocity in units of keplerian frequency. This leads to $t_{so}\sim200/\Omega_K(R) = 30 T_K(R)$ this corresponds quite well with the values measured in Fig.~(\ref{Fig:Rspatime}). Comparing the self organisation time scale with the ones deduced from section \ref{sec:transport} we find $t_{so}\ll t_{\mathrm{acc}}\simeq t_\Psi$. Self-organisation into ring-like structures is not impeded by the global transport mass or magnetic field, it happens on shorter time scales.
 
As mentioned above, the self-organisation is due to radial motions which are driven by the radial component of the turbulent stress tensor. It is tempting to interpret this as a viscous instability, as proposed by \cite{lightman1974}.
Indeed, \cite{lightman1974} showed that, for the system to be unstable to a viscous type instability, the torque needs to be a decreasing function of the surface density, $\pdv{\log\overline{T_{tu,\,r\varphi}}}{\log \Sigma}<0$ . In the case of a strongly diffusive field, the limit where the field is steady, this is equivalent to $\overline{\alpha_\mathrm{tu}}\propto\mean{\beta_{\mathrm{mid}}}^{-1}$, where we define 
\begin{equation}
    \overline{\alpha_\mathrm{tu}} = \frac{\overline{T_{tu,\,r\varphi}}}{\overline{P}},
\end{equation}
and
\begin{equation}
    \mean{\beta_{\mathrm{mid}}}=\mean{\beta_{p}}(r,\theta=\pi/2).
\end{equation}

\begin{figure}[h!]
  \centering
  \includegraphics[width=\hsize]{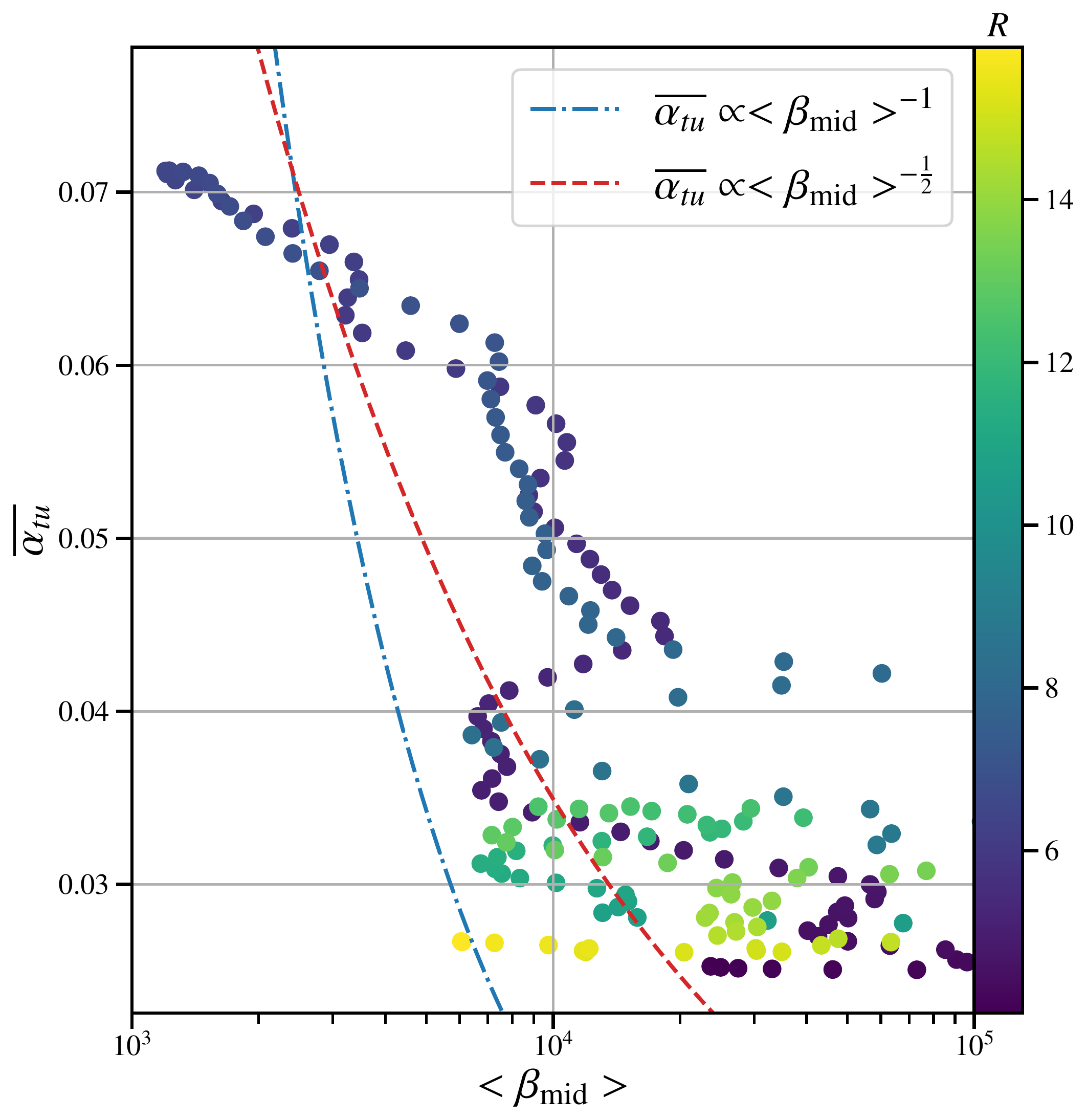}
      \caption{Vertically (within the disk region) averaged turbulent angular momentum transport coefficient, $\overline{\alpha_\mathrm{tu}}$ as a function of the mean poloidal plasma, $\mean{\beta_{\mathrm{mid}}}$, evaluated in the disk mid-plane for different cylindrical radii, $R=[4,16]$. We also show different scaling laws for  $\overline{\alpha_\mathrm{tu}}$ to compare them with our simulation.
              }
         \label{Fig:scaling}
 \end{figure}

However, in MRI turbulence, the measured scaling is closer to $\overline{\alpha_\mathrm{tu}}\propto\mean{\beta_{\mathrm{mid}}}^{-1/2}$ \citep{Salvesen_16}. For the sake of completeness we check the scaling for the turbulent transport of angular momentum in our simulation. In Fig.(\ref{Fig:scaling}) we sample the values of the averaged turbulent parameter and the local mean plasma beta at the disk mid-plane for radii ranging from $R=4$ to $R=16$. We expect $\overline{\alpha_\mathrm{tu}}$ to not only be a function of $\mean{\beta_{\mathrm{mid}}}$,  indeed in Fig.(\ref{Fig:scaling}) we have the same value of $\overline{\alpha_\mathrm{tu}}$ for different plasma beta. This is probably a consequence of the dependence of $\overline{\alpha_\mathrm{tu}}$ on the local shear, $\diff \log \Omega/\diff \log R$. Indeed, self-organisation induces local deviation from the Keplerian shear, which in turn influences the efficiency of the MRI-driven angular momentum transport \citep{gres15,pessah2007}, weaker shear leading to lower values of $\overline{\alpha_\mathrm{tu}}$. It can be seen that the scaling of the turbulent angular momentum transport coefficient, $\overline{\alpha_\mathrm{tu}}$, verifies a scaling closer to $-1/2$ than $-1$.

The $-1/2$ scaling is inconsistent with a simple viscous type instability, hence the self-organisation of our simulation into a ring-like system is not purely viscous, and must probably be somehow connected to magnetic field transport in the disk. 

In conclusion, the fact that the self-organisation is driven by radial motions of matter tends to disfavour a wind-driven instability \citep{riols2019} and tends to favour some kind of viscously-driven instability \citep{hawley2001,bai2014}.
However, as explained above, a simple viscous instability could not explain the disk instability since the scaling of $\overline{\alpha_\mathrm{tu}}$ on $\mean{\beta_{\mathrm{mid}}}$ is not steep enough.  {Finally, we note that the radial density structure of the outflow is unaffected by the ring-like structures found in the turbulent disk. The radial structures disappear in the upper layers (LA or TA), probably because the rings are washed out by the supersonic accretion in those regions. }

\cite{bai2014} construct a phenomenological model for ring formation, that does not require a steep scaling for $\overline{\alpha_\mathrm{tu}}$, based on the evolution of the vertical magnetic field driven by an anisotropic anti-diffusivity. However, it is not evident that an anisotropic anti-diffusion emerges from MRI turbulence. Furthermore, recent measures of the mean field diffusivities of MRI turbulence do not agree with a negative anisotropic diffusivity \citep{gres15}.

Finally, \cite{johan2009} explore the possibility that stochastic perturbations driven by an inverse turbulent cascade of the magnetic components could be driving the self-organisation of the disk into ring-like structures as well as the formation of pressure bumps. However, their modelling concentrates on unstratified shearing box simulations with a zero-net flux. This leads to their ring-like structures being short lived, lasting at most 50 local orbits while our inner ring-like structure are still stable at 75 local orbits and show no sign of weakening. Nonetheless, a stochastic model taking into account the feedback of the vertical field evolution may be needed to explain the ring formation observed here.


\section{Parameter exploration}
\label{sec:param}

To have a comprehensive understanding of how the properties described in section \ref{sec:fiducial} and section \ref{sec:ring} evolve with different fundamental parameters of the system we have produced 4 additional simulations, see Tab.~(\ref{tab:simu}). In these simulations we explore the effects of the initial magnetic field strength (SB2, SB3, SB4), the geometrical thickness (SEp) and the limited extent of the toroidal domain (S2pi). In this section we compute the averages defined in section \ref{sec:Method} and Appendix \ref{A:RA_MHD} for all simulations between $t_1$ and $t_2=T_\mathrm{end}$ (see tab \ref{tab:simu}) with a $\Delta t = 0.16T_\mathrm{in}$. 

We start our analysis by confirming that all of our simulations, except for simulation SB2, reach a final steady state quantitatively similar to the one described in section \ref{sec:Global_picture}: a turbulent disk, an atmosphere with super-sonic accretion and a super-fast wind . Therefore we address the weak-field simulations first, while the steady state of simulation SB2 as well as its properties will be described at the end of this section.

\subsection{Weak-field simulations: S2pi, SB3, SEp}
\label{sec:weakfield}
We start by validating the wedge approximation, $\Delta \varphi<2\pi$, for global simulations, S2pi harbours no significant differences when compared to SB4. Indeed, since S2pi and SB4 are identical with respect to the properties of their vertical structure, we ignore S2pi while we discuss such properties.

We confirm that similarly to simulation SB4, all weak-field simulations reach a  equilibrium within the disk and a magnetostatic equilibrium within laminar and turbulent atmospheres (section \ref{sec:dyn_equi}). The role of the turbulent magnetic pressure is as essential for the equilibrium and for the transitions (section \ref{sec:dyn_equi})  within those simulations as it is for SB4. The turbulence, in the same ways as for SB4, is also a consequence of strong field compressible MRI. This has been verified by checking the growth rate achieved within the turbulent regions (disk and atmosphere) using the procedure detailed in section \ref{sec:dis_turb}.  
\begin{figure}[h!]
  \centering
  \includegraphics[width=0.9\hsize]{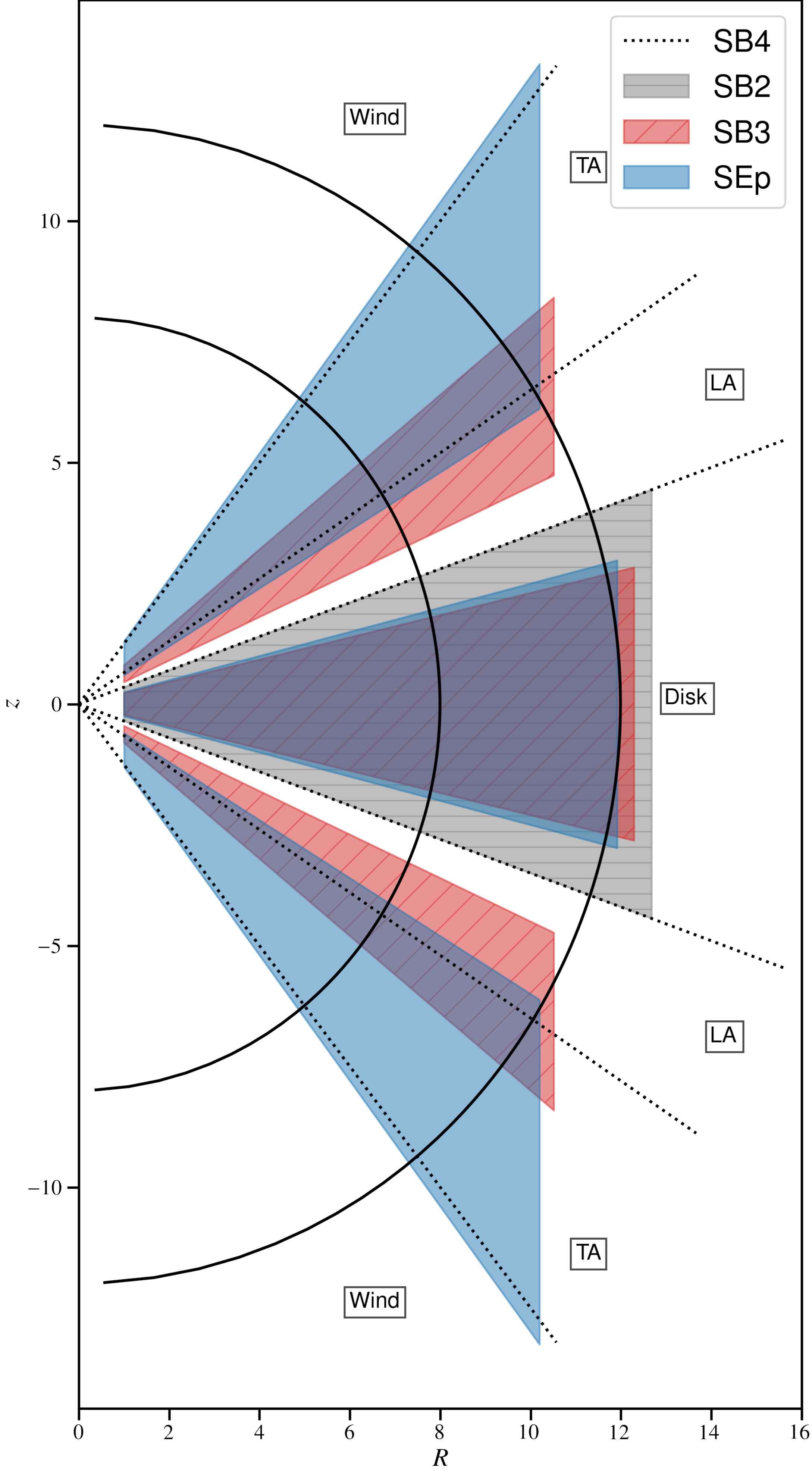}
      \caption{Extent of the turbulent zones (disk and atmosphere) found in the different simulations, see Tab.~(\ref{tab:simu}). As the magnetic field increases, the turbulent layer goes down and eventually merges with the turbulent disk.
              }
         \label{Fig:scale_height}
 \end{figure}

To quantify the evolution of the turbulent stratification as a function of the different parameters we compare the heights and extents of the turbulent disk, the laminar atmosphere and the turbulent atmosphere for the different simulations. We quantify the extent and height of the region by looking at $\cos\psi$ the angle between the poloidal velocity and the poloidal magnetic field (section \ref{sec:Global_picture}). Then, by comparing the definition of the vertical domains (TA and LA) given by $\cos\psi$ with the one given by the ratio between the turbulent and laminar torques, $|\alpha_\mathrm{tu}/\alpha_\mathrm{la}|$, we cross check the validity of the domains. 

We summarise our findings in Fig.~(\ref{Fig:scale_height}), where we show the turbulent regions (disk and atmosphere) as the shaded regions for the different simulations. We see that when the magnetic field increases (from SB4 to SB3) the disk region decreases in size. This is because the quenching of the MRI is only possible when $\mean{\beta_p}\sim1$ (section \ref{sec:dis_turb}). Hence, starting at a lower $\mean{\beta_\mathrm{mid}}$ limits the vertical extent of the disk and push the transition to the laminar atmosphere at lower altitudes. 

\begin{figure}[t]
  \centering
  \includegraphics[width=0.9\hsize]{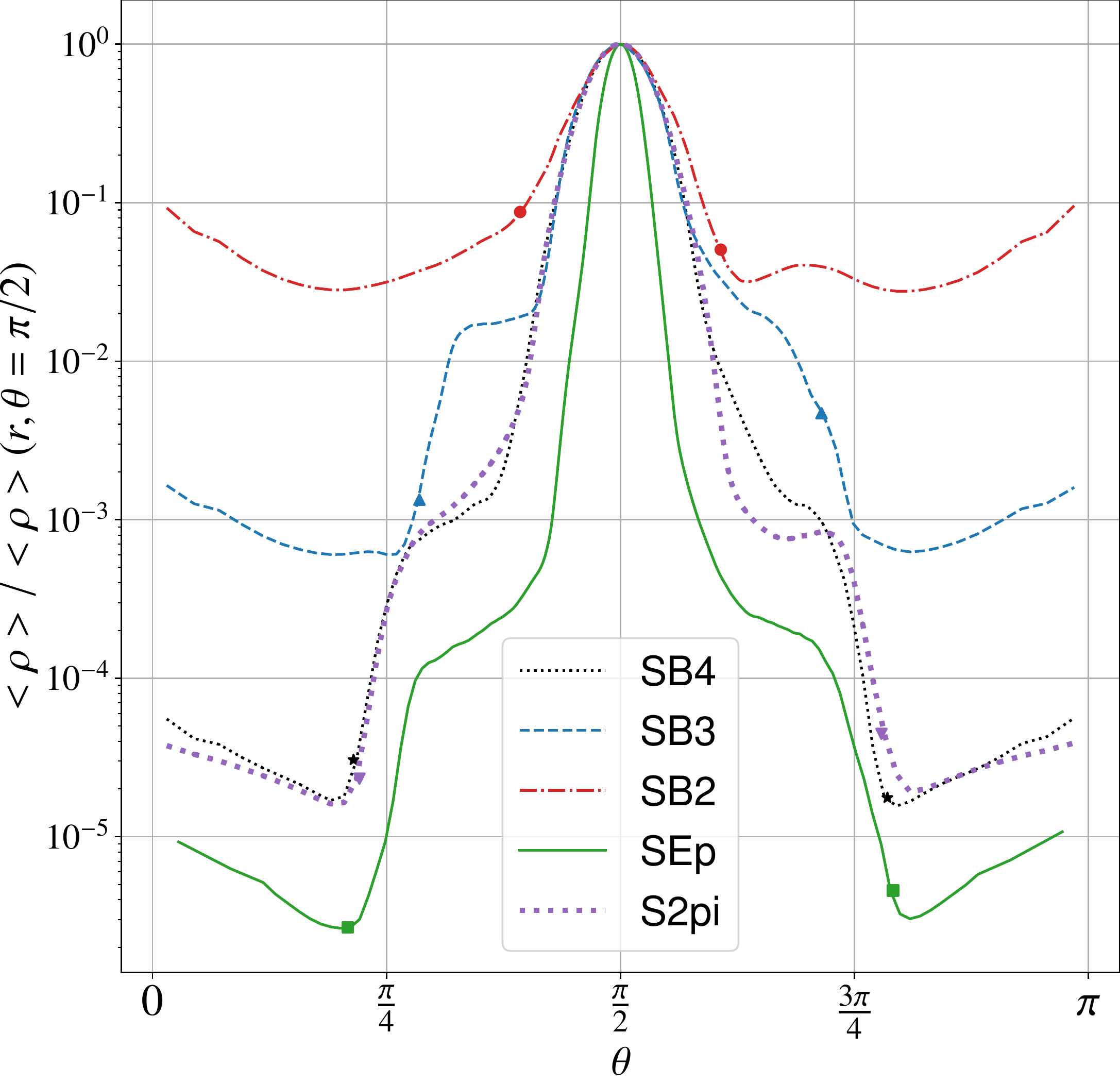}
      \caption{Mean density of the different simulations normalised to its value at the disk mid-plane, see Tab.~(\ref{tab:simu}). The symbols correspond to the height at which the flow reaches the SM speed. The mean profile are also averaged radially between $r=8$ and $r=12$.
              }
         \label{Fig:densities}
 \end{figure}

The altitude of the jet launching surface is tightly related to the density profile. In Fig.~(\ref{Fig:densities}) we show the density profiles of the different simulations normalised to their value at the disk mid-plane. We see that the simulations SB4 and S2pi converge to a similar density configuration. Moreover, we observe in Fig.~(\ref{Fig:densities}) that as $\beta_\mathrm{ini}$ decreases the density profile becomes shallower and as a consequences more mass is fed into the TA and the outflow.  This is possibly related to the enhanced mass loading fulfilled by the turbulent magnetic pressure. Indeed, decreasing $\beta_\mathrm{ini}$ leads to a decrease of $\mean{\beta_{\mathrm{mid}}}$ which increases the turbulence strength \citep{Salvesen_16,mish2020}. This increase in turbulent strength with the decrease of the plasma beta is also verified in our simulations.

From Fig.~(\ref{Fig:scale_height},\ref{Fig:densities}), we see that, as is expected when the disk geometrical thickness decreases, the vertical extent of the turbulent disk decreases. However, it is striking that the turbulent atmosphere stays at roughly the same height compared to SB4. This shows that the turbulent atmosphere and its properties mostly depend on the magnetisation and not on the disk geometrical thickness. However, the density profile is affected, becoming less dense within the turbulent atmosphere. It is not clear if an even smaller $\epsilon$ could lead to the disappearance of the turbulent and laminar atmosphere.

\begin{figure*}[t]
  \centering
  \includegraphics[width=0.49\hsize]{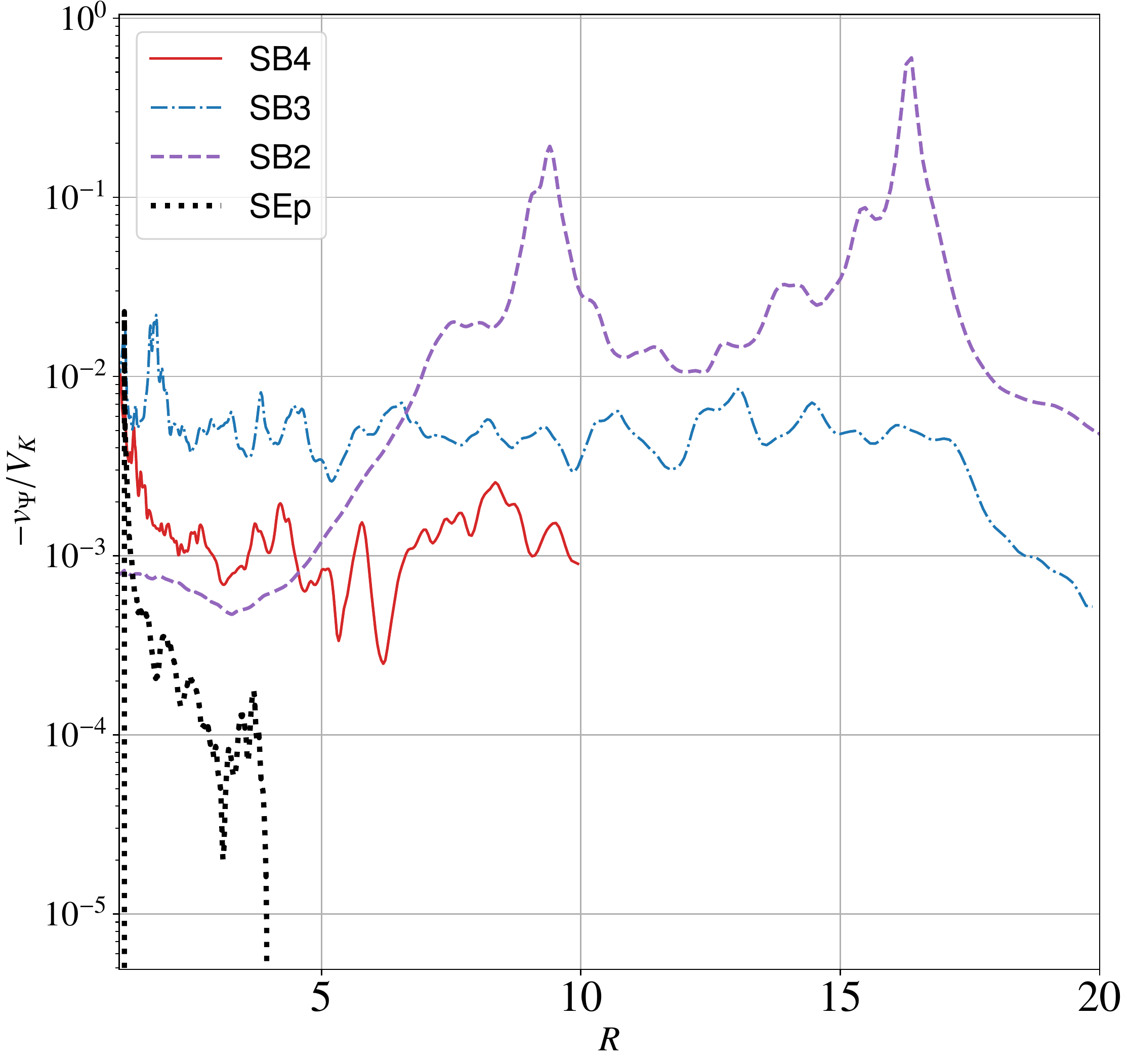}
  \includegraphics[width=0.49\hsize]{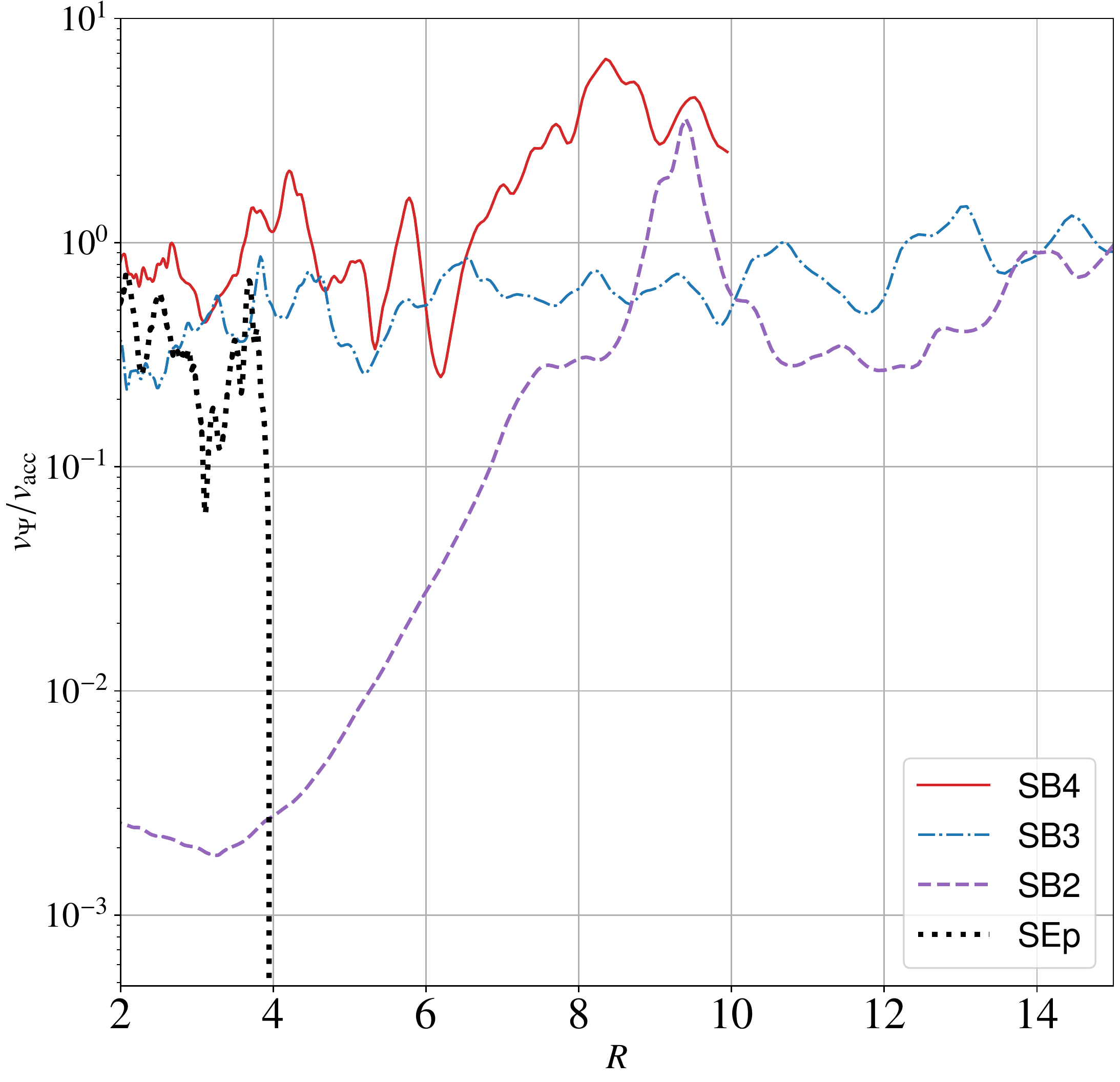}
      \caption{(left) Field advection velocity as a function of radius normalised to the Keplerian velocity. (right) Ratio of field and mass advection velocities as a function of radius. We truncate the advection velocity where we can no longer measure the drift of the magnetic field lines. The truncation radii corresponds to $R=3.5$, $R=10$, $R=20$, for SEp, SB4 and SB3 respectively. 
              }
         \label{Fig:vpsi}
 \end{figure*}

The initial magnetic field as well as the disk geometrical thickness influence the transport of mass, angular momentum and magnetic field. To understand the impact of the different parameters on accretion, we compute the mass weighted accretion velocity, $v_\mathrm{acc}$, from Eq.~(\ref{eq:glob_acc}) and then we average it between $t_a=318 T_\mathrm{in}$ and $t_b=955T_\mathrm{in}$ with temporal resolution of $\Delta t = 0.8T_\mathrm{in}$ for all simulations. 

We summarise the values in Tab.~(\ref{tab:transport}). We see that, as the magnetic field increases, $v_{\mathrm{acc}}$ also increases. Since the mean radial velocity in the accreting atmosphere does not vary a lot between simulations ($\mean{u_r}\simeq 0.3 V_K$), the dependency of $v_{\mathrm{acc}}$ must be related to the evolution of the density profile. This is confirmed by Fig.~(\ref{Fig:densities}): when the magnetic field strength increases, the density profile becomes shallower and the accreting atmosphere becomes denser. Similarly, the decrease of $v_{\mathrm{acc}}$ as $\epsilon$ decreases is also a consequence of the density profile. Hence, the mass weighted accretion velocity is tightly related to the vertical density structure.

All simulations feature magnetic field transport towards the inner boundary (section \ref{sec:transport} and Appendix \ref{A:all_Mag_flux}).
To quantify the magnetic flux transport we compute $v_\Psi$ defined in Eq.~(\ref{eq:Psi_def}) for all our simulations and then we time average it in the same manner as $v_{\mathrm{acc}}$.
We show $v_\Psi$, normalised to the Keplerian velocity, as a function of radius in Fig.~(\ref{Fig:vpsi}) (left). We confirm that $v_\Psi$ follows a radial scaling close to the Keplerian velocity. It is clear, that as the $\beta_{\mathrm{ini}}$ decreases $v_\Psi/V_K$ increases. Indeed, $v_\Psi/V_K$ seems to be proportional to $\mean{\beta_\mathrm{mid}}^{-1}$. Moreover, $v_\Psi/V_K$ strongly decreases when we decrease the disk geometrical thickness.  

In Fig.~(\ref{Fig:vpsi}) (right) we show $v_\Psi/v_{\mathrm{acc}}$ as a function of the radial coordinate. We observe that, except for the inner regions of SB2, all simulations follow approximately the same dependency,
\begin{equation}
\label{eq:v_psipropv_acc}
    v_\Psi \simeq v_{\mathrm{acc}}.
\end{equation}
This is remarkable and must be related to the evolution of the vertical structure since, as explained above, the evolution of $v_{\mathrm{acc}}$ is tightly related to it. 

In Tab.~(\ref{tab:transport}) we show the accretion time scale as well as the magnetic field transport time scale computed from their respective velocities. In all simulation the secular time scales $t_\Psi$ and $t_\mathrm{acc}$ remain longer than the dynamical time scale $T_K$. Nonetheless, a word of caution is appropriate since for simulation SB2 the accretion time scale $t_\mathrm{acc}$ is of the order of the dynamical time scale, while $t_\Psi$ is comparable for the outer regions of SB2.
 \begin{table}[h!]
\centering{
\begin{tabular}{|l|l|l|l|l|}
\hline
Name & $-v_{\mathrm{acc}}$ [$V_K$] & $t_{\mathrm{acc}}$ [$T_K$] & $t_{\Psi}$ [$T_K$] & $\mean{\beta_\mathrm{mid}}$ \\ \hline
SB4  & $ 1.1 \times 10^{-3}$        & $144$           & $160$ & $10^4$           \\ \hline
SB3  & $ 1 \times 10^{-2}$        & $16$           & $31$ & $10^3$              \\ \hline
SEp  & $ 4 \times 10^{-4}$          & $400$           & $1.1\times10^{3}$   &     $10^4$    \\ \hline
SB2 ($R>8$)  & $ 4 \times 10^{-2}$        & $4$           & $5$    & $4\times10^2$         \\ \hline
SB2 ($R<4$)  & $ 2 \times10^{-1}$        & $1$           & $265$ &     $1$          \\ \hline
\end{tabular}
\caption{ Values of the mass weighted accretion velocity defined in section \ref{sec:transport} in units of $V_K$ for the different simulations. We also show the accretion time scale calculated from the first column in units of the local $T_K(R) = 2\pi/\Omega_K(R)$ as well as the magnetic advection time scale in the same units, calculated from Fig.~(\ref{Fig:vpsi}) (left).  Finally, we include the mean poloidal $\mean{\beta_p}$ at the disk mid-plane. Quantities are temporally averaged between $t_a=318 T_\mathrm{in}$ and $t_b=955T_\mathrm{in}$.
              }
\label{tab:transport}
              }
\end{table}

To understand the impact of the different parameters on the outflows, we summarise in Tab. (\ref{tab:inv}) the values of the MHD invariants, defined in section \ref{sec:wind}, computed from a field line originating at $R_0=6$ for the different simulations. We find that the invariants do not vary from one simulation to another. This may seem surprising. However, it should be kept in mind that the wind invariants depend on the flow dimensionless properties ($\mean{u_r}/V_K$, $\mean{u_\varphi}/V_K$, $\mean{B_r}/\mean{B_\theta}$, $\mean{B_r}/\mean{B_\varphi}$) at the wind launching point, that is the top of the turbulent atmosphere. Since these dimensionless properties are similar between all weak-field simulations, the invariants should naturally be independent of $\mean{\beta_{\mathrm{mid}}}$. 

 \begin{table}[h!]
\centering{
\begin{tabular}{|l|l|l|l|l|}
\hline
Name & $\omega$ & $\lambda$ & $\kappa$ & $e$ \\ \hline
SB4  & $0.8$        & $5$           & $0.2$        & $6$     \\ \hline
SB3  & $0.9$        & $4$           & $0.4$          & $6.5$     \\ \hline
SB2  & $0.9$        & $4.5$           & $0.75$          & $6$     \\ \hline
SEp  & $0.8$          & $5$           & $0.2$        & $6$     \\ \hline
\end{tabular}
\caption{Values of the MHD invariants for all simulations, measured in the upper 'hemisphere' for a field line originating at $R_0=6$.}
\label{tab:inv}
              }
\end{table}
  \begin{figure}
    \centering
   \includegraphics[width=\hsize]{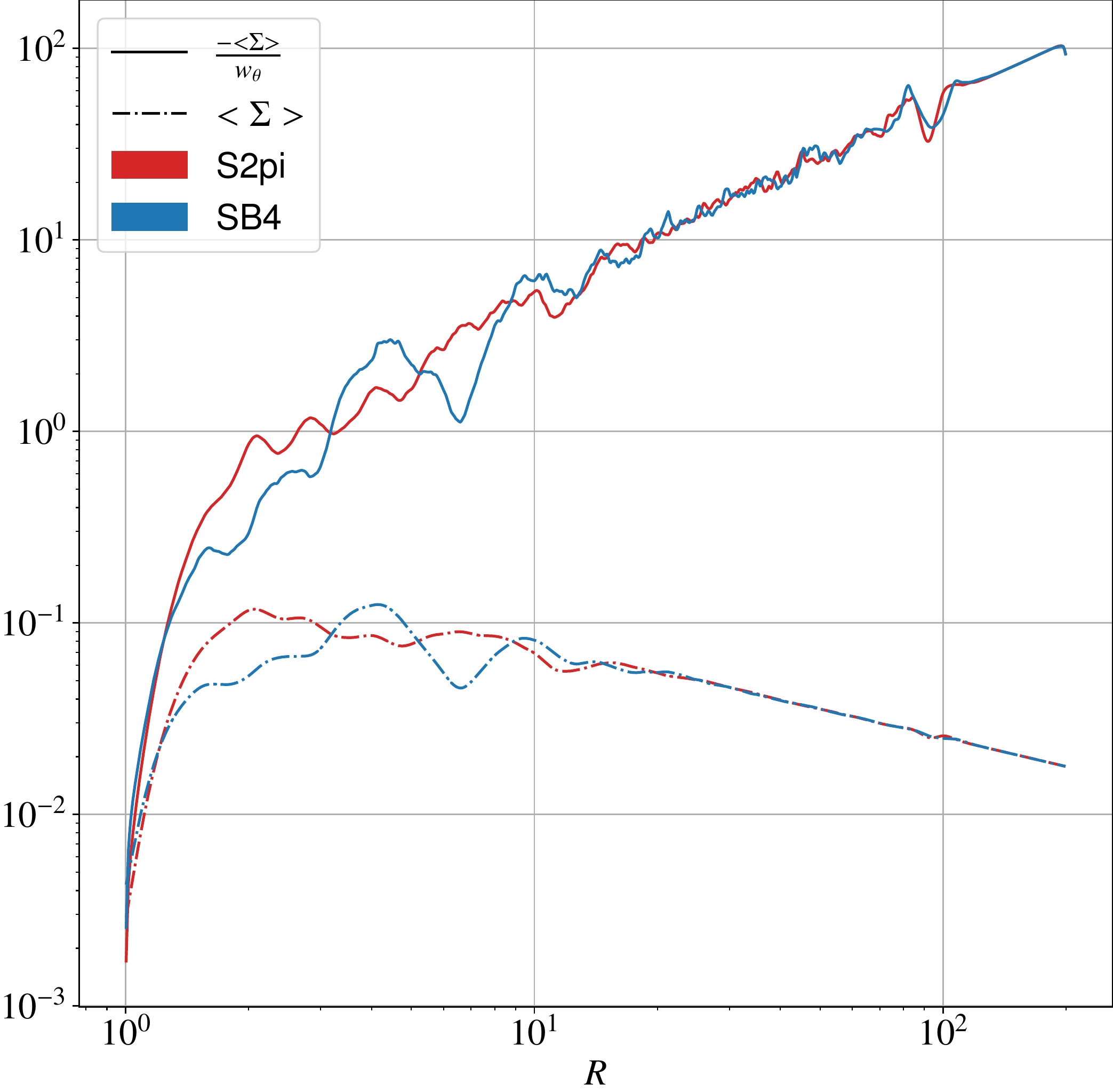}
    \caption{Inverse of potential vorticity and surface density as functions of the radial coordinate, for simulations SB4 and S2pi, $w_\theta=\rot \vmean{u}|_\theta$.}
    \label{fig:vorten}. 
\end{figure}
Ring-like structures are also found in all weak-field simulations. They are always a consequence of converging radial mass flows. In general the rings in the weak-field simulations feature a similar time-scale of self-organisation, $t_{so}\sim 200/\Omega_K(R)=32T_K(R)$, consistent with the one measured in SB4. However, there are some quantitative differences:
\begin{itemize}
    \item The ring-like structures in run SB3 are accreted towards the inner region a lot faster than for the fiducial run SB4, $v_{\mathrm{ring}}\simeq3\times 10^{-3} V_K(R)$. This velocity is around a factor of 3 smaller than the local $v_\mathrm{acc}(R)$ in this simulation. Moreover, the accreted rings are sufficiently far away ($R\sim4$ and $R\sim6$) from the inner boundary for us to believe that radial migration is not a bias of the inner boundary, in contrast with SB4. The ratios between the maxima and minima of density are very similar to the ones in the fiducial run, $\Sigma_{max}/\Sigma_{min} \simeq 2$. 
    \item The ring-like structures in run S2pi are less pronounced than in run SB4, $\Sigma_{max}/\Sigma_{min} \simeq 1.5$. This may be related to the limited toroidal extent of simulation SB4. Similarly to SB4, the radial migration of the ring-like structures for S2pi is inexistent or negligible.
    \item The ring-like structures in run SEp are very similar to the ones found in run SB4, the ratios between minima and maxima are similar, $\Sigma_{max}/\Sigma_{min} \simeq 2$. However, run SEp features many more ring-like structures, by about a factor 2. This is consistent with a viscous like instability bounded at small scale by the disk scale-height, in which the spacing between ring-like structures is expected to scale like $h$. Finally, radial migration of ring-like structures is also negligible in this simulation.
\end{itemize}
{The Rossby Wave Instability (RWI) is known to produce non-axisymmetric structures, like Rossby vortices or spiral shocks which smear out local density maxima. The RWI is triggered\footnote{Where for simplicity, we neglect the entropy profile. In practice, it will not change the location of the local maxima of $\mathcal{F}(r)$ due to our prescribed cooling.} when the inverse of the potential vorticity}
\begin{equation}
    \mathcal{F}(r) = \frac{\mean{\Sigma}}{\rot \vmean{u}|_\theta},
\end{equation}
{has a local maximum \citep{love1999}. To check whether our ring-like structures are unstable to the RWI, we show the inverse of potential vorticity as a function of radii in Fig.~(\ref{fig:vorten}) for simulations S2pi and SB4. We find that for SB4, the local maxima of the inverse of the vorticity potential are well correlated with the local maxima  of the column density profile. This correlation is less clear for run S2pi, as the ring-like structures are less pronounced. One would therefore conclude that SB4 rings should be RWI unstable and therefore develop non-axisymmertic structures. However, no clear spiral structures are visible in SB4, while non-axisymmetric features similar to the ones shown by \cite{mish2020} are appreciable in S2pi. It is therefore temping to think that the RWI in SB4 is stabilised by some mechanism. One possibility is that the RWI is stabilised because of turbulent diffusion which damps RWI modes. Another one is that the limited azimuthal extension of SB4 ($\Delta \varphi=\pi/2$) filters out some of the most destructive low-$m$ modes excited by the RWI. The presence of strong spiral waves in S2pi is in favour of the second hypothesis, but a proper study of the ring stability is needed to ascertain this conclusion, which is outside the scope of this work.}

\subsection{Strong field simulation: SB2}
\label{sec:strongfield}
  \begin{figure}
    \centering
   \includegraphics[width=\hsize]{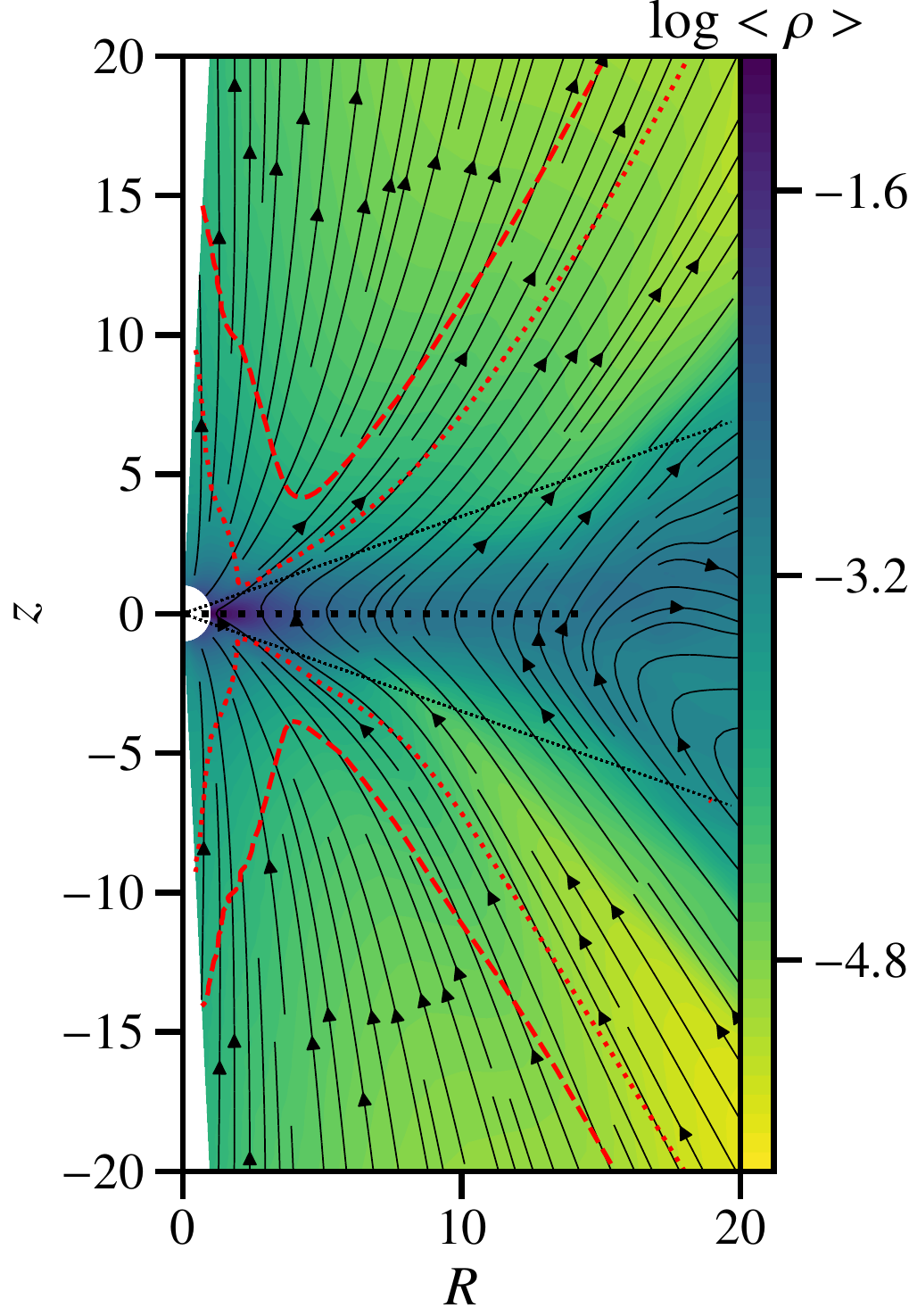}
    \caption{Appearance of the SB2 strong field simulation: Logarithm of the density (background colour), mean poloidal magnetic field lines (black solid lines), Alfv\'en (red dotted line) and fast magneto-sonic (red dashed line) critical surfaces. The black dotted lines represent the end of the disk region defined in Fig.(\ref{Fig:scale_height}). The quantities are averaged between $t_1= 1719 T_\mathrm{in}$ and $t_2= 1910T_\mathrm{in}$ with a resolution of $\Delta t = 0.16 T_\mathrm{in}$.}
    \label{fig:Field_lines_SB2}. 
\end{figure}

The main difference between SB2 and all weak-field simulations is that SB2 takes a much longer time to reach a reasonable steady state, the whole disk being subject to a drastic reorganisation of the vertical magnetic field. At $t_\mathrm{end}=1910T_\mathrm{in}$, only the region below $R\simeq 7$ has reached a steady state, while the outer regions continue to struggle readjusting the magnetic field distribution. The final stage of our simulation shows an internal quasi-stationary highly magnetised region with $\mean{\beta_\mathrm{mid}} \simeq 1$ and an outer evolving region with $\mean{\beta_\mathrm{mid}} \sim \beta_\mathrm{ini}$. Hereafter, we define $R_J$ as the end of the equipartition field region. 

The reorganisation as well as the lack of stationarity in the outer regions are a consequence of the enormous $v_\Psi$ in the outer regions. As shown in Fig.(\ref{Fig:vpsi}), $v_\Psi$ follows the same scaling (Eq.\ref{eq:v_psipropv_acc}) as for the weak field simulations, driving in this case a much faster field advection. This magnetic flux accumulation reaches a saturation in the inner region with $R_J\simeq 7$, where turbulent field diffusion balances advection leading to $v_\Psi\rightarrow0$. Because of the presence in our simulation of an important magnetic flux, this transition radius is increasing in time as more magnetic field is being brought in. This can be clearly seen in the lower left panel of Fig.~\ref{Fig:Psi_all}, where we measure a radial drift speed of $R_J$ to be $\dot R_J \sim 10^{-3}V_K(R_\mathrm{in})$. This increase in $R_J$ is highly time variable, involving bursts and dramatic changes in the disk within an extended transition region beyond $R_J$. Describing in details this transition region is beyond the scope of this work (see discussion next Section). We note however that the saturation value of the magnetic field reached in the inner steady zone is near equipartition. For a larger magnetic field strength, \citet{ferr95} showed that no vertical equilibrium could be found anymore because of the overwhelming magnetic compression. This is consistent with our own findings and advocates for a maximal mean magnetic field value set by the vertical equilibrium.

In both radially distinct regions the turbulent atmosphere collapses into the disk, following the same trend as the weak field simulations (Fig.~\ref{Fig:scale_height}). This leads to the disappearance of the laminar atmosphere and the formation of a thicker disk when compared to SB3. The merging of the turbulent disk with the turbulent atmosphere is directly related to the density profile (see Fig.~\ref{Fig:densities}). As described in the previous section, the turbulent magnetic pressure determines the steepness of the density profile. Thus, as the disk magnetisation $\mean{\beta_\mathrm{mid}}^{-1}$ increases the amount of mass lifted by the turbulent magnetic pressure increases, leading to a massive turbulent atmosphere that merges with the disk. 

Figure~(\ref{fig:Field_lines_SB2}) shows a zoomed-in outcome of our SB2 simulation, averaged between $t_1= 1719T_\mathrm{in}$ and $t_2= 1910T_\mathrm{in}$. The shape of the poloidal magnetic field lines (black solid lines) anchored in the inner region is clearly different from that seen in weak field simulations (Fig.\ref{Fig:field_stream}). Its structure is remarkably similar to that originally invoked in \cite{blan82} and later computed by \cite{ferr97}. The background colour is the density and the dotted black lines delineate the turbulent disk (as in Fig.~\ref{Fig:scale_height}). The turbulence is consistent with the strong field compressible MRI (section \ref{sec:dis_turb}) and leads to a turbulent disk region in magnetostatic equilibrium, much alike the turbulent atmosphere of weak-field simulations. Below $R_J \sim 7$, accretion occurs within the turbulent disk and the mass weighted accretion speed is supersonic. The disk rotation is near keplerian with $\mean{u_\varphi}\simeq 0.8 v_K$, namely a deviation of the order of $\epsilon$ as expected in near equipartition disks. A cold outflow is launched from the disk surface at $z\simeq 3.5h$, becoming soon a super Alfv\'enic (red dotted curve) and then super fast magnetosonic (red dashed curve) jet. 

Overall, this inner accretion-ejection structure shares most properties with the Jet Emitting Disk model \citep{ferr97}. The main difference is the amount of ejected mass as the outflows obtained here are quite massive (Fig.\ref{Fig:densities}). As discussed above, this is a consequence of the turbulence itself since the turbulent magnetic pressure plays a major role in the magnetostatic vertical equilibrium and launching of the outflow. Finally, we report the lack of ring-like structures in our strong field simulation. This is presumably due to the fact that accretion is now too fast, $t_{acc}\ll t_{so}$,with the $t_{so}$ as obtained from the weak field simulations. Mass has no time to accumulate locally and lead to the formation of rings and gaps.

\section{Discussion and conclusion}
\label{sec:discussion}

We performed 3D ideal MHD simulations of global accretion disks threaded by a large scale vertical magnetic field. Our simulations evolved for more than thousand orbits at the inner radius, allowing our accretion-ejection configuration to achieve a reasonable steady-state. We found that the disk structure is drastically dependent on its magnetisation $\mean{\beta_\mathrm{mid}}^{-1}$. 


(1) At disk magnetisations lower than $10^{-3}$, a non trivial vertical stratification sets in on a scale that can go up to $z\sim 10 h$, where an outflow that becomes super fast magnetosonic is launched. The disk is highly turbulent, mostly in hydrostatic equilibrium and the presence of an important magnetic turbulent pressure allows to lift up a strong amount of mass into a laminar ideal MHD atmosphere. This lifted mass is basically falling in towards the central object, dragging in the magnetic field which, in turn, provides a torque that transfers its angular momentum back to the underlying disk. When the magnetic field achieves equipartition in these upper layers, MRI is re-ignited and drives a turbulence. In this levitating turbulent atmosphere, transsonic accretion is achieved mostly through the turbulent torque and a small fraction of the mass is then loaded into the open field lines where a magnetically driven cold wind is launched. In contrast with shearing box simulations, the bipolar wind configuration is found to be fairly symmetric. We report also the formation in the disk of rings and gaps, on time scales shorter than the accretion time scale. 

(2) The ambient magnetic field is found to be always dragged {inwards} in the disk, at a velocity which increases with the disk magnetisation. The disk vertical structure is also seen to be deeply affected, with a progressive decrease in height of the disk turbulent atmosphere. However, the outflow properties, as described by the usual MHD invariants, remain remarkably constant. This shows that the outflow is determined by the physical conditions at the disk upper layers, which remain mostly constant, rather than by the disk mid-plane properties.  
  
(3) However, beyond a threshold on the disk magnetisation, located between $10^{-3}$ and $10^{-2}$, the global accretion-ejection configuration undergoes a drastic readjustment. The magnetic field is being accumulated into the central regions until a global equilibrium is achieved involving both the disk and its magnetosphere. The inner disk reaches a steady-state (balance between field advection and diffusion) when its magnetisation achieves unity and no more gaps and rings are found. Despite this strong field regime, the disk is turbulent and drives a super fast-magnetosonic outflow right from its surface $z \sim 3 h$. The size of this inner region keeps on increasing in time, as more magnetic flux is being added from the outer regions. The mass-weighted accretion speed is supersonic in the inner region. The transition region, where the disk material must become transsonic, is quite complex and its investigation is postponed for future work.

Some of our results have already been obtained or discussed in the literature. Previous numerical works \citep{Zhu_Stone,mish2020} found the same stationary states as in our simulations. However, these authors did not analyse the appearance of the second upper turbulent zone and how its properties are related to the vertical structure of the system. Another major difference is related to the magnetic field advection. \cite{Zhu_Stone} report a small amount of magnetic flux accumulation through their inner boundary albeit without providing a clear analysis. We show that the field advection speed  $v_\Psi$ follows a Keplerian scaling and increases with the disk magnetisation. The strong field simulations of \cite{mish2020} are somewhat different as they do not detect considerable magnetic flux evolution. This is in strong contrast with our own results. Our guess is that it may be a consequence of their disk being two times thinner than ours ($\epsilon=0.05)$. Indeed, Fig.~(\ref{Fig:vpsi}) clearly shows that the geometrical thickness has a tremendous impact on the magnetic field advection speed (see e.g. \cite{lubow1994}). This is an interesting aspect with potentially strong astrophysical consequences and deserves further investigations.

One of the main results is the realisation of the utmost importance of the MRI in the strong field regime \citep{kim2000} for accretion-ejection structures. While it is widely believed that the MRI can only exist in high $\beta$ plasmas, \cite{kim2000} have shown that the MRI exist for arbitrary $\beta$ provided that arbitrary long wavelength perturbations are allowed. In the $\beta<1$ regime, the MRI can be quenched when the toroidal field gets too strong. We have shown that the turbulent and laminar regions of the atmosphere, which is a $\beta<1$ region, is naturally explained by the MRI in this peculiar regime. 

Magnetic field dragging seems to be a generic property of ideal MHD accretion disks threaded by a large scale magnetic field (provided that $\epsilon\sim 0.1$ or larger). Indeed, it is also reported in GRMHD simulations of accretion disks around a Kerr black hole (e.g. \cite{Mckinney2012,White2019} and references therein, see also \cite{Liska2020}). In these simulations, the magnetic field is seen to be gradually advected and concentrated into the central region, leading to the build up of what has been termed a Magnetically Choked Accretion Flow \citep{Mckinney2012}. As in our SB2 simulation, magnetic field dragging stops when the vertical field reaches an equipartition value, defining a transition radius with an outer disk at lower magnetisation. As more magnetic flux is being advected, a violent relaxation occurs at this radius involving a magnetic Rayleigh-Taylor instability. It leads to the expulsion of the magnetic flux excess and a progressive increase in time of the transition radius, as long as some magnetic flux remains available in the simulation \citep{Mckinney2012}. This situation is quite nicely consistent with the hybrid disk configuration proposed for transition disks \citep{comb08,wang17}  and around black holes \citep{Ferreira_2006Xray}. Within this framework, an inner jet emitting disk (JED, \cite{ferr97}) is established until a transition radius $R_J$, beyond which an outer Standard Accretion Disk (SAD, \cite{shak73}) is settled. Because of its supersonic accretion speed associated with the launching of powerful jets, a JED provides the physical conditions allowing to explain most observational properties of X-ray Binaries (see \cite{marcel2019} and references therein).

We further note that our weakly magnetised disks (i.e. with a magnetisation smaller than $10^{-3}$) provide a highly subsonic mass-weighted accretion speed as expected in a SAD. However they differ from the SAD model in two important aspects: (i) the presence of outflows and (ii) the existence of an upper turbulent accreting zone.   

As stated above, the properties of the super fast-magnetosonic jets found in our simulations are remarkably independent of the disk magnetisation. 
Indeed, our outflows exhibit a magnetic lever arm $\lambda\sim 4-5$ valid for 4 orders in magnitude in $\mean{\beta_\mathrm{mid}}^{-1}$ (see Tab.~\ref{tab:inv}). The best observational constraints on jet kinematics are provided by proto-planetary systems, where the magnetic lever arm has been recently characterised in molecular observations of several outflows. In HH212 \citet{tabo2017} measure $\lambda\simeq 5.5$ for an outflow launching zone extending from $R_{0,1}=0.1\,\, \mathrm{au}$ to $R_{0,2}=40\,\, \mathrm{au}$. In HH30, \citet{louvet2018} estimate $\lambda \simeq1.6$ and a launching radius $R_0 \simeq 1.5\,\, \mathrm{au}$, while  \citet{devalon2020} measure $\lambda\simeq1.6$ and $R_0 \simeq 2\,\, \mathrm{au}$ for DG Tau B. This discrepancy in $\lambda$ with our simulations could be explained by several aspects. If some additional heat is deposited at the base of the outflow, the disk ejection efficiency is enhanced which leads to the decrease of the magnetic lever arm \citep{cass00b}. Such heating could be a natural outcome of irradiation from the central object and local turbulent heating. Moreover, these molecular  outflows  are  emitted  from  regions  beyond 1 au. At those large distances, disk ionisation becomes an issue and non ideal MHD effects come into play. The inclusion of ambipolar diffusion and Hall effect, which are believed to be present in the outer regions of proto-planetary disks, may change the outflow properties \citep{beth17}.

The second important difference with the  Shakura-Sunyaev SAD model is the existence of a turbulent accreting zone levitating above a weakly magnetised disk. Our simulations show that the disk has almost no accretion taking place within the densest body and an elevated supersonic accretion zone. This is somewhat reminiscent of the layered accretion model invoked in the outer regions of proto-planetary disks \citep{gamm1996,Fle2003,bai13}. However, the physical reason is completely different. In our case, the disk is fully ionised and turbulent whereas { in the dead zone picture non-ideal effects quench turbulence and lead to the formation of a MRI stable ('dead') zone in the disk}. The interesting part is that, in both cases and as long as the vertical field magnetic is weak, accretion takes place at the disk 'surface'. { This is a feature that is drastically different from the conventional view of Shakura-Sunyaev accretion disks. Moreover, this may lead to direct observational consequences on the emitted spectra of weakly magnetised accretion disks}. Indeed, mass accretes across the magnetic field lines and requires thereby some magnetic (turbulent) diffusivity \citep[see however][]{Zhu2020}. As a consequence, a fraction of the accretion energy must also be dissipated locally in this low-density region. In the context of compact objects, we expect the production of an optically thin emission, usually referred to as a coronal emission. This 'elevated accretion' disk configuration could thus provide an explanation for the 'warm corona' component often seen in soft X rays in some active galactic nuclei \citep{done2012,petrucci2018}.

{It is important to note that, to make the problem numerically tractable we have decided to ignore the effects of the energy equation. We did this by forcing a locally isothermal structure with a fixed disk scale, $\epsilon = 0.1$ or $\epsilon=0.05$. Optically thick disks in AGN and X-ray binaries  are expected to be geometrically thin with $\epsilon=0.01$ or less, which is incompatible with our simulation setup. Nonetheless, we believe that the accreting atmosphere should perdure at lower $\epsilon$, as long as the disk magnetisation is small. To properly study the impact of the energy equation, one would need to include the effects of radiative transfer into our simulations, which is clearly beyond the scope the present paper.  Irradiation from the central object or disk regions \citep[as in][for instance]{begel1983,higginbottom015,gressel2020}  could lead to a quantitative modification of our picture. Indeed,  including heat deposition by external irradiation will probably lead to more massive outflows \citep{cass00b} and modify the vertical stratification leading to the turbulent accreting layer. The study of such thermo-magnetic winds is postponed for future work. }

The parameter space explored in our simulations overlaps with the one explored by \cite{jacquemin-ide_magnetically-driven_2019} using self-similar solutions. The outflow invariants computed in our simulations (Tab.~\ref{tab:inv}) are found consistent with the borders of the parameter space they derive semi-analytically. However, while in our simulations $\lambda\sim 4-5$ is roughly constant, they obtain values ranging from $2$ to $100$. As shown in their Figs.(4) and (5), the magnetic lever arm is independent of the disk magnetisation and is strongly affected by the mass load $\kappa$. The reason for their wide range in $\lambda$ stems therefore only from their capability to produce outflows with a mass load $\kappa$ ranging from a few $10^{-3}$ to almost unity, while our 3D simulations always provide $\kappa \sim 0.2-0.7$. This discrepancy is related to our different vertical stratification. In our simulations, the levitating turbulent atmosphere acts as a buffer between the proper turbulent disk and the ideal MHD outflow. As discussed in Section \ref{sec:dyn_equi}, the existence of a dense, first laminar then turbulent, atmosphere is a consequence of the disk turbulence itself.  
This turbulent accreting atmosphere is actually missing in all works where turbulence is not self-consistently computed and must therefore be prescribed: self-similar studies \citep{ferr97,jacquemin-ide_magnetically-driven_2019}, 2.5D MHD simulations done with alpha prescriptions \citep{murp10,Stepa2016} and other semi-analytical models \citep{guilet2013}. Our guess is that including the complex vertical stratification should allow to recover with these approaches an outflow mass load consistent with full 3D simulations. This requires however the use of numerically 'educated' profiles for all relevant turbulent effects. This is postponed for future work.

Finally, weakly magnetised disks are subject to a secular instability which leads to the formation of long-lived ring-like structures in the disk surface density distribution, the magnetic field being concentrated in the gaps. These ring-like structures are also visible in the surface density profiles of \cite{Zhu_Stone} and, like for us, they also seem to increase in number when the disk geometrical thickness is decreased. This is consistent with a viscous like instability bounded at small scale by the disk scale-height, for which the spacing between ring-like structures is expected to scale like $h$. On the other hand, we showed that this global self-organisation is unlikely due to the wind instability proposed by \cite{riols2019}. Yet, it remains unclear if it is purely a viscous instability, as the turbulent viscosity scaling $\overline{\alpha}_{tu}(\mean{\beta_\mathrm{mid}})$ is not steep enough to develop a viscous instability.  When the disk magnetisation increases above the threshold, these ring-like structures are smeared out and disappear as the accretion speed becomes supersonic. Our suspicion is that the ring-formation mechanism and the magnetic field transport could be related to each other. {Finally, we show that the limited toroidal extent of our simulation SB4 and SB3 could have an impact on the long-term evolution of the ring-like structures. Indeed, it is not clear if the rings are destroyed by non-axis-symmetric instabilities, like the RWI.}

In any case, our results advocate for the existence of a dichotomy in ideal MHD accretion disks. In disks where the magnetisation remains below a threshold value, around $10^{-3} - 10^{-2}$, ring-like structures should be present, and absent in disks with equipartition magnetic fields. It could thus be an observationally convenient way to probe the existence and strength of large scale magnetic fields in proto-planetary disks (see for instance \cite{devalon2020}).

\begin{acknowledgements}
This work was granted access to the HPC resources of TGCC under the allocation A0080402231 attributed by GENCI (Grand Equipement National de Calcul Intensif). GL acknowledges support from the European Research Council (ERC) under the European Union Horizon 2020 research and innovation programme (Grant agreement No. 815559 (MHDiscs)). J.F. acknowledges support from the CNES french agency and CNRS PNHE. This research made use of the SciPy and NumPy libraries \citep{oliphant2006guide,van2011numpy,2020SciPy-NMeth}.
\end{acknowledgements}

%
%

\bibliographystyle{aa}
\bibliography{biblio}

\begin{appendix}
\section{Reynolds Averaged MHD equations}
\label{A:RA_MHD}
In this appendix we compute the Reynolds averages of the MHD equations.
We proceed by decomposing the different terms into fluctuating (turbulent) terms plus the laminar (averaged) terms, using the averages defined in section \ref{sec:Ave_proc}. Then we can average Eq.~(\ref{Eq:Mass_Con}-\ref{Eq:Induc}) with respect to time and $\varphi$.
We start with the equation for the conservation of mass, Eq.~(\ref{Eq:Mass_Con}), thanks to its simple form this is straightforward
\begin{equation}
    \label{Eq:MMass_con}
    \tdiff{\phimean{\de \rho}} + \divi{\mean{\rho}\left[\mean{\vec{u}}+\rhomean{\de \vec{u}}\right]}=0.
\end{equation}
It is important to note that the toroidal component of the divergence vanishes thanks to Eq.~(\ref{Eq:dev_azi}). Indeed, the  averaged quantities are only function of the latitudinal, $\theta$, and radial, $r$, coordinates. The term on the left hand side of the equation is associated with the non-steadiness of the turbulence, if the simulation is quasi stationary this term should be negligible. 
Then we average the toroidal component of the induction equation, Eq.~(\ref{Eq:Induc}):
\begin{equation}
    \label{Eq:Minduc_poi}
    \tdiff{\phimean{\de B_\varphi}} = \left.\rot\left[\vmean{u}\times\vmean{B}+\vec{\mathcal{E}}\right]\right\vert_\varphi
\end{equation}
where the turbulent emf is defined as 
\begin{equation}
    \vec{\mathcal{E}} = \mean{\vec{\de u}\times\vec{\de B}}.
\end{equation}
We write the poloidal component of the induction equation as the toroidal component of the equation for the vector potential
\begin{equation}
    \pdv{A_\varphi}{t} = \vec{u_p}\times\vec{B_p} + \frac{1}{R}\pdv{ G}{\varphi},
\end{equation}
where $A$ is the vector potential of the magnetic field, $\rot \vec{A} = \vec{B}$ and $G$ emerges to satisfy the gauge condition.
Averaging this equation leads to
\begin{equation}
     \tdiff{\phimean{\de A_\varphi}} = \vmean{u_p}\times\vmean{B_p} + \vec{\mathcal{E}}_\varphi,
\end{equation}
where the term $\vgrad G$ is averaged out thanks to the condition of  Eq.~(\ref{Eq:dev_azi}).
We project, then we rearrange the equation of latitudinal momentum using $u_\theta=\mean{u_\theta}+\de u_\theta$ and the conservation of mass to find a not so usual expression:
\small
\begin{align}
    &\pdv{r\de u_\theta \rho}{t}+\rho\upo\cdot\vgrad{r\mean{u_\theta}} + \divi{r\left[\rho \de u_\theta \vec{u}-\frac{1}{4\pi}B_\theta\vec{B}\right]} =\nonumber \\ &\cot\theta \left[\rho u_\varphi^2 - \frac{1}{4\pi}B_\varphi^2 \right] -\pdv{}{\theta}\left(\frac{B^2}{8\pi}+P\right),
    \label{Eq:LatMom}
\end{align}
 \normalsize
where $\cot\theta=\cos\theta/\sin\theta$. We  decompose the rest of the quantities into turbulent and laminar terms and average with respect to time and the toroidal coordinate to find:
\small
\begin{align}
    &\tdiff{r\phimean{\rho\de u_\theta}}+\mean{\rho}\rhomean{\vec{u_p}}\cdot\vgrad r\mean{u_\theta} -\frac{1}{4\pi}\vmean{B_p}\cdot\vgrad r \mean{B_\theta}\nonumber \\ 
    &+\divi r \left[\mean{\rho}\rhomean{\de u_\theta \vec{u_p}}
    -\frac{1}{4\pi}\mean{\de B_\theta \vec{\de B_p}}\right]=\nonumber \\
    &-\pdv{}{\theta}\left[\mean{P}+\frac{1}{8\pi}\left(\mean{B}^2+\mean{\de B^2}\right)\right] \nonumber \\ &+\cot\theta\left[\mean{\rho}\left(\mean{u_\varphi}^2+\mean{\de u_\varphi^2}\right)-\frac{1}{4\pi}\left(\mean{B_\varphi}^2+\mean{\de B_\varphi^2}\right)\right]
    \label{eq:MLatMom}
\end{align}
 \normalsize
where $\rhomean{\vec{u}}=\vmean{u}+\rhomean{\de \vec{u}}$ and
\begin{equation}
    \rhomean{\de u_\theta \vec{ u_p}}=\rhomean{\de u_\theta \de\vec{ u_p}}+\vmean{u_p}\rhomean{\de u_\theta}.
\end{equation}
In the derivation of Eq.~(\ref{eq:MLatMom}) we have used
\begin{equation}
    \mean{\divi{\vec{B}B_i}} = \vmean{B_p}\cdot\vgrad{\mean{B_i}}+\divi{\mean{\de \vec{B_p}\de B_i}}
\end{equation}
In the same way as before the derivative with respect to the toroidal component vanishes thanks to Eq.~(\ref{Eq:dev_azi}).
We repeat the same procedure for the radial projection of Eq.~(\ref{Eq:Mom_Con}) to find
\small
\begin{align}
    &\pdv{\de u_r \rho}{t}+\rho\upo\cdot\vgrad{\mean{u_r}} + \divi{\left[\rho \de u_r \vec{u}-\frac{1}{4\pi}B_r\vec{B}\right]} =\nonumber \\ &\frac{1}{r} \left[\rho\left(u_\varphi^2+u_\theta^2\right) - \frac{1}{4\pi}\left(B_\varphi^2+B_\theta^2\right) \right] -\pdv{}{r}\left(\frac{B^2}{8\pi}+P\right)-\rho g,
    \label{Eq:radMom}
\end{align}
\normalsize
as before we then average the equation and find:
\small
\begin{align}
    &\tdiff{\phimean{\rho\de u_r}}+\mean{\rho}\rhomean{\vec{u_p}}\cdot\vgrad{\mean{u_r}} -\frac{1}{4\pi}\vmean{B_p}\cdot\vgrad{}\mean{B_r}\nonumber \\
    &+\divi \left[\mean{\rho}\rhomean{\de u_r \vec{u_p}}
    -\frac{1}{4\pi}\mean{\de B_r \vec{\de B_p}}\right]=
    -\pdv{}{r}\left[\mean{P}+\frac{1}{8\pi}\left(\mean{B}^2+\mean{\de B^2}\right)\right] \nonumber \\ &+\frac{\mean{\rho}}{r}\left[\mean{u_\varphi}^2+\mean{\de u_\varphi^2}+\mean{u_\theta}^2+\mean{\de u_\theta^2}\right] -\mean{\rho} g \nonumber \\
    &-\frac{1}{4\pi r}\left[\mean{B_\varphi}^2+\mean{\de B_\varphi^2}+\mean{B_\theta}^2+\mean{\de B_\theta^2}\right]
    \label{eq:MradMom}
\end{align}
\normalsize
We project and rearrange the equation for toroidal momentum transport to find the usual equation of angular momentum transport:
\begin{equation}
\label{Eq:AngMom}
    \pdv{Ru_\varphi \rho}{t} + \divi{R\left[\rho u_\varphi \vec{u}-\frac{1}{4\pi}B_\varphi \vec{B}\right]} = -\pdv{}{\varphi}\left(\frac{B^2}{8\pi}+P\right)
\end{equation}
we use the conservation of mass and then repeat the averaging procedure to find:
\small
\begin{align}
    \label{eq:MAngMom}
    &\tdiff{R\phimean{\rho\de u_\varphi}}+\mean{\rho}\rhomean{\vec{u_p}}\cdot\vgrad R\mean{u_\varphi} +\nonumber \\ &\divi R \left[\mean{\rho}\rhomean{\de u_\varphi \vec{ u_p}}
    -\frac{1}{4\pi}\mean{\de B_\varphi \vec{\de B_p}} - \frac{1}{4\pi}\mean{ B_\varphi} \mean{\vec{B_p}}\right]=0,
\end{align}
\normalsize
\section{Computation of Fluxes}
\label{sec:Comp_flux}
In this section we show how we compute the different fluxes for the equation of conservation of angular momentum and the equation of conservation of mass. First we define the following useful average
\begin{equation}
    \overline{X} = \int\limits_{\theta_1}^{\theta_2}r \sin\theta \,X\diff{\theta}.
\end{equation}
We can then integrate Eq.~(\ref{Eq:Mass_Con}) with respect to time and space to find
\begin{equation}
    \Delta M = F_{r_2}-F_{r_1}+F_{\theta_1}-F_{\theta_2}
\end{equation}
where we have introduced
\begin{align}
    F_{r_i} =& -\int\limits_{t_1}^{t_2} r_i\left.\left[\overline{\phimean{\rho u_r}}\right]\right\vert_{r_i}\diff{t},\\
    F_{\theta_i} =& \int\limits_{t_1}^{t_2}\int\limits_{r_1}^{r_2}r \sin\theta\left.\phimean{\rho u_\theta}\right\vert_{\theta_i}\diff{r}\diff{t}\\
    \Delta M =& \int\limits_{r_1}^{r_2}\left[r\overline{\phimean{\rho}}\right]^{t_2}_{t_1}\diff{r},
\end{align}
we can redo the same exercice for Eq.~(\ref{Eq:AngMom}) to find
\begin{equation}
    F_{M_r}+F_{M_\theta} = F_{r_2}-F_{r_1}+F_{\theta_1}-F_{\theta_2}
\end{equation}
There is an additional term $F_{M_\theta}$ when compared to the cylindrical case due to the spherical geometry. The angular momentum fluxes are defined with respect to an equilibrium velocity, $R\Upomega$. We define
\begin{equation}
    v_\varphi = u_\varphi - R\Upomega(r,\theta),
\end{equation}
the deviations to the equilibrium velocity,  we are free to choose the function $\Upomega(r,\theta)$. Indeed, a proper choice of the equilibrium velocity, $R\Upomega$, leads to a simplification of this term. we define the angular momentum fluxes as
\begin{align}
    F_{r_i} =& \int\limits_{t_1}^{t_2} r_i^2\left.\left[\overline{\phimean{\rho v_\varphi u_r}\sin\theta}-\overline{\frac{\phimean{B_\varphi B_r}\sin\theta}{4\pi}}\right]\right\vert_{r_i}\diff{t},\\
    F_{\theta_i} =& \int\limits_{t_1}^{t_2}\int\limits_{r_1}^{r_2}r^2 \sin^2\theta\left.\left[\phimean{\rho u_\theta v_\varphi}-\frac{1}{4\pi}\phimean{B_\varphi B_\theta}\right]\right\vert_{\theta_i}\diff{r}\diff{t},\\
    F_{M_r} =& \int\limits_{t_1}^{t_2}\int\limits_{r_1}^{r_2}r\overline{\phimean{\rho u_r}}\pdv{\Upomega r^2\sin^2\theta}{r}\diff{r}\diff{t},\\
    F_{M_\theta} =& \int\limits_{t_1}^{t_2}\int\limits_{r_1}^{r_2}\overline{\phimean{\rho u_\theta}}\pdv{\Upomega r^2\sin^2\theta}{\theta}\diff{r}\diff{t},
\end{align}
We can simplify the term $F_{M_\theta}$ by choosing $\Upomega = \Omega_K/\sin^2\theta$. This leads to a mathematical form similar to the one obtained in the cylindrical coordinates.
It is important to note that the function  $\Upomega(r,\theta)$ is different from the mean angular velocity of the flow $\mean{\Omega}$,
\subsection{Mass flux}
\label{sec:Mass_flux}
To get a global understanding of mass accretion and ejection, let us define the fluxes through the different disk surfaces. Using the regions defined above (\S \ref{sec:Global_picture}), we compute the fluxes through the boundaries of each regions, delimited radially by $r_1=8$ and $r_2=14$ (see figure \ref{Fig:ratio}, top). We then integrate Eq.~(\ref{Eq:Mass_Con}) and get
\begin{equation}
    \Delta M = F_{m,r_2}-F_{m,r_1}+F_{m,\theta_1}-F_{m,\theta_2},
\end{equation}
 \begin{figure}[h!] 
  \centering
  \includegraphics[width=\hsize]{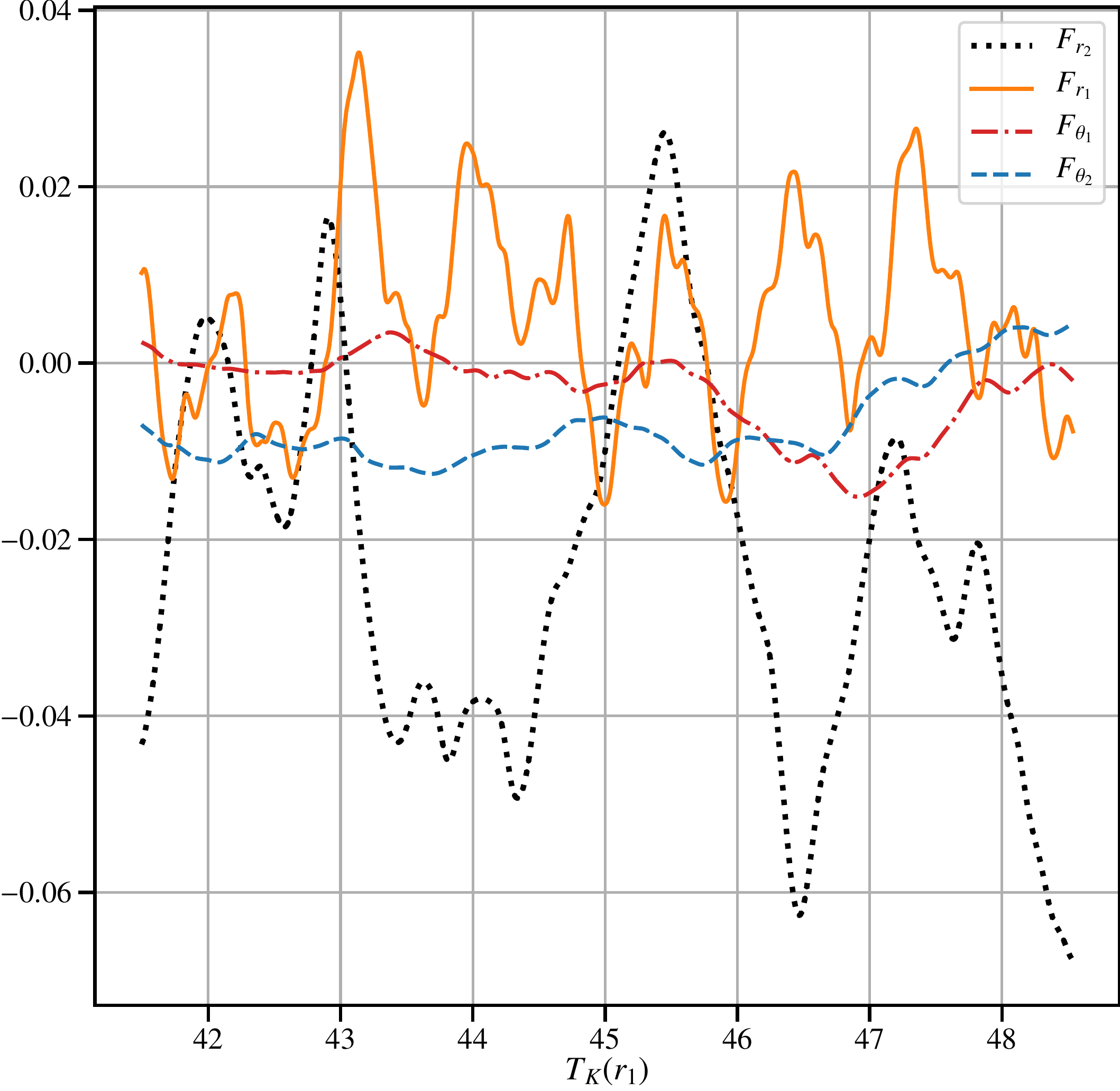}
      \caption{Mass fluxes inside the disk, defined for $|z|<3h$ as function of the local orbital time at $r_1=8$. The different fluxes are defined in appendix \ref{sec:Comp_flux} , negative fluxes corresponding to mass being removed from the disk.}
         \label{Fig:Mdisk_budget}
 \end{figure}
where the mass fluxes are time-averaged on a duration $\Delta t$ which spans several local orbital periods. A strict steady state would translate into $\Delta M = 0$.
To illustrate time-variability, we first average the mass budget on a sliding window of length $\Delta t =200/\Omega_{K}(R_{\mathrm{in}})$. We show in Figure~\ref{Fig:Mdisk_budget} the fluxes calculated within the disk region. First, we find that there is a coherent latitudinal mass flux removing mass from the disk and transferring it to the atmosphere. In addition, the radial mass flux seems quasi-periodic, with a period of the order of the local orbital period. This is most probably due to the stochastic excitation of spiral density waves by MRI-driven turbulence \citep{hein09a,hein09b}.
 \begin{figure}[h!]
  \centering
  \includegraphics[width=0.9\hsize]{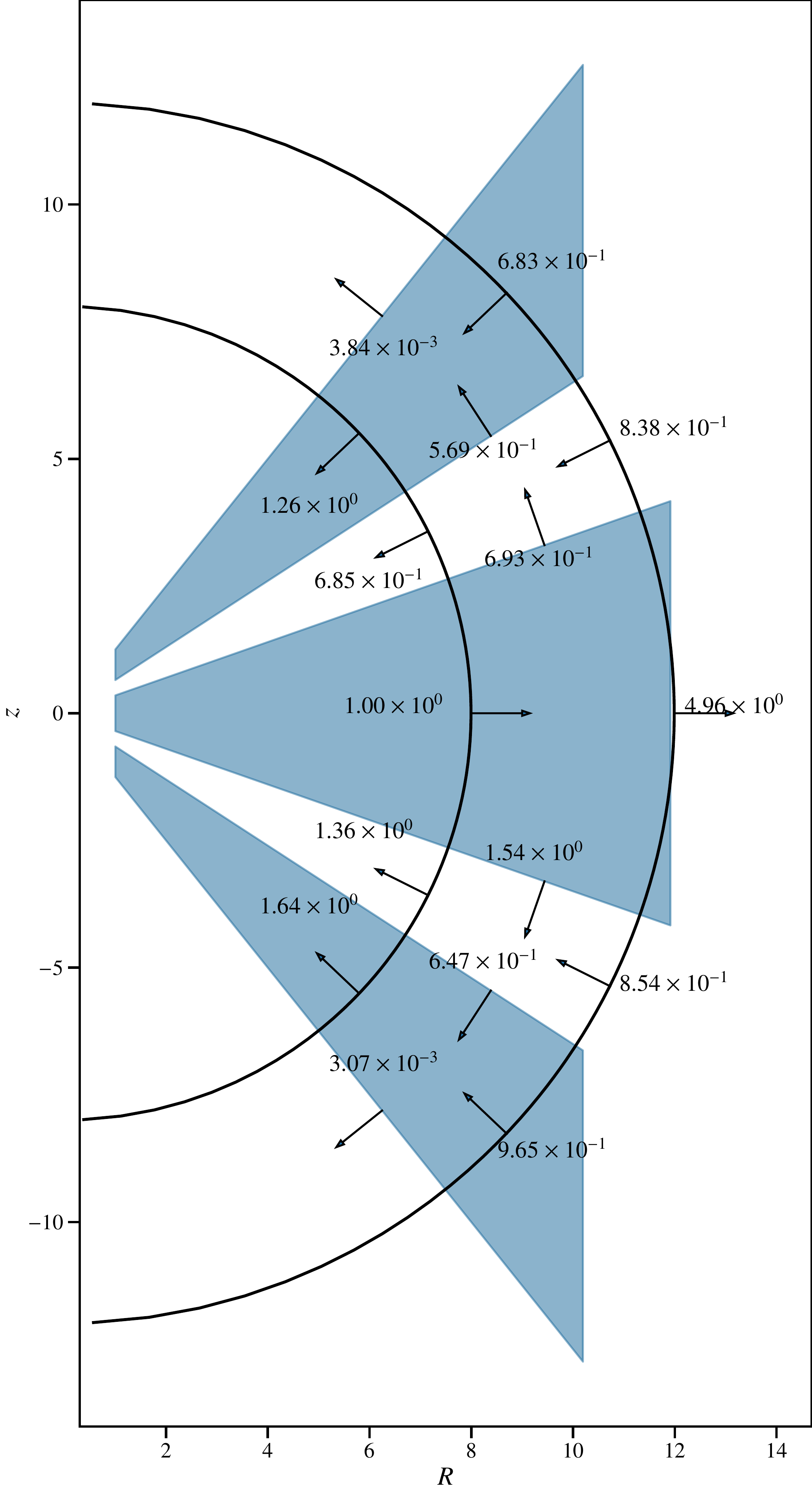}
      \caption{Mass fluxes inside of the whole domain, the regions shaded blue are turbulent regions. The different fluxes are defined in appendix \ref{sec:Comp_flux}. The arrows represent the direction of the fluxes. Since we are representing the different mass fluxes of a region of the disk, their sum (taking into account their direction) should be equal to the mass difference between $t_1$ and $t_2$.
              }
         \label{Fig:Mflux_res}
 \end{figure}
 
The large variability of the radial fluxes on orbital timescales makes their secular interpretation obtuse. To characterise the mass budget inside the disk, we therefore average the mass fluxes over the whole time domain from $t_1=923T_\mathrm{in}$ to $t_2=1114T_\mathrm{in}$ and plot them as simple scalars (Fig.~\ref{Fig:Mflux_res}). We recover the mass transfer from the disk to the atmosphere, but we now also have access to the secular radial mass flux, indicating that mass is flowing on average outwards and that the disk is decreting. Interestingly, the flux budget indicates that $\Delta M=- 6.19 <0$, implying that the disk portion is not in a stationary state and is loosing mass.
This related to the appearance of ring-like structures within the disk region, a form of secular self-organisation. The domain, between $r_1=8$ and $r_2=14$, is located on a two distinct gaps, one at $R=9.5$ and another at $R=15$. The formation of ring-like structures is discussed in section \ref{sec:ring}. 

Turning now to the regions above the disk, we find that the laminar atmosphere is accreting, but  less efficiently than the turbulent atmosphere, despite the fact that both regions receive approximately the same amount of mass from the outer radial boundary. This trend can be understood from the shape of the stream lines (Fig.~\ref{Fig:field_stream}), since the mass transported within the laminar atmosphere inevitably ends up being accreted in the turbulent atmosphere at a smaller radius. This is confirmed by Fig.~\ref{Fig:Mflux_res}  which shows that the turbulent atmosphere is being fed mass by the laminar atmosphere. Eventually, most of the mass received by the turbulent atmosphere is radially accreted, the mass ejected being almost negligible. 
The mass budget $\Delta M$ for the turbulent and laminar atmospheres is approximately null, corresponding to a steady system.

Overall, our analysis suggests that mass is being channeled from a decreting turbulent disk, through the laminar atmosphere along the field lines, and is accreted once in the turbulent atmosphere. To understand this exotic configuration, we need to look at the angular momentum transport in our domain.

\subsection{Angular momentum flux }
\label{sec:Ang_flux}
\begin{figure}[h!]
  \centering
  \includegraphics[width=0.9\hsize]{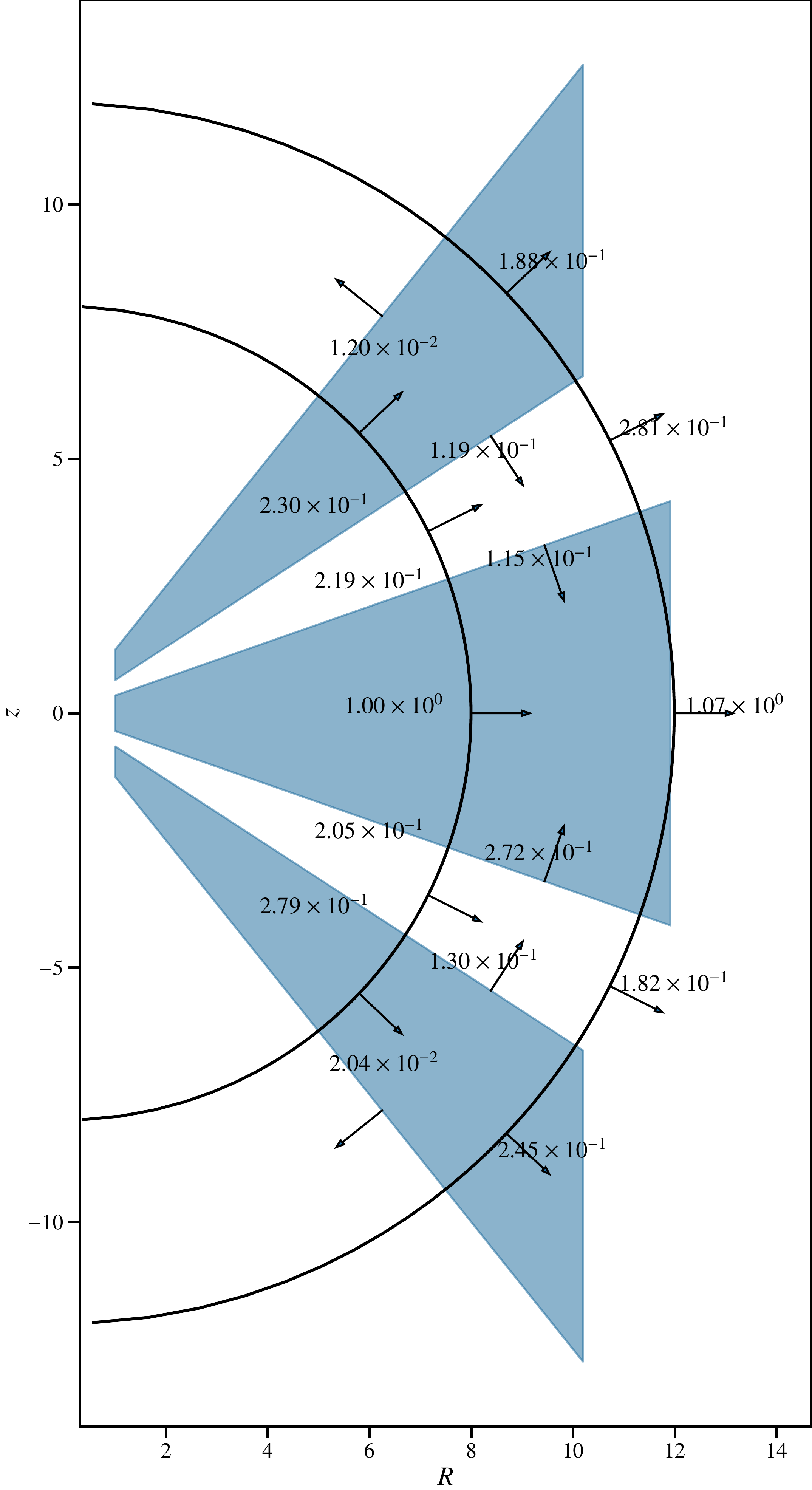}
      \caption{Angular momentum fluxes accross the whole domain, the regions shaded blue are turbulent regions. The different fluxes are defined in appendix \ref{sec:Comp_flux} and they are integrated between $t_1$ and $t_2$. The arrows represent the direction of the fluxes. In this figure we are only representing the fluxes that emerge from the Reynolds and magnetic stresses $F_{L,r_2},F_{L,r_1},F_{L,\theta_1},F_{L,\theta_2}$ not the flux of angular momentum though accretion. However, the sum of these fluxes (taking in to account their direction) leads to the value of the angular momentum being transported by the accretion.
              }
         \label{Fig:Lflux_res}
 \end{figure}

We integrate the equation of conservation of angular momentum with respect to time and space, as we did for the conservation of mass, to find
\begin{equation}
    F_{L,M_r} = F_{L,r_2}-F_{L,r_1}+F_{L,\theta_1}-F_{L,\theta_2},
\end{equation}
where the fluxes definition can be found in appendix \ref{sec:Comp_flux}. The left-hand side denotes angular momentum transported inwards by the accreting flow, while the right-hand side measure the angular momentum flux due to the torques (turbulent {and} laminar). The angular momentum fluxes are quasi-stationary when compared to the mass fluxes. Hence, we skip the examination of the time variability of these fluxes and average them across the whole time domain ($t_1=5800/\Omega_{K}(R_{\mathrm{in}})$ to $t_2=7000/\Omega_{K}(R_{\mathrm{in}})$). 
We summarise their values in figure \ref{Fig:Lflux_res}.  

We can see that the angular momentum is transported radially outwards by the torques in all of the regions. Surprisingly, in the latitudinal direction, angular momentum is transported downwards: from the atmosphere to the disk. The disk is therefore azimuthally accelerated by the atmosphere, as already seen by \cite{Zhu_Stone}. Overall, this latitudinal flux of angular momentum implies that the disk is {gaining} angular momentum, hence decretion must occur\footnote{We note that in the absence of any incoming latitudinal flux, the disk would be loosing angular momentum because of the sole radial fluxes, hence accretion would occur}.
Overall, one concludes that the latitudinal transport of angular momentum from the laminar atmosphere to the disk leads to the mass decretion within the disk found in the previous section.
Higher up, the laminar atmosphere gets its angular momentum from the inner radius and the turbulent atmosphere. Indeed, the angular momentum is transported radially outwards. However, it is important to note that in this region the transport is dominated by laminar stresses (Fig.~(\ref{Fig:ratio}).

Finally, below the wind basis, the turbulent atmosphere looses angular momentum from all sides because of the torques. The wind itself however carries less than 10\% of the angular momentum extracted from this region, so most of it is actually transported radially outwards or transferred down to the laminar atmosphere. In contrast to the laminar atmosphere however, the transport is here due to turbulent torques (Fig.~\ref{Fig:ratio}). This is be consistent with an MRI-turbulent region, as shown in section \ref{sec:dis_turb}.

\section{MRI in the low $\beta$ regime}
\label{A:MRI_lowbeta}
We rederive here in a simplified manner some of the results shown by \cite{kim2000} for the MRI in the strong field (or cold plasma) limit. Consider an ideal, isothermal plasma in differential rotation around a central object and set up a cylindrical coordinates system $(R,\varphi,z)$. Assume that the plasma is initially at equilibrium, with angular velocity $\Omega(R)$ constant on cylinders, and constant density $\rho_0$, pressure $c_s^2\rho_0$ (where $c_s$ is the isothermal sound speed) and magnetic field $\bm{B}_0$. Take a fiducial radius $R_0$ and attach a cartesian frame $(x,y,z)$ centred on $R_0$ and rotating at $\Omega=\Omega(R_0)$  so that $x$ corresponds to the local radial direction $R-R_0$ and $y$ the local azimuthal direction $\varphi$. Let us then add a small perturbation to this plasma so that $\rho=\rho_0+\delta \rho$ and $\bm{B}=\bm{B}_0+\delta b$. In order to characterise the evolution of this perturbation, we use the displacement vector field $\bm{\xi}$, for which the equation of motion in the rotating frame $(x,y,z)$ reads

\begin{align}
   \nonumber\rho_0 \frac{D^2\bm{\xi}}{Dt^2}=-c_s^2\bm{\nabla} \delta \rho-\frac{1}{4\pi}\bm{\nabla} &(\bm{B}_0\cdot \delta \bm{b})+\frac{1}{4\pi}\bm{B}_0\cdot\bm{\nabla}\delta \bm{b}\\
   &-2\rho_0 \bm{\Omega\times }\frac{D\bm{\xi}}{Dt}-2\rho_0\Omega \frac{d\Omega}{d\log R}\xi_x\bm{e}_x,
\end{align}
where we have introduced the rotation vector $\bm{\Omega}=\Omega\bm{e}_z$. One recognises in the last two terms the Coriolis force and the radial effective potential. The density and magnetic perturbations are easily deduced from (\ref{Eq:Mass_Con}) and (\ref{Eq:Induc})
\begin{align*}
\delta\rho&=-\rho_0\bm{\nabla}\cdot\bm{\xi},\\
\delta\bm{b}&=\bm{B}_0\cdot\bm{\nabla\xi}-\bm{B}_0(\bm{\nabla}\cdot\nabla{\xi}),
\end{align*}

which can be combined to the equation of motion on $\xi$ to get
\begin{align*}
    \frac{D^2\bm{\xi}}{Dt^2}& = (c_s^2+\VA^2)\bm{\nabla}(\bm{\nabla}\cdot \bm{\xi})-\bm{\nabla}\Big[\VbA\cdot\bm{\nabla}(\VbA\cdot\bm{\xi})\Big]\\
    &+\VbA\cdot\bm{\nabla}\Big[\VbA\cdot\bm{\nabla \xi}\Big]
    -\VbA\Big[\VbA\cdot\bm{\nabla}(\bm{\nabla}\cdot\bm{\xi})\Big]\\
    &-2\bm{\Omega\times }\frac{D\bm{\xi}}{Dt}+2q\Omega^2 \xi_x\bm{e}_x.
\end{align*}
In this equation, we have introduced the Alfv\'en velocity vector $\VbA\equiv\bm{B}_0/\sqrt{4\pi\rho_0}$ and $q\equiv -\mathrm{d}\log \Omega/\mathrm{d}\log R$. This equation is reminiscent of the traditional stability analysis of ideal plasmas \citep[e.g.][]{Friman60}, with the addition of inertial and gravitational forces.

Although this equation is in principle general, we simplify it by considering a background field with only toroidal and vertical components $\bm{B}_0=B_{0,\varphi}\bm{e}_y+B_{0,z}\bm{e}_z$, and perturbations with only a vertical spatial dependency. We then seek for growing (unstable) solutions of the form 
$\bm{\xi}(z,t)=\hat{\bm{\xi}}\exp(\gamma t+ik z)$. One eventually gets the dispersion relation
\begin{align}
   \nonumber c_s^2k^2\Big[\gamma^4+\big(2\VAz^2k^2+\kappa^2\big)\gamma^2+\VAz^2&k^2(\VAz^2k^2-2q\Omega^2)\Big]\\
   \nonumber +\gamma^6+\gamma^4\Big[(\VA^2&+\VAz^2)k^2+\kappa^2\Big]\\
   \label{Eq:theMeat} +\gamma^2\VA^2k^2&\left(\VAz^2k^2+4\Omega^2\frac{\VAy^2}{\VA^2}- 2q\Omega^2\right)=0.
\end{align}
where we have introduced the epicyclic frequency $\kappa^2\equiv 2\Omega^2(2-q)$. Although general solutions can be derived for this dispersion relation, they are not really enlightening. It is more instructive to look at the weakly magnetised and strongly magnetised limits. 

In the weakly magnetised case ($\VA/c_s\rightarrow 0$) the first line of (\ref{Eq:theMeat}) dominates, and we recover\footnote{We note that we also get sound waves propagating vertically from the $\gamma^6$ and $\gamma^4 c_s^2k^2$ terms. These however have no importance on the stability criterion.} the usual dispersion relation for the MRI in the incompressible limit \citep{balb91}. The stability condition is then given by the constant term: $\VAz^2k^2-2q\Omega^2<0\rightarrow\mathrm{instability}$. The usual thinking is that this stability condition implies that $\beta<1$ disks are stable for the MRI. This statement is correct provided that $1/k<H\simeq c_s/\Omega$, that is the vertical perturbation wavelength fits in the hydrostatic disk scale height. {However}, in the disk atmosphere, the vertical scale height is {not} set by the hydrostatic equilibrium, and is actually much larger than this, so longer wavelength perturbations can in principle exist and be unstable in the atmosphere.

If we now take the strong field limit $\VA/c_s\rightarrow \infty$, another picture starts to emerge. The first line of the dispersion relation then becomes negligible, and the stability condition now reads $\VAz^2k^2+4\Omega^2\VAy^2/\VA^2- 2q\Omega^2<0\rightarrow\mathrm{instability}$. Obviously, a weak enough $\VAy$ is required for the MRI to survive {independently of $k$}, in sharp contrast to the weak field case. More specifically, a necessary condition for instability is $\VAy^2/\VA^2<q/2$. Let us finally point out that in this limit, equation (\ref{Eq:theMeat}) reduces to equation (46) in \cite{kim2000}, which corresponds to their cold plasma limit.

Overall, we find that in the strongly magnetised regime, where $\VA\gg c_s$, the condition of existence for the MRI is affected by the strength of the toroidal field, a strong enough field stabilising the plasma. This is confirmed by computing the growth rate $\gamma$  in the general case from the 6th order polynomial (\ref{Eq:theMeat}). We illustrate this in Fig.~\ref{fig:growth}, which shows the growth rate (maximised on $k$) as a function of the field strength $\VAy/\VAz$ for the Keplerian case $q=3/2$. We believe this is the main explanation to why turbulence disappears from the laminar atmosphere and then reappears once $\VAy/\VAz$ is sufficiently low.

\section{Transport of Magnetic field and magnetic flux}
\label{A:all_Mag_flux}
 \begin{figure*}
    \centering{
   \includegraphics[width=0.495\hsize]{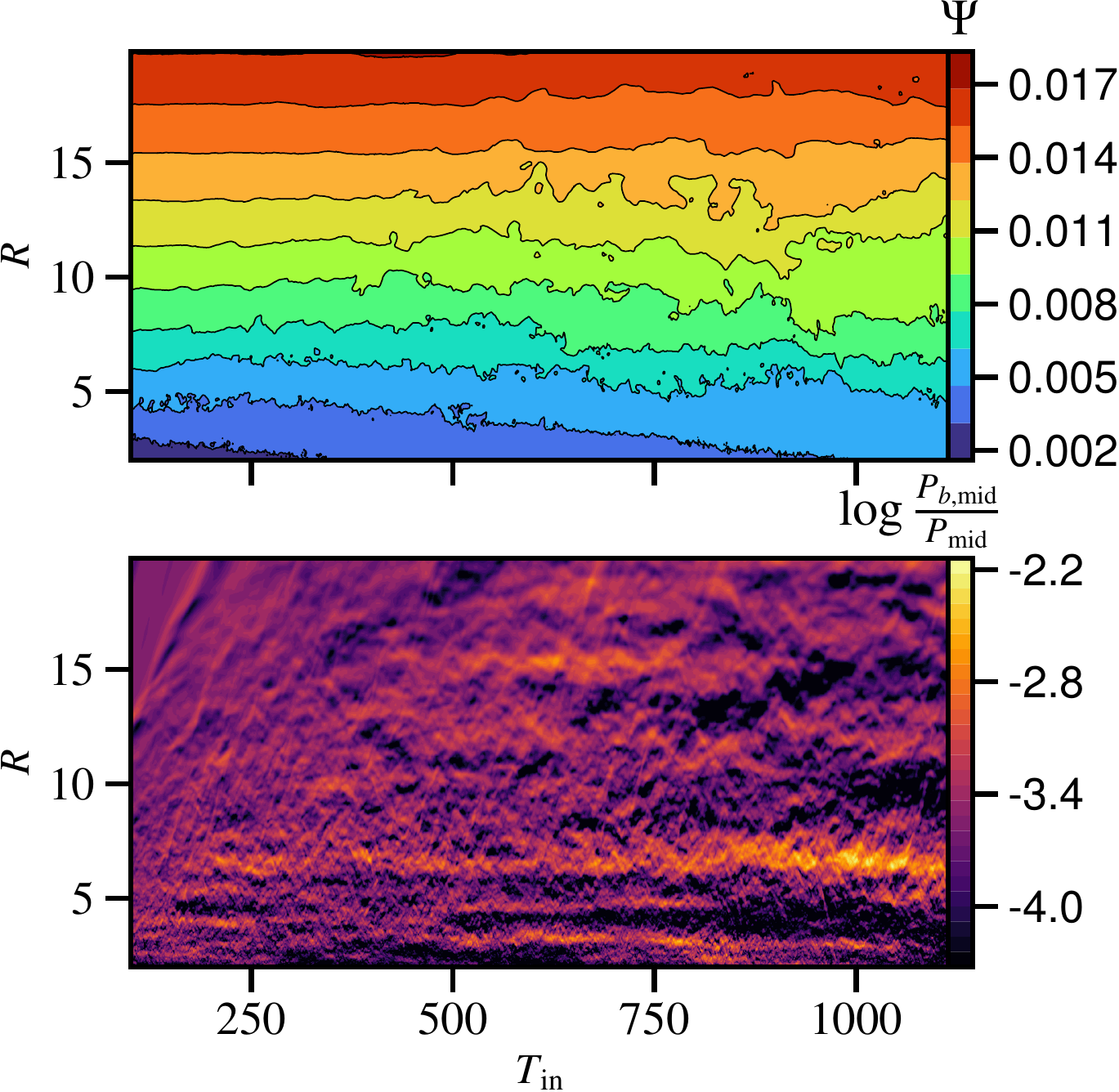}
   \includegraphics[width=0.485\hsize]{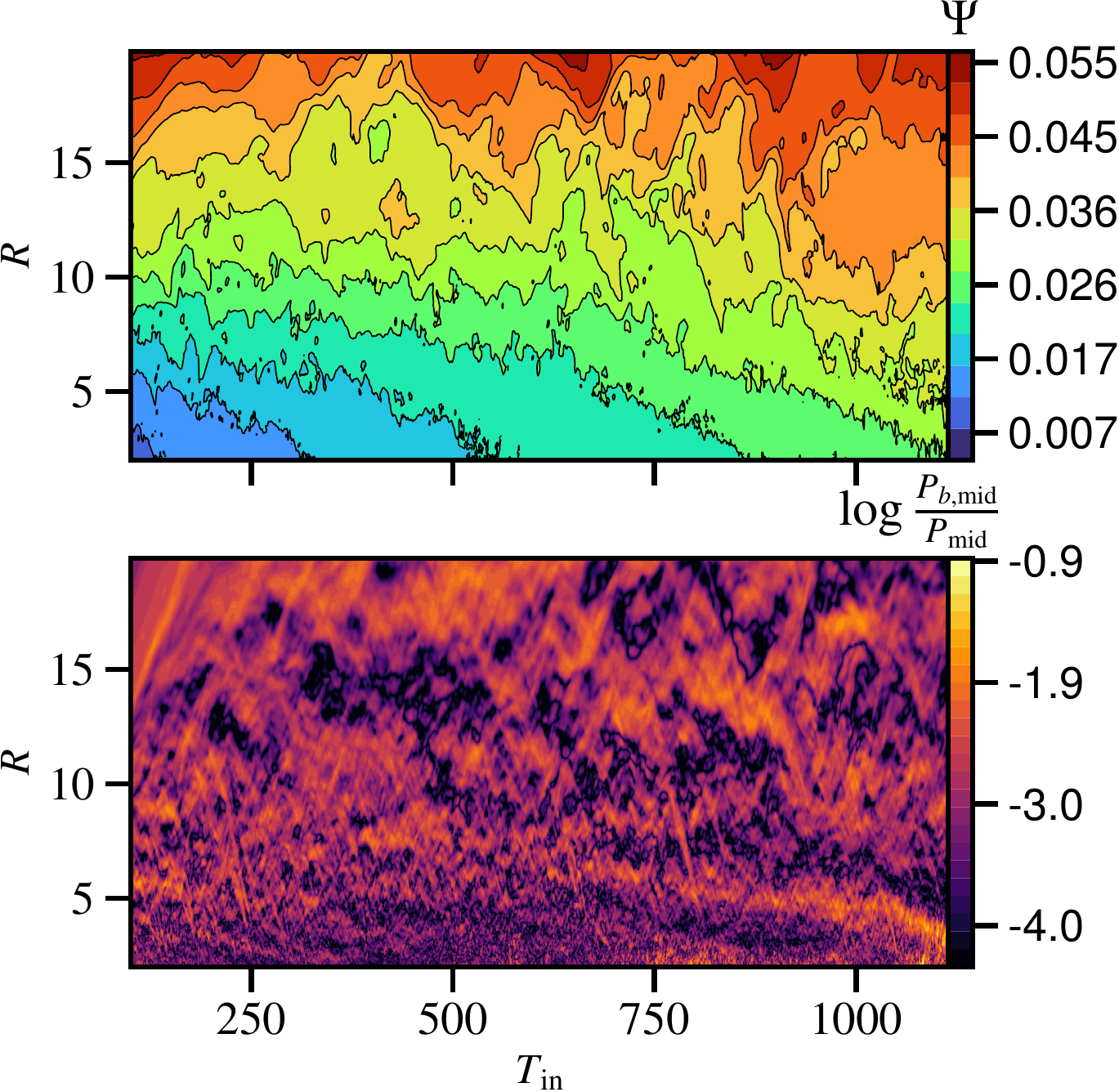}\\
   \includegraphics[width=0.495\hsize]{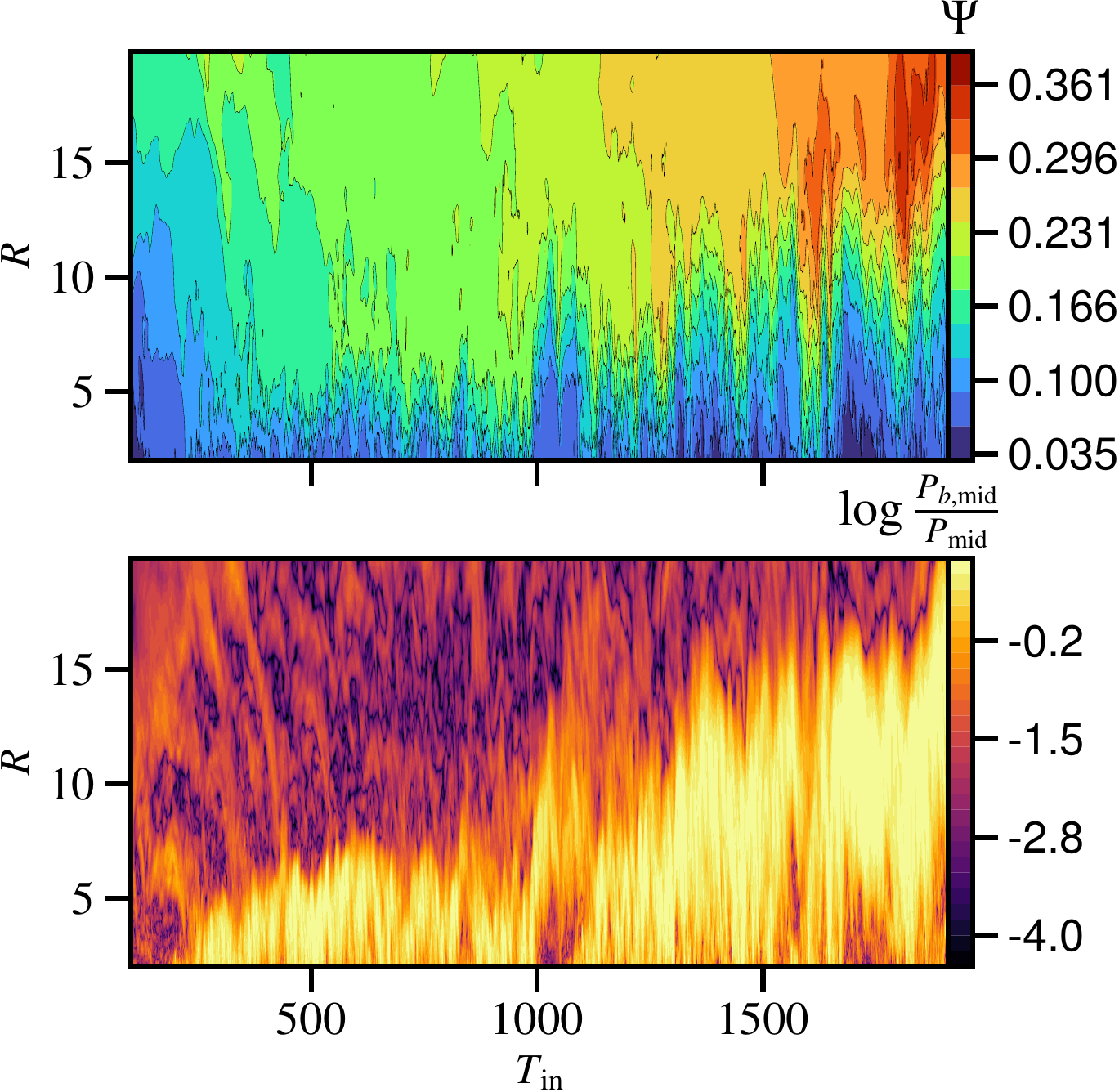}
   \includegraphics[width=0.485\hsize]{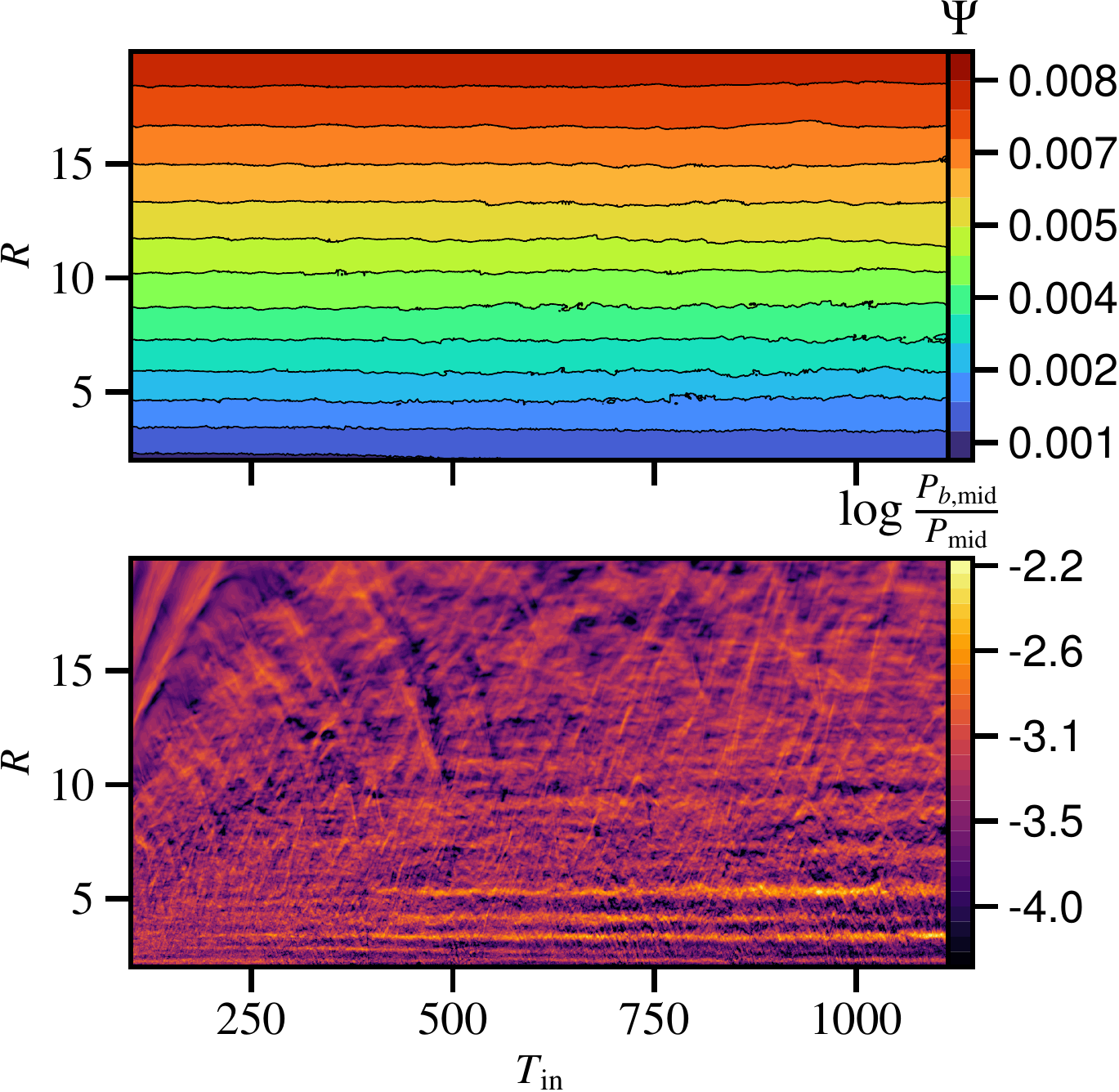}}
      \caption{Magnetic field flux $\Psi$ defined in Eq.~(\ref{eq:Psi_def}) and ratio $P_{b,\mathrm{mid}}/P_{\mathrm{mid}}$ (see text) as functions of time and the radial coordinate for the different simulations: SB4 (top left), SB3 (top right), SB2 (bottom left), and SEp (bottom right).  
      We note that $P_{b,\mathrm{mid}}/P_{\mathrm{mid}} > \mean{\beta_\mathrm{mid}}^{-1}$ since the former contains also the turbulent magnetic pressure.
              }
         \label{Fig:Psi_all}
 \end{figure*}
We show in Fig.~(\ref{Fig:Psi_all}) for various simulations the space-time diagrams of the magnetic flux $\Psi$ (defined in section \ref{sec:transport}) and, for practical reasons, the ratio $P_{b,\mathrm{mid}}/P_{\mathrm{mid}}$, where
\begin{align}
    P_{b,\mathrm{mid}} &=\frac{1}{8\pi}\left[\frac{1}{2h_\mathrm{disk}} \int\limits_{\theta_{d1}}^{\theta_{d2}}r \sin\theta \,\mean{B_z}_\varphi\diff{\theta}\right]^2,\\
    P_{\mathrm{mid}} &= \frac{1}{2h_\mathrm{disk}} \int\limits_{\theta_{d1}}^{\theta_{d2}}r \sin\theta \,\mean{P}_\varphi\diff{\theta},
\end{align}
and $\theta_{d2,d1} =\pi/2 \pm \arctan(h_\mathrm{disk}/R)$. It is important however to note that the ratio $P_{b,\mathrm{mid}}/P_{\mathrm{mid}}$ is larger than $\beta_\mathrm{mid}^{-1}$ since it contains the turbulent field contribution $\mean{\delta B^2}$. 

While the turbulent magnetic pressure measured at the mid plane in weakly magnetised disks remains always sub-thermal (see Fig. \ref{Fig:pressure}), it becomes supra-thermal at large magnetisations. For instance, in the case of a disk with $\mean{\beta_\mathrm{mid}}^{-1} \simeq 1$, we measure $\mean{\delta B^2}(r,\theta=\pi/2)/8\pi\sim 10 P_0$. Thus, the yellow disk region shown in the lower left panel in Fig.~(\ref{Fig:Psi_all}), which is associated with the SB2 strong field simulation, corresponds in fact to a region where the disk magnetisation $\mean{\beta_\mathrm{mid}}^{-1}$ is around unity.

\end{appendix}

\end{document}